\documentclass[twoside,11pt]{report}

\usepackage{times}
\usepackage{epsfig}
\usepackage{amsmath}
\usepackage{amssymb}
\usepackage{latexsym}
\usepackage{pstricks}
\usepackage{my_jeep}
\usepackage{a4wide}    
\usepackage{my_fancyheadings}

\begin{document}

\marginparwidth 1.5cm
\setlength{\hoffset}{-1cm}
\newcommand{\mpar}[1]{{\marginpar{\hbadness10000%
                      \sloppy\hfuzz10pt\boldmath\bf\footnotesize#1}}%
                      \typeout{marginpar: #1}\ignorespaces}
\def\mda{\mpar{\hfil$\downarrow$\hfil}\ignorespaces}
\def\mua{\mpar{\hfil$\uparrow$\hfil}\ignorespaces}
\def\mla{\marginpar[\boldmath\hfil$\rightarrow$\hfil]%
                   {\boldmath\hfil$\leftarrow $\hfil}%
                    \typeout{marginpar: $\leftrightarrow$}\ignorespaces}

\providecommand{\SP}{\scriptscriptstyle}

\providecommand{\stl}{\tilde{t}_{\SP L}}
\providecommand{\str}{\tilde{t}_{\SP R}}
\providecommand{\ste}{\tilde{t}_1}
\providecommand{\stz}{\tilde{t}_2}
\providecommand{\sti}{\tilde{t}_i}
\providecommand{\stj}{\tilde{t}_j}

\providecommand{\stez}{\tilde{t}_{1,2}}

\providecommand{\sq}{\tilde{q}}
\providecommand{\sql}{\tilde{q}_{\SP L}}
\providecommand{\sqr}{\tilde{q}_{\SP R}}
\providecommand{\sqlr}{\tilde{q}_{\SP LR}}
\providecommand{\st}{\tilde{t}}
\providecommand{\gt}{\tilde{g}}
\providecommand{\sbx}{\tilde{b}}

\providecommand{\tb}{\bar{t}}
\providecommand{\sqb}{\bar{\tilde{q}}}
\providecommand{\stb}{\bar{\tilde{t}}}
\providecommand{\steb}{\bar{\tilde{t}}_1}
\providecommand{\stzb}{\bar{\tilde{t}}_2}
\providecommand{\stib}{\bar{\tilde{t}}_i}
\providecommand{\stjb}{\bar{\tilde{t}}_j}

\providecommand{\sten}{\tilde{t}_{1 0}}
\providecommand{\stzn}{\tilde{t}_{2 0}}
\providecommand{\stln}{\tilde{t}_{{\SP L} 0}}
\providecommand{\strn}{\tilde{t}_{{\SP R} 0}}
\providecommand{\stin}{\tilde{t}_{i 0}}
\providecommand{\stjn}{\tilde{t}_{j 0}}

\providecommand{\NX}{{\mathfrak N}}
\providecommand{\NY}{{\mathfrak N'}}

\providecommand{\nni}{\tilde{\chi}_i^0}
\providecommand{\nnj}{\tilde{\chi}_j^0}
\providecommand{\nne}{\tilde{\chi}_1^0}
\providecommand{\nnz}{\tilde{\chi}_2^0}
\providecommand{\nnd}{\tilde{\chi}_3^0}
\providecommand{\nnv}{\tilde{\chi}_4^0}
\providecommand{\cpi}{\tilde{\chi}_i^+}
\providecommand{\cpj}{\tilde{\chi}_j^+}
\providecommand{\cpe}{\tilde{\chi}_1^+}
\providecommand{\cpz}{\tilde{\chi}_2^+}
\providecommand{\cmi}{\tilde{\chi}_i^-}
\providecommand{\cmj}{\tilde{\chi}_j^-}
\providecommand{\cme}{\tilde{\chi}_1^-}
\providecommand{\cmz}{\tilde{\chi}_2^-}

\providecommand{\mse}{m_{\tilde{t}_{\SP 1}}}
\providecommand{\msz}{m_{\tilde{t}_{\SP 2}}}
\providecommand{\msi}{m_{\tilde{t}_{\SP i}}}
\providecommand{\msj}{m_{\tilde{t}_{\SP j}}}
\providecommand{\mni}{m_{\tilde{\chi}_{\SP i}}}
\providecommand{\mnj}{m_{\tilde{\chi}_{\SP j}}}

\providecommand{\msez}{m_{\tilde{t}_{\SP 1,2}}}
\providecommand{\msen}{m_{\tilde{t}_{\SP 1 0}}}
\providecommand{\mszn}{m_{\tilde{t}_{\SP 2 0}}}
\providecommand{\msin}{m_{\tilde{t}_{\SP i 0}}}
\providecommand{\msjn}{m_{\tilde{t}_{\SP j 0}}}

\providecommand{\mni}{m_{\tilde{\chi}_{\SP i}}}
\providecommand{\mnj}{m_{\tilde{\chi}_{\SP j}}}
\providecommand{\mne}{m_{\tilde{\chi}_{\SP 1}^{\SP 0}}}
\providecommand{\mnz}{m_{\tilde{\chi}_{\SP 2}^{\SP 0}}}
\providecommand{\mnd}{m_{\tilde{\chi}_{\SP 3}^{\SP 0}}}
\providecommand{\mnv}{m_{\tilde{\chi}_{\SP 4}^{\SP 0}}}
\providecommand{\mce}{m_{\tilde{\chi}_{\SP 1}^{\SP +}}}
\providecommand{\mcz}{m_{\tilde{\chi}_{\SP 2}^{\SP +}}}

\providecommand{\mst}{m_{\tilde{t}}}
\providecommand{\mn}{m_{\tilde{\chi}}}
\providecommand{\mg}{m_{\tilde{g}}}
\providecommand{\ms}{m_{\tilde{q}}}
\providecommand{\mt}{m_t}
\providecommand{\mz}{m_{\SP Z}}
\providecommand{\mw}{m_{\SP W}}
\providecommand{\mbin}{m_{\SP {\tilde{B}}}}
\providecommand{\mwin}{m_{\SP {\tilde{W}}}}
\providecommand{\ml}{\lambda}

\providecommand{\tmixn}{\tilde{\theta}_0}
\providecommand{\tmix}{\tilde{\theta}}
\providecommand{\set}{s_{\tmix}}
\providecommand{\cet}{c_{\tmix}}
\providecommand{\szt}{s_{2\tmix}}
\providecommand{\czt}{c_{2\tmix}}
\providecommand{\svt}{s_{4\tmix}}
\providecommand{\cztq}{c^2_{2\tmix}}
\providecommand{\sztq}{s^2_{2\tmix}}

\providecommand{\br}{{\rm BR}}
\providecommand{\gev}{{\rm \, GeV}}
\providecommand{\tev}{{\rm \, TeV}}
\providecommand{\pb}{{\rm pb}}
\providecommand{\fb}{{\rm fb}}
\providecommand{\order}{{\cal O}}
\providecommand{\ZM}{Z^{1/2}}
\providecommand{\MM}{{\cal M}^2}
\providecommand{\M}{{\cal M}}
\providecommand{\D}{{\cal D}}
\providecommand{\V}{{\cal V}}
\providecommand{\W}{{\cal W}}
\providecommand{\R}{{\cal R}}

\providecommand{\PX}{{\cal P}_{12}}

\providecommand{\as}{\alpha_s}
\providecommand{\swz}{\sin^2 \theta_W} 
\providecommand{\msbar}{\overline{{\rm MS}}}
\providecommand{\drbar}{\overline{{\rm DR}}}
\providecommand{\ee}{\epsilon}
\providecommand{\fac}{{\cal N}}

\providecommand{\Bp}{\dot{B}}
\providecommand{\real}{\,\text{Re\,}}
\providecommand{\ntp}{\negthickspace}

\providecommand{\xzt}{\sigma_{2\tmix}}
\providecommand{\mugte}{\mu_{\tilde{g} t 1}}
\providecommand{\mugtz}{\mu_{\tilde{g} t 2}}
\providecommand{\muteg}{\mu_{t 1 \tilde{g}}}
\providecommand{\mutzg}{\mu_{t 2 \tilde{g}}}
\providecommand{\muget}{\mu_{\tilde{g} 1 t}}
\providecommand{\muegt}{\mu_{1 \tilde{g} t}}
\providecommand{\muzgt}{\mu_{2 \tilde{g} t}}
\providecommand{\MB}{|{\cal M}_B|^2}
\providecommand{\pg}{p_{\tilde{g}}}
\providecommand{\pt}{p_t}
\providecommand{\ps}{p_{\tilde{t}_{\SP 1}}}

\providecommand{\Ac}{{A^c}}


\providecommand{\eg}{{\it e.g.}\;}
\providecommand{\ie}{{\it i.e.}\;}
\providecommand{\etal}{{\it et al.}\;}
\providecommand{\ibid}{{\it ibid.}\;}

\newcommand{\zpc}[3]{${\rm Z. Phys.}$ {\bf C#1} (19#2) #3}
\newcommand{\npb}[3]{${\rm Nucl. Phys.}$ {\bf B#1} (19#2)~#3}
\newcommand{\plb}[3]{${\rm Phys. Lett.}$ {\bf B#1} (19#2) #3}
\newcommand{\prd}[3]{${\rm Phys. Rev.}$ {\bf D#1} (19#2) #3}
\newcommand{\prl}[3]{${\rm Phys. Rev. Lett.}$ {\bf #1} (19#2) #3}
\newcommand{\prep}[3]{${\rm Phys. Rep.}$ {\bf #1} (19#2) #3}
\newcommand{\fp}[3]{${\rm Fortschr. Phys.}$ {\bf #1} (19#2) #3}
\newcommand{\nc}[3]{${\rm Nuovo Cimento}$ {\bf #1} (19#2) #3}
\newcommand{\ijmp}[3]{${\rm Int. J. Mod. Phys.}$ {\bf #1} (19#2) #3}
\newcommand{\jcp}[3]{${\rm J. Comp. Phys.}$ {\bf #1} (19#2) #3}
\newcommand{\ptp}[3]{${\rm Prog. Theo. Phys.}$ {\bf #1} (19#2) #3}
\newcommand{\sjnp}[3]{${\rm Sov. J. Nucl. Phys.}$ {\bf #1} (19#2) #3}
\newcommand{\cpc}[3]{${\rm Comp. Phys. Commun.}$ {\bf #1} (19#2) #3}
\newcommand{\mpl}[3]{${\rm Mod. Phys. Lett.}$ {\bf #1} (19#2) #3}
\newcommand{\cmp}[3]{${\rm Commun. Math. Phys.}$ {\bf #1} (19#2) #3}
\newcommand{\jmp}[3]{${\rm J. Math. Phys.}$ {\bf #1} (19#2) #3}
\newcommand{\nim}[3]{${\rm Nucl. Instr. Meth.}$ {\bf #1} (19#2) #3}
\newcommand{\prev}[3]{${\rm Phys. Rev.}$ {\bf #1} (19#2) #3}
\newcommand{\el}[3]{${\rm Europhysics Letters}$ {\bf #1} (19#2) #3}
\newcommand{\ap}[3]{${\rm Ann. of~Phys.}$ {\bf #1} (19#2) #3}
\newcommand{\jetp}[3]{${\rm JETP}$ {\bf #1} (19#2) #3}
\newcommand{\jetpl}[3]{${\rm JETP Lett.}$ {\bf #1} (19#2) #3}
\newcommand{\acpp}[3]{${\rm Acta Physica Polonica}$ {\bf #1} (19#2) #3}
\newcommand{\science}[3]{${\rm Science}$ {\bf #1} (19#2) #3}
\newcommand{\vj}[4]{${\rm #1~}$ {\bf #2} (19#3) #4}
\newcommand{\ej}[3]{${\bf #1}$ (19#2) #3}
\newcommand{\vjs}[2]{${\rm #1~}$ {\bf #2}}
\newcommand{\hep}[1]{${\tt hep\!-\!ph/}$ {#1}}
\newcommand{\hex}[1]{${\tt hep\!-\!ex/}$ {#1}}
\newcommand{\desy}[1]{${\rm DESY-}${#1}}

\def\openone{\leavevmode\hbox{\small1\kern-3.8pt\normalsize1}}%

\def\slpa{\slash{\pa}}                      
\def\slin{\SLLash{\in}}                     
\def\bo{{\raise-.3ex\hbox{\large$\Box$}}}   
\def\cbo{\Sc [}                             
\def\pa{\partial}                           
\def\de{\nabla}                             
\def\dell{\bigtriangledown}                 
\def\su{\sum}                               
\def\pr{\prod}                              
\def\iff{\leftrightarrow}                   
\def\conj{{\hbox{\large *}}}                
\def\ltap{\raisebox{-.4ex}{\rlap{$\sim$}} \raisebox{.4ex}{$<$}} 
\def\gtap{\raisebox{-.4ex}{\rlap{$\sim$}} \raisebox{.4ex}{$>$}} 
\def\face{{\raise.2ex\hbox{$\displaystyle \bigodot$}\mskip-2.2mu \llap {$\ddot\smile$}}}                            
\def\dg{\dagger}                            
\def\ddg{\ddagger}                          


\def\sp#1{{}^{#1}}                           
\def\sb#1{{}_{#1}}                           
\def\oldsl#1{\rlap/#1}                       
\def\slash#1{\rlap{\hbox{$\mskip 1 mu /$}}#1}
\def\Slash#1{\rlap{\hbox{$\mskip 3 mu /$}}#1}
\def\SLash#1{\rlap{\hbox{$\mskip 4.5 mu /$}}#1}  
\def\SLLash#1{\rlap{\hbox{$\mskip 6 mu /$}}#1}   
\def\wt#1{\widetilde{#1}}                    
\def\Hat#1{\widehat{#1}}                     
\def\Bar#1{\overline{#1}}                    
\def\bra#1{\left\langle #1\right|}           
\def\ket#1{\left| #1\right\rangle}           
\def\VEV#1{\left\langle #1\right\rangle}     
\def\abs#1{\left| #1\right|}                 
\def\leftrightarrowfill{$\mathsurround=0pt \mathord\leftarrow \mkern-6mu
        \cleaders\hbox{$\mkern-2mu \mathord- \mkern-2mu$}\hfill
        \mkern-6mu \mathord\rightarrow$}       
\def\dvec#1{\vbox{\ialign{##\crcr
        \leftrightarrowfill\crcr\noalign{\kern-1pt\nointerlineskip}
        $\hfil\displaystyle{#1}\hfil$\crcr}}}       
\def\dt#1{{\buildrel {\hbox{\LARGE .}} \over {#1}}} 
\def\dtt#1{{\buildrel \bullet \over {#1}}}          
\def\der#1{{\pa \over \pa {#1}}}                
\def\fder#1{{\d \over \d {#1}}}                 


\def\partder#1#2{{\partial #1\over\partial #2}} 
\def\parvar#1#2{{\d #1\over \d #2}}             
\def\secder#1#2#3{{\partial^2 #1\over\partial #2 \partial #3}}
\def\on#1#2{\mathop{\null#2}\limits^{#1}}       
\def\bvec#1{\on\leftarrow{#1}}                  
\def\oover#1{\on\circ{#1}}                      


\catcode`@=11 \def\citer{\@ifnextchar
  [{\@tempswatrue\@citexr}{\@tempswafalse\@citexr[]}}

\def\@citexr[#1]#2{\if@filesw\immediate\write\@auxout{\string\citation{#2}}\fi
  \def\@citea{}\@cite{\@for\@citeb:=#2\do
    {\@citea\def\@citea{--\penalty\@m}\@ifundefined {b@\@citeb}{{\bf
          ?}\@warning
       {Citation `\@citeb' on page \thepage \space undefined}}%
\hbox{\csname b@\@citeb\endcsname}}}{#1}}
\catcode`@=12

\pagestyle{empty}
\begin{titlepage}

\begin{center}
\vspace*{0.5cm}
\huge\sc
 Production of\\ 
 Supersymmetric Particles\\
 at High-Energy Colliders\\
\vfill
\large\rm
 Dissertation \\
 zur Erlangung des Doktorgrades\\
 des Fachbereichs Physik\\
 der Universit\"at Hamburg\\
\vfill
 vorgelegt von \\
 Tilman Plehn\\
 aus Siegen\\
\vfill
 Hamburg\\
 1998
\vspace*{1cm}
\end{center}

\newpage 

\vspace*{13cm}
\vfill
\small 

\begin{tabular}{ll}
\begin{minipage}[t]{6.0cm}
Gutachter der Dissertation:
\end{minipage}  & 
\begin{minipage}[t]{5.0cm}
Prof.~Dr.~P.M.~Zerwas\\
Prof.~Dr.~J.~Bartels\\
\end{minipage} \\
\bigskip
\begin{minipage}[t]{6.0cm}
Gutachter der Disputation:
\end{minipage} & 
\begin{minipage}[t]{5.0cm}
Prof.~Dr.~P.M.~Zerwas\\
Prof.~Dr.~G.~Kramer\\
\end{minipage} \\
\bigskip
\begin{minipage}[t]{6.0cm}
Datum der Disputation:
\end{minipage} & 
\begin{minipage}[t]{5.0cm}
13.~Juli 1998
\end{minipage} \\
\bigskip
\begin{minipage}[t]{6.0cm}
Sprecher des Fachbereichs Physik \\
und Vorsitzender des \\
Promotionsausschusses:
\end{minipage} & 
\begin{minipage}[t]{5.0cm}
${}$ \\
${}$ \\
Prof.~Dr.~B.~Kramer\\
\end{minipage} 
\end{tabular}

\newpage
\normalsize

\begin{center}
\begin{minipage}[t]{15cm}
  {\begin{center}\sc{Abstract}\end{center}} 
  
  The production and decay of supersymmetric particles is presented in
  this thesis. The search for light mixing top squarks and
  neutralinos/charginos will be a major task at the upgraded Tevatron
  and the LHC as well as at a future $e^+e^-$ linear collider. The
  dependence of the hadro-production cross section for weakly and
  strongly interacting particles on the renormalization and
  factorization scales is found to be weak in next-to-leading order
  supersymmetric QCD. This yields an improvement of derived mass
  bounds or the measurement of the masses, respectively, of
  neutralinos/charginos and stops at the Tevatron and at the LHC.
  Moreover, the next-to-leading order corrections increase
  the predicted neutralino/chargino cross section by +20$\%$ to
  +40$\%$, nearly independent of the mass of the particles. The
  consistent treatment as well as the phenomenological implications of
  scalar top mixing are presented.  The corrections to strong and weak
  coupling induced decays of stops, gluinos, and heavy neutralinos
  [including mixing stop particles] are strongly dependent on the
  parameters chosen. The decay widths are defined in a renormalization
  scheme for the mixing angle, which maintains the symmetry between
  the two stop states to all orders.  The correction to the stop
  production cross sections, depending on the fraction of incoming
  quarks and gluons, varies between --10$\%$ to +40$\%$ for an
  increasing fraction of incoming gluons. The dependence of the cross
  section on all parameters, except for the masses of the produced
  particles, is in contrast to the light-flavor squark case
  negligible. The calculation of the stop production cross section was
  applicable to the Tevatron search for particles which could be
  responsible for the HERA anomaly.
  \\[10ex]

  {\begin{center}\sc{Zusammenfassung}\end{center}}
  
  In dieser Arbeit wird die Produktion und der Zerfall
  supersymmetrischer Teilchen betrachtet. Die Suche nach leichten
  mischenden top--Squarks und Neutralinos/Charginos ist eine der
  wichtigsten Aufgaben am Tevatron und LHC ebenso wie an einem
  zuk\"unftigen $e^+e^-$--Linearbeschleuniger. Die Abh\"angigkeit der
  Wirkungsquerschnitte f\"ur die Produktion stark und schwach
  koppelnder Teilchen von der Renormierungs-- und
  Faktorisierungsskala ist in n\"achstf\"uhrender Ordnung reduziert.
  Dies erlaubt verbesserte Massenschranken oder eine verbesserte
  Massenbestimmung f\"ur Neutralinos/Charginos und Stops am Tevatron
  und am LHC. Dar\"uber hinaus vergr\"o{\ss}ern die Korrekturen den
  vorhergesagten Wirkungsquerschnitt f\"ur Neutralinos/Charginos um
  +20$\%$ bis +40$\%$, nahezu unabh\"angig von der Masse der
  produzierten Teilchen.  Weiterhin wird eine konsistente Behandlung
  der Stop--Mischung vorgestellt und deren ph\"anomenologische
  Konsequenzen untersucht.  Die Korrekturen zu starken und schwachen
  Zerf\"allen von top--Squarks, Gluinos und schweren Neutralinos ---
  insofern sie top--Squarks ent\-halten --- h\"angen signifikant von den
  gew\"ahlten Parametern ab. Die Zerfallsbreiten enthalten die
  Definition des Stop--Mischungswinkels, welche die Symmetrie zwischen
  den beiden Stop--Zust\"anden in beliebiger Ordnung St\"orungstheorie
  erh\"alt.  Die Korrekturen zu den Wirkungsquerschnitten f\"ur
  Stop--Produktion variieren zwischen --10$\%$ und +40$\%$ und wachsen
  mit dem Anteil einlaufender Gluonen gegen\"uber Quarks. Im Gegensatz
  zur Produktion massenentarteter Squarks ist die Abh\"angigkeit
  von Parametern \"uber die Stop--Masse hinaus vernachl\"assigbar. Die
  Berechnung des Wirkungsquerschitts f\"ur die Stop--Produktion konnte
  am Tevatron auf die Suche nach Teilchen, die f\"ur die
  HERA--Anomalie verantwortlich sein k\"onnten, angewandt werden.
 
\end{minipage}
\end{center}

\newpage

\setlength{\parindent}{0pt}
\vspace*{3cm}

Vernunft und Wissenschaft gehen oft verschiedene Wege\smallskip 


\vspace{1cm}
\hspace{5cm} Paul K. Feyerabend, Wider den Methodenzwang

\setlength{\parindent}{16pt}

\end{titlepage}

\pagenumbering{roman}
\tableofcontents
\cleardoublepage

\pagestyle{fancyplain}
\pagenumbering{arabic}
\setlength{\parindent}{16pt}
\chapter*{Introduction}
\addcontentsline{toc}{chapter}{Introduction}

A fundamental element of particle physics are symmetry principles.
The electroweak as well as the strong interaction, combined to the
Standard Model, are based on the gauge symmetry group
SU(3)$\times$SU(2)$\times$U(1). The extension~\cite{sohnius} of this
concept to a theory incorporating global or local supersymmetry is a
well-motivated step for several reasons:\bigskip

The Standard Model has been well-established by the discovery of the
gluon and the weak gauge bosons, and by precision measurements at LEP
and at the Tevatron, as well as at HERA.  Currently, there is no
experimental compulsion to modify the Standard Model at energy scales
accessible to these colliders, provided the predicted Higgs boson will
be found at LEP or at a future hadron or electron collider.  However,
a set of conceptual problems cannot be solved in the Standard Model
framework: The mass of the only fundamental scalar particle, the Higgs
boson, is not stable under quantum fluctuations, \ie loop
contributions to the Higgs mass term become large at high scales and
have to be absorbed into the counter terms for the physical Higgs
mass.  This hierarchy problem leads to fine tuning of the parameters
in the Higgs potential, to avoid the breakdown of perturbative weak
symmetry breaking.

Possible grand unification scenarios are based on a gauge group at
some high unification scale, which contains the different Standard
Model gauge groups. Simple unification groups are the
SU(5)~\cite{susy_su5} or SO(10)~\cite{susy_so10}, the latter favored
in scenarios with massive neutrinos. Non-minimal scenarios may yield
intermediate symmetries and threshold effects, but as long as they
include a simple unifying gauge group, the three running Standard
Model couplings have to meet in one point at the unification scale.
The requirement of one unification point and additional bounds from
the non-observation of the proton decay lead to difficulties in the
Standard Model, when it is embedded into a grand desert scenario, and
most likely restrict the validity of the Standard Model to scales 
around the weak scale.\bigskip

In supersymmetric extensions of the Standard Model the masses of
scalar particles remain stable even for very large scales, as required
by grand unification scenarios. Quantum fluctuations due to fermions
and bosons cancel each other; the leading singularities also vanish in
softly broken supersymmetric theories. The hierarchy problem does
therefore not occur in the extended supersymmetric Higgs sector.
Including an intermediate supersymmetry breaking scale, the minimal
supersymmetric extension of the Standard Model may be valid up to a
grand unification scale without any fine tuning, being compatible with
grand desert unification scenarios. Given the strong and the Fermi
coupling constant at low scales, it predicts the weak mixing angle in
very good agreement with the measured value~\cite{thetaw_mssm}. For a
large top quark mass the renormalization group evolution can drive the
electroweak symmetry breaking at low scales. The minimal
supersymmetric Higgs sector consists of two doublets, in order to give
masses to up and down type quarks while preserving supersymmetry and
gauge invariance. Hence, after breaking the weak gauge symmetry, five
physical Higgs bosons occur. The non-diagonal CP even current
eigenstates yield a light scalar Higgs boson with a strong theoretical
upper bound on its mass.  In some regimes of the supersymmetric
parameter space this particle is accessible to LEP2, and the
dependence of the theoretical mass bound on low-energy supersymmetry
parameters can be used to constrain the fundamental mixing parameter
$\tan\beta$.

In supersymmetric $R$ parity conserving models the lightest
supersymmetric particle is stable. This LSP, which in many scenarios
turns out to be the lightest neutralino, is a possible candidate for
cosmological cold dark matter.\medskip

In analogy to the gauge symmetries one may extend the global to a
local supersymmetry. This invariance gives rise to higher spin states
in the Lagrangean: a massless spin-2 graviton field and its spin-3/2
gravitino partner appear~\cite{sugra}. The general
Einstein-gravitation is implemented into a theory of the strong and
electroweak interaction.  The so-obtained K\"ahler potential can in
simple cases be derived by superstring
compactification~\cite{string}.\bigskip

The breaking of exact supersymmetry is reflected in the observed mass
difference between the Standard Model particles and their partners.
Due to the current mass limits, this mass difference is, in case of
strongly interacting particles, much larger than the typical mass
scale of the Standard Model particles. Assuming no mixing for
light-flavor squarks, there are stringent mass limits on the squarks
and gluinos from the direct search at the
Tevatron~\cite{tev_search,roland}. Due to large Yukawa couplings, the
partners of the third generation Standard Model particles may mix.
Since more parameters of the supersymmetric Lagrangean enter through
the non-diagonal mass matrices and the couplings, the mass limits for
these third generation sfermions are weakened. Moreover, all
supersymmetric partners of the electroweak gauge bosons and the
extended Higgs boson degrees of freedom mix. The search for these
weakly interacting particles, neutralinos and charginos, at hadron
colliders~\cite{tev_search} has not reached its limitations and will
complete the limits obtained from the search at
LEP2~\cite{lep_search}. The search for strongly interacting and also
for light weakly interacting supersymmetric particles is one major
task for the upgraded Tevatron and the LHC. The investigation of
mixing effects in the strong and weak coupling sector requires
precision measurements at hadron as well as at lepton
colliders.\smallskip

The reconstruction of supersymmetric particles from detector data is
difficult in $R$ parity conserving theories, since two LSPs leave the
detector unobserved. Moreover, hadron colliders do not have an
incoming partonic state with well-defined kinematics, but the partonic
cross sections have to be convoluted with parton density functions.
The derivation of mass bounds or the mass determination, respectively,
has to be performed by measuring the total hadronic cross section, if
rather specific final state cascades cannot be used to determine the
mass.  Especially for strongly interacting final state particles, the
cross sections depend on the factorization and renormalization scales
through the parton densities and the running QCD coupling. The scale
dependences lead to considerable uncertainties in the determination of
mass bounds. The next-to-leading order cross sections will improve the
mass bounds not only by their accuracy but also by their size. These
hadronic cross sections for mixing supersymmetric particles at the
upgraded Tevatron as well as at the LHC will be given in this
thesis.\bigskip

Similarly to light-flavor squarks and gluinos, the search for top
squarks with a non-zero mixing angle will lead to stringent mass
bounds, which are essentially independent of the mixing parameters and
the masses of other supersymmetric particles. However, it will most
likely be impossible to measure the mixing angle at hadron colliders
directly, since the cross sections for the production of a mixed stop
pair are strongly suppressed. The analysis of mixing effects in the
stop sector will be completed by the direct measurement of the mixing
angle in $e^+e^-$ collisions~\cite{lincol_rep}.\smallskip

Regarding certain decay channels the direct search for gauginos and
higgsinos at hadron colliders resembles the search for weak gauge
bosons. Although in most supergravity inspired scenarios not all
gauginos and higgsinos are light enough to be found at the upgraded
Tevatron, the search for light neutral and charged gauginos is
promising and could improve the LEP2 results at the upgraded Tevatron
and at the LHC. Even if the leading order cross sections are
independent of the QCD coupling, they depend on the factorization
scale through the parton densities. The next-to-leading order
predictions will again considerably improve the bounds derived for
masses and couplings.\bigskip

However, all search strategies for supersymmetric particles depend on
cascade decays leading to leptons, jets, and LSPs in the final state,
the latter provided $R$ parity is conserved. As long as the masses of
the particles under consideration are not known, the analysis of these
multiple decay channels does not give strong limits, \eg on mixing
parameters involved. But for a sufficiently large sample of events
including supersymmetric particles, the whole variety of possible
decays and couplings will help to determine the mass and mixing
parameters of the supersymmetric extension of the Standard Model. The
measurement of low energy parameters can then be used to search for
universal parameters, predicted by grand unification or supergravity
inspired scenarios.

\subsubsection{Outline of the Thesis}

Since the supersymmetric observables presented in the following
analyses can, from a phenomenological point of view, be treated
independently, technical features are covered with their first
appearance.\medskip

The general physics background is described in the first chapter. A
short introduction into supersymmetric extensions of the Standard
Model is complemented by the discussion of special aspects concerning
mixing particles; the next-to-leading order treatment of the mixing
angle~\cite{ourdecay} in the CP conserving stop sector is presented,
and the regularization prescriptions used for supersymmetric gauge
theories are summarized. The supersymmetric Feynman rules and a
complete set of formulae considered useful for the detailed
understanding of the calculations are given in the appendices.

The production cross sections for neutralinos and charginos at hadron
colliders are treated in chapter~\ref{chap_neut}. They include the
virtual and real next-to-leading order corrections, the latter
calculated using the dipole subtraction method. The treatment of
on-shell singularities is described in detail. The possible
improvement of the current analysis by using the next-to-leading order
cross section is pointed out.

In chapter~\ref{chap_decay} the decay widths including mixing stop
particles in next-to-leading order supersymmetric QCD~\cite{ourdecay}
are given. They include weak and strong coupling stop decays as well
as gluino and heavy neutralino decays to a light stop. The treatment
of the mixing angle follows the theoretical description in
chapter~\ref{chap_susy}.  The complete analytical results for the
next-to-leading order stop decay width is presented in the appendix.

The study of scalar top quarks is continued in
chapter~\ref{chap_stops}, where the production cross section at hadron
colliders is given for both of the mass eigenstates in next-to-leading
order supersymmetric QCD~\cite{my_stop}. One crucial point is the
influence of the mixing angle and supersymmetric parameters, which are
present in the virtual corrections, on the experimental analysis and
on the mass bounds.  The real gluon emission is calculated using the
cut-off method.

For a light stop the production cross section at hadron colliders can
be adapted to $R$ parity violating scenarios~\cite{my_leptohadro}. The
resonance cross section for the production of $R$ parity violating
squarks in $ep$ collisions is calculated in next-to-leading
order~\cite{my_leptohera}, and the search results for these particles
at HERA and at the Tevatron are combined. The influence of other
search strategies for $R$ parity violating squarks is
reviewed.\smallskip

The analytical calculations have been performed using the symbolic
manipulation program FORM~\cite{form}, for the numerical integration
routine VEGAS~\cite{vegas} was chosen, and the parton cross
sections have been calculated using the CTEQ4~\cite{cteq} parton
densities in leading and next-to-leading order.

\chapter{Supersymmetry} 
\label{chap_susy}

\section{Supersymmetric Extensions of the Standard Model}

Global supersymmetry is a possible extension of the set of symmetries
appearing in flat space-time gauge theories. The most general
extension of a Poincar\'e invariant theory would be an $N$-extended
super-Poincar\'e Algebra containing central
charges~\cite{haag,sohnius}. The supersymmetry generators $Q^i \;
[i=1,...,N]$ and their complex conjugate $\overline{Q}^i$ transform
fermionic into bosonic fields and {\it vice versa}, therefore obeying
an anticommutation relation. These anticommutators lead to a ${\mathbb
  Z}_2$ graded Lie algebra, containing the supersymmetry as well as
the Poincar\'e group generators and circumventing the No-Go
theorem~\cite{no_go}\footnote{Any Lie group containing the Poincar\'e
  group and a compact inner symmetry group factorizes, \ie the
  generators of the Poincar\'e group and the inner symmetry group
  commute with each other. The extended Lie algebra becomes trivial.}.
The dimension of the extension $N$ determines the maximum spin present
in the particle spectrum of the theory.  Renormalizability requires a
maximum spin of one for global supersymmetry, which is equivalent to
$N\le4$.  Including the graviton results a maximum spin two for local
supersymmetry, supergravity, and renders $N\le8$.  For $N=1$ this
super-Poincar\'e algebra becomes particularly simple, since the
central charges vanish and the generators $Q,\overline{Q}$ anticommute
with themselves.  Extended supersymmetric theories have some
remarkable features: for $N=2$ the particle spectrum can be calculated
non-perturbatively~\cite{witten}, $N=4$ leads to a completely finite
theory, and $N\ge5$ contains gravitation. However, the observed low
energy particle spectrum and CP violation are only compatible with
($N$=1) global supersymmetry. We will make use of the incorporation of
global supersymmetry into local supergravity only by assuming certain
characteristics of the mass spectrum at high scales, where unification
is required.\medskip

Since supersymmetric theories by definition contain scalar particles
not only in the Higgs sector, the behavior of scalar masses is of
importance: In the Standard Model the scalar Higgs boson mass suffers
from UV divergent radiative corrections, proportional to $g^2
\Lambda^2$ where $g$ is a gauge coupling and $\Lambda$ is an UV
cut-off parameter. This cut-off parameter could be fixed by some scale
where new physics appears. Assuming the Standard Model not being an
effective theory for mass scales around the weak gauge boson mass,
\eg leads to a physical scalar mass of the order of the weak scale and
higher order loop contributions of the order of the cut-off, which
could be the Planck scale. These corrections have to be absorbed,
using fine-tuning of mass and coupling counter terms in the
Lagrangean. The large corrections in the Standard Model originate from
gauge boson and top quark loops. In supersymmetric extensions
additional corrections arising from the supersymmetric partners enter
with a minus sign and weaken the UV degree of divergence to a
logarithmic behavior [$\delta m/m \propto \log \Lambda^2$]. For broken
supersymmetry another term proportional to the mass difference between
the Standard Model loop particles and their supersymmetric partners
arises.  Assuming \eg a grand desert SU(5) scenario\footnote{Though
  not all matter fields can be unified in one SU(5) multiplet. A more
  generic GUT model would be supersymmetric SO(10), directly broken to
  the Standard Model gauge group. In contrast to SU(5), SO(10)
  unification with non-zero neutrino masses may lead to the observed
  baryon asymmetry~\cite{pluemi}. The numerical analyses \eg of gauge
  coupling unification in grand desert SU(5) and SO(10) scenarios are
  similar.}  the natural shift of the scalar masses between the weak
and the unification scale is limited to less than one order of
magnitude~\cite{murayama}.\medskip

\begin{figure}[t] \begin{center}
\epsfig{file=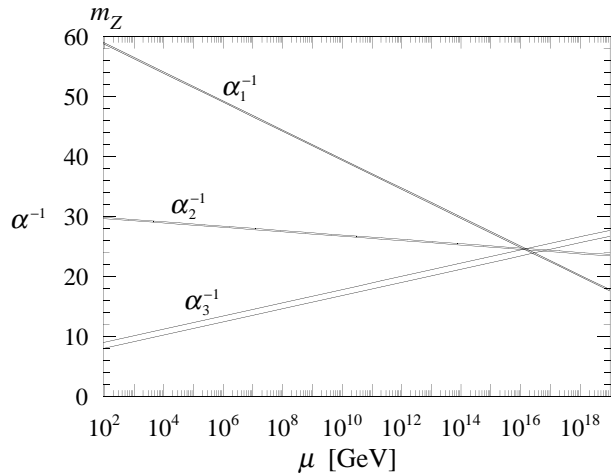,width=8cm}
\end{center}
\vspace*{-0.5cm}
\caption[]{\it The renormalization group evolution of the three 
  Standard Model gauge couplings, assuming  
  a simple SU(5) GUT gauge group and a grand desert~\cite{murayama}.
  \label{fig_susy_gauge}}
\end{figure}

Assuming a supersymmetric extension of the Standard Model, the
evolution of the gauge couplings can be evaluated, based on different
scenarios. Embedding the Standard Model into a simple GUT gauge group
does neither fix the gauge group nor possible intermediate scenarios,
\ie SU(5) unification with a grand desert is only one possibility.
The evolution of the gauge couplings depends on threshold effects and
masses in intermediate unification models.  However, it can be shown,
that, using supersymmetry, the gauge couplings unify up to a certain
accuracy, Fig.~\ref{fig_susy_gauge}.  The unification of the three
Standard Model gauge couplings determines one of the three parameters
involved [$\alpha,s_w^2 \equiv \sin^2 \theta_w, \alpha_s$]
theoretically.  This prediction has to be compared to the measured
value \eg for the weak mixing angle in the $\msbar$ scheme.  In
contrast to the Standard Model, which for reasons described above will
hardly be valid up to a large unification scale, the predicted value
for the minimal supersymmetric extension of the Standard Model
$s^2_w(\mz)=0.2334\pm0.005$ agrees very well with the measured value
of $0.2316\pm0.0003$~\cite{thetaw_mssm}. Any specific GUT scenario
fixes the renormalization group evolution of all the masses and
couplings.  The determination of $\alpha_s(\mz)$ from $\alpha$ and
$s^2_w$ again reflects the improvement of the Standard Model one-scale
GUT, which gives $\alpha_s(\mz)=0.073\pm0.002$, to the supersymmetric
one-scale GUT, which yields $\alpha_s(\mz)=0.129\pm0.010$. However,
the measured value of $\alpha_s(\mz)=0.118\pm0.004$ indicates, that
the minimal supersymmetric GUT prefers a slightly larger value;
threshold effects may be responsible for the difference. Very light
gauginos and very heavy squarks might even for the SU(5) GUT model
lead to the measured value of $\alpha_s$~\cite{murayama}.

\section{The Minimal Supersymmetric Standard Model}

The Minimal Supersymmetric Standard Model [MSSM] adds a minimal set of
supersymmetric partner fields to the Standard Model [SM]. These fields
contain scalar partners -- sleptons and squarks -- of all chiral
eigenstates of the Dirac fermions, incorporated into chiral
supermultiplets. Absorbing Standard Model gauge fields into vector
supermultiplets leaves Majorana fermion\footnote{Majorana fermions are
  defined as their own anti-particles, \ie the Majorana spinor is
  constructed by combining two Weyl spinors. Some arbitrariness may
  arise from the different treatment of electric and color charge,
  which leads to Majorana neutralinos and gluinos, but Dirac
  charginos. But it is just a name for the particles. Dirac charginos
  also yield fermion number violating vertices.}  partners of the
neutral U(1), SU(2), SU(3) gauge fields and Dirac fermion partners of
the charged SU(2) gauge fields, called gauginos.  The SU(3) ghost
fields are defined by re-writing the Fadeev-Popov determinant;
therefore they do not receive supersymmetric partners, which enter by
requiring the original Lagrangean being invariant under a global
supersymmetry transformation.

For reasons described later in this chapter the supersymmetric scalar
potential cannot include conjugate fields. Hence, at least two
different complex Higgs doublets have to be introduced to give masses
to up and down type quarks. They form five physical Higgs particles
after breaking SU(2)$\times$U(1) invariance. This extension of the
Higgs sector is not generically supersymmetric. However these scalar
Higgs degrees of freedom have to develop partner fields. This yields
neutral and charged Majorana/Dirac fermions with the same quantum
numbers as the SU(2) gauginos.  \medskip

The supersymmetric Lagrangean in superspace can be constructed by
extending the integration over the Lagrange density from space-time to
a superspace integration, \ie by adding two Grassmann dimensions
$\{\theta,\bar{\theta}\}$. All superfields can be written as a finite
power series in these Grassmann variables, containing the component
fields in the coefficients. Only two elements enter the supersymmetric
Lagrangean: (i) the so-called $F$ term of a chiral supermultiplet,
denoted as $\Phi_{\theta^2}$ in the expansion of the superfield in the
Grassmann variable $\theta$; (ii) the $D$ term of a vector multiplet
$V_{\theta^2 \bar{\theta}^2}$~\cite{sohnius}.  The kinetic real vector
supermultiplet is defined as the product of a chiral supermultiplet
and its conjugate $\Bar{\Phi}_j\Phi_j$, its $D$ terms contain the $F$
components of the chiral multiplets $F_j^*F_j$, which is absorbed into
the scalar potential.\medskip

The most general ansatz for a superpotential formed by chiral
supermultiplets relies on the fact, that the product of two chiral
supermultiplet is again chiral:
\begin{equation}
\W(\{\Phi\}) = m_{ij} \Phi_i \Phi_j + \lambda_{ijk} \Phi_i \Phi_j \Phi_k
\label{eq_susy_superpot}
\end{equation}
Higher orders in the polynomial would give mass dimensions bigger than
four and therefore spoil renormalizability. The superpotential occurs
in the Lagrangean as ($\W+\Bar{\W}$). The scalar potential in the
component-field Lagrangean\footnote{A product of a chiral superfield
  and a conjugate is not a chiral but a vector multiplet, as the
  kinetic superfield. The superpotential therefore does not contain
  conjugate superfields and neither does the scalar potential contain
  conjugate component Higgs fields.} contains, after integration over
the Grassmann variables, the non-Yukawa terms arising from the
superpotential $\W$; it is defined as
\begin{equation}
\V = - \left(  F_j^*F_j + \frac{\partial\W(A)}{\partial A_j} F_j
        + \frac{\partial\Bar{\W}(\Bar{A})}{\partial\Bar{A}_j} F_j^* 
       \right)
\end{equation}
The Euler-Lagrange equations yield $F_j^* = -\partial \W(A)/\partial
A_j$ where $A_j$ are the sfermion fields in the supermultiplet. This
fixes the most general scalar potential including chiral
supermultiplets in the matter sector of the Lagrangean:
\begin{equation}
\V = \sum_j \left| F_j \right|^2
\end{equation}\medskip 

Including the gauge sector for a non-abelian gauge group leads to
vector multiplets containing the gauge fields and their partners. The
scalar potential will also contain the $D$ auxiliary component
field terms of the gauge multiplet\footnote{ The supermultiplet
  constructed from the gauge vector multiplet and including the field
  strength component field is chiral. But its $F$ term contains the
  $D$ component fields of the vector gauge multiplet.}
\begin{equation}
\V = 
 \sum_j \left| F_j \right|^2 + \frac{1}{2} \sum_a \left( D^a \right)^2
   = \sum_j \left| F_j \right|^2 
   + \frac{g^2}{2} \sum_a \left( S^* T^a S \right)^2  
\label{eq_susy_scalarpot}
\end{equation}
The $D$ term is written for a general non-abelian SU(N) gauge group.
$S$ are the scalar fields transforming under the fundamental
representation of the corresponding gauge group, and $T^a$ the
generators of the underlying gauge group.

\subsection{$\mathbf{R}$ Parity}
\label{sect_susy_rpar}

The most general superpotential as given in
eq.(\ref{eq_susy_superpot}) contains trilinear couplings of chiral
matter supermultiplets, like the Higgs, the quark, and the lepton
supermultiplet. Couplings between the different Higgs fields or
between the Higgs field and corresponding lepton or quark
supermultiplets are needed to construct the two doublet Higgs sector
in the scalar potential. Although these vertices conserve the over-all
fermion number, they may violate the baryon and lepton number and would
lead to the same effects as leptoquarks, \eg proton
decay~\cite{dreiner}. In extensions of the Standard Model these
operators are forbidden by gauge invariance, as long as their dimension
is less than six. The MSSM either needs to suppress the different
couplings or remove the whole set by applying a new ${\mathbb Z}_2$
symmetry which changes the sign of the Grassmann variables in the
Lagrangean.  The corresponding conserved charge is defined as
\begin{equation}
R = (-1)^{3B+L+2S}
\end{equation}
where $B$ is the baryon number, $L$ the lepton number, and $S$ the
spin of the particle. This number is chosen to give $(+)$ for Standard
Model particles and $(-)$ for supersymmetric partners. The Higgs
particles in the two doublet model are all  described by $R=+1$.
Accounting for $R$ symmetry in the supersymmetric Lagrangean removes
trilinear chiral supermultiplet vertices containing no Higgs
superfield. The general superpotential eq.(\ref{eq_susy_superpot}) can
be separated into an $R$ parity conserving and an $R$ parity violating
part, which read for one generation of quarks and leptons
\begin{alignat}{9}
\W =& \; 
  \W_{R} + \W_{\slash{R}} 
  \notag \\
\W_{R} =& \; 
  \lambda^E \overline{E} H_1^j L^j + \lambda^D \overline{D} H_1^j Q^j
  + \lambda^U \overline{U} H_2^j Q^j - \mu H_1^j H_2^j \notag \\ 
\W_{\slash{R}} =& \; 
  \frac{1}{2} \lambda L^j L^j \overline{E} 
  + \lambda' L^j Q^j \overline{D}
  + \frac{1}{2} \lambda'' \overline{U} \, \overline{D} \, \overline{D}
\label{eq_susy_rviol}
\end{alignat}
The contraction of two indices is defined by the antisymmetric
($2\times2$) matrix $\epsilon_{ij}$; $L,Q$ are electron and quark
doublet superfields, $E,D,U$ are the singlet superfields for the
electron, $d$ and $u$ type quark; $\lambda^{E,D,U}$ are the Yukawa
coupling matrices, and $\mu$ is the Higgs mass parameter, which also
defines the higgsino mass [see appendix~\ref{chap_app_feynman}]. The
Yukawa couplings $\lambda,\lambda'$ violate lepton number, $\lambda''$
violates baryon number.\medskip

The combination $(\lambda' \cdot \lambda'')$ leads to proton decay via
an $s$ channel $d$ type leptoquark and therefore has to be strongly
suppressed. The conservation of $R$ is therefore a sufficient
condition for the stability of the proton. However, this symmetry has
been introduced {\it ad hoc} for weak scale supersymmetry as a less
rigid substitute for the conservation of some combination of $B$ and
$L$. The exact vanishing of $(\lambda' \cdot \lambda'')$ is not a
necessary condition for a stable proton; \ie if $\W_{\slash{R}}$ is
not removed by hand by demanding $R$ symmetry in the supersymmetry
Lagrangean, then many different constraints can be imposed on
combinations of couplings and masses in the $R$ parity violating
sector. The limits from direct production at HERA as well as from rare
decays generically determine $\lambda \cdot \lambda/m^2$, dependent on
the flavor of the squark considered. The same holds for atomic parity
violation.  The bounds from neutral meson mixing influence $\lambda
\cdot \lambda/m$, and the direct searches at hadron colliders and LEP
are only sensitive to the mass, except for the analysis of specific
decay channels~\cite{r_theo,r_exp}.\medskip

Phenomenologically, exact $R$ parity conservation leads to the
existence of a stable lightest supersymmetric particle (LSP), and
allows for the production of supersymmetric particles only in pairs.
For cosmological reasons this LSP has to be charge and color neutral,
which restricts the choices in the MSSM framework to the lightest
neutralino or the sneutrino. In GUT models exact $R$ parity
conservation is not necessary to obtain low-energy $R$ parity
conservation. For broken $R$ parity the unstable 'LSP' could therefore
be long-living and charged, allowing for charginos, sleptons and even
stops, as long as the lifetime is small enough to circumvent the
cosmological constraints.

\subsection{Soft Breaking}
\label{sect_susy_soft}

If supersymmetry would be exact, the squarks and sleptons were
mass degenerate with the Standard Model particles. Since the gauge
couplings have to respect supersymmetry in order to cancel the
quadratic divergences, breaking supersymmetry means enforcing a mass
difference between Standard Model particles and their supersymmetric
partners.  The mechanism of introducing mass terms by soft
breaking~\cite{soft_breaking} at a given scale has to respect gauge
symmetry, weak-scale $R$ parity, stability of scalar masses, and
experimental bounds \eg on FCNC. Soft breaking terms can be added to
the superpotential eq.(\ref{eq_susy_superpot}) at any given scale.
They exhibit the generic form
\begin{equation}
{\cal L}_{\rm soft} =
 - \left( m_0^2 \right)_{ij} C^*_i C_j 
 - \frac{1}{2} \left[ \left( m_{1/2} \right)_j \lambda_j \lambda_j
                      + {\rm h.c.} \right]
 - \left[  \frac{1}{6} A_{ijk} C_i C_j C_k
         + B \mu H_1 H_2 
         + {\rm h.c.} \right]
\end{equation}
The component fields involved are generic scalars $C$, Majorana
fermions $\lambda$ and the Higgs fields $H_1,H_2$, which are again
contracted using $\epsilon_{ij}$. The possible set of parameters
consists of:

\begin{itemize}
\item[--] Scalar mass matrices $(m_0^2)_{ij} \; [i,j=1,...n]$ for
  squarks and sleptons with $n$ generations. The diagonal masses can be
  chosen real, since $(\W+\Bar{\W})$ enters the Lagrangean.
\item[--] Three real gaugino masses $(m_{1/2})_j \; [j=1,2,3]$.
\item[--] 27 complex trilinear couplings $A_{ijk} [i,j,k=1,2,3]$ which
  conserve the $R$ charge.
\item[--] Two masses for the Higgs scalars and a complex Higgs mass
  parameter $\mu B H_1^j H_2^j$.
\end{itemize}
Evolving soft breaking mass terms by means of the renormalization
group equations can lead to breaking of the U(1)$\times$SU(2)
symmetry by driving one mass squared negative. This generalization of
the Coleman-Weinberg mechanism~\cite{rad_symm} links the large top
Yukawa coupling to electroweak symmetry breaking.

\subsection{Supersymmetric QCD}
\label{sect_susy_susyqcd}

\subsubsection{Particle Content}

The search for directly produced supersymmetric particles at hadron
colliders is dominated by strongly interacting final states.  In these
production processes the quantum corrections in next-to-leading order
are expected to be significant. Moreover, the corrections to the
production of weakly interacting particles at hadron colliders are
dominated by strong coupling effects. Although the parton picture and
thereby the incoming state is not affected by the heavy supersymmetric
partners of quarks and gluons, a consistent description of virtual
particle effects requires the inclusion of these particles.\smallskip

The supersymmetric extension of the QCD part of the Standard Model is
straightforward, since the SU(3) invariance is unbroken. One chiral
mass superfield $Q$ contains the left handed quark doublets $(u_{\SP
  L},d_{\SP L})$ and their squark partners $(\tilde{u}_{\SP
  L},\tilde{d}_{\SP L})$. Two more superfields $\Bar{U},\Bar{D}$
connect the quark singlet fields $(u^c_{\SP R},d^c_{\SP R})$ to their
partners $(\tilde{u}^*_{\SP R},\tilde{d}^*_{\SP R})$. The ${\rm
  SU}(3)_C \times {\rm SU}(2)_L \times {\rm U}(1)_Y$ quantum numbers
for quarks and squarks are identical. Whereas the $\sql$ is a SU(3)
triplet, the $\sqr$ is an anti-triplet and couples with $(-T^a)$ to
the quark and gluino, as can be seen in
Fig.~\ref{fig_app_feynsusyqcd}. The gluon vector superfield mirrors
the gluons to gluinos ($\gt$), which are real Majorana fermions and
therefore carry two degrees of freedom\footnote{The matching of the
  degrees of freedom is a subtlety in dimensional regularization,
  see section \ref{sect_susy_ward}.}. The number of generations is not
restricted by supersymmetry. The CKM matrix for the quarks will in the
following be assumed to be the unity matrix. The same holds for the
squark CKM matrix, which is not fixed by first principles to be either
diagonal or equal to the quark matrix.

The general mass matrix for up-type squarks is given by
\begin{eqnarray}
{\cal M}^2 \!&=&\! 
  \left( \begin{array}{cc} 
  m_Q^2 \! + \! m_q^2 \! + \! 
  \left( \frac{1}{2}-\frac{2}{3}s_w^2 \right) \mz^2 \cos(2\beta) &
 -m_q \left( A_q + \mu \cot \beta \right) \\ 
 -m_q \left( A_q + \mu \cot \beta \right)  &  
  m_U^2 \! + \! m_q^2 \! + \! 
  \frac{2}{3} s_w^2 \mz^2 \cos(2\beta) 
           \end{array}  \right)  
\label{eq_susy_squarkmass}
\end{eqnarray} 
For down type squarks $\cot \beta$ in the off-diagonal element has to
be replaced by $\tan \beta$. The entries $m_{\SP Q},m_{\SP U},A_q$ are
the soft breaking masses. In the diagonal elements the quark mass
still appears, as in exact supersymmetry. The $\mz$ contributions
arise from the different SU(2) quantum numbers of the scalar partners
of left and right-handed quarks. For light-flavor squarks this matrix
can be assumed being diagonal, since the chirality flip Yukawa
interactions are suppressed. For the top flavor these off-diagonal
elements cannot be disregarded. Taking into consideration bottom-tau
unification the ratio of the Higgs vacuum expectation values
$\tan\beta$ has to be either smaller than $\sim$2.5 or larger than
$\sim$40. In the second case, a large value for $\tan\beta$ compensates
for the small bottom quark mass and yields a strongly mixing sbottom
scenario. The results for the stop mixing may be generalized to the
sbottom case.

Neglecting additional mixing from a CKM like matrix, the chiral squark
eigenstates are equal to the mass eigenstates for the light flavors.
If we furthermore assume the soft breaking mass being dominant and
invariant under SU(2), then the light-flavor mass matrix is
proportional to the unity matrix, \ie the masses of the ten light
flavor squarks are equal.  As long as only strong coupling processes
are considered, we will have to deal with ten identical particles.
This will not be the case for the scalar top sector as will be shown
in section \ref{sect_susy_stop}.

\subsection{Mixing Stop Particles}
\label{sect_susy_stop}

\subsubsection{Diagonalization of Mass Matrices}

For scalar top quarks the off-diagonal elements of the squark mass
matrix eq.(\ref{eq_susy_squarkmass}) are large. Any real symmetric
mass matrix of the form
\begin{eqnarray}
\MM =  \left( \begin{array}{cc} 
\MM_{LL} & \MM_{LR} \\ \MM_{LR} & \MM_{RR} 
       \end{array}  \right)  
\end{eqnarray} 
can be diagonalized by a real orthogonal transformation, \ie a
uniquely defined real rotation matrix.  The eigenvalues are
\begin{equation}
m^2_{\SP 12} =
     \frac{1}{2} \left[ 
       {\rm Tr}(\MM) 
       \mp \left[ {\rm Tr}^2(\MM) - 4 {\rm Det}(\MM) \right]^{1/2}
                              \right]
\end{equation}
The cosine of the mixing angle can be chosen positive
$-\pi/4<\theta<\pi/4$:
\begin{equation}
\cos (2\theta) = \frac{\left| \MM_{LL}-\MM_{RR} \right|}
                      {\sqrt{{\rm Tr}^2(\MM)-4 {\rm Det}(\MM)}}
 \qquad \qquad
\sin (2\theta) = \frac{2 \MM_{LR}}
                      {\sqrt{{\rm Tr}^2(\MM)-4 {\rm Det}(\MM)}}
\end{equation}
There is no flat limit from different to equal mass eigenvalues for
this diagonalization procedure, since the diagonalized matrix would be
proportional to the unity matrix and therefore commute with any
rotational matrix.

\subsubsection{Stop Mixing}

 In the scalar top sector the unrenormalized chiral eigenstates are
$\stln,\strn$. The chirality-flip Yukawa interactions give rise to
off-diagonal elements in the mass matrix eq.(\ref{eq_susy_squarkmass})
\ie the bare mass eigenstates $\sten$ and $\stzn$ are obtained by a
leading-order rotation, as described above. 
\begin{eqnarray}
\left( \begin{array}{c} \sten \\ \stzn \end{array} \right)
= 
\left(  \begin{array}{cc} 
           \cos\tmixn & \ \sin\tmixn \\ -\sin\tmixn & \ \cos\tmixn
        \end{array}  \right)  
\left( \begin{array}{c} \stln \\ \strn  \end{array}  \right)
\end{eqnarray}

\begin{figure}[t] \begin{center}
\epsfig{file=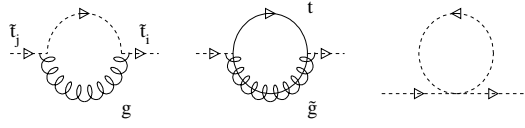,width=8cm}
\end{center}
\caption[]{\it Feynman diagrams for the stop self energy in NLO, 
  including mixing in the second and third diagram, the top-gluino 
  loop and the pure squark tadpole. 
  \label{fig_susy_feyn}}
\end{figure}

The mass eigenvalues and the leading-order rotation angle $\tmixn$ can
be expressed by the elements of the mass matrix. However, SUSY-QCD
corrections, involving the stop and gluino besides the usual particles
of the Standard Model, modify the stop mass matrix and the stop
fields. The Feynman diagrams are given in Fig.~\ref{fig_susy_feyn}. As
described in appendix~\ref{chap_app_feynman}, the coupling to a quark
and a gluino as well as the coupling between four squarks can switch
the chirality state and therefore contribute not only to the diagonal
but also to the off-diagonal matrix elements\footnote{It can be shown
  that a correction to the mass matrix renders the NLO mass matrix
  complex symmetric and not hermitian, as long as CP is conserved, \ie
  only imaginary parts from the absorptive scalar integrals arise.}.
This gives rise to the renormalization of the masses and of the wave
functions [$\stin=\ZM_{ij}\,\stj$]. Any leading-order observables
concerning the mixing top squarks are linked by a re-rotation of
$\pi/2$, denoted by $\PX$, eq.(\ref{eq_app_perm}). In next-to-leading
order this symmetry is broken by the mixing stop self energy. In order
to restore this symmetry in any order perturbation theory\footnote{Any
  observable containing only one kind of external stop particles can
  be transformed by exchanging the stop masses and adding (-) signs to
  $\sin(2\tmix)$ and $\cos(2\tmix)$.  This prescription $\PX$ will be
  used for stop decay widths and for the hadronic production cross
  section in LO and NLO later and is defined in
  eq.(\ref{eq_app_perm}).}, we choose a real wave-function
renormalization matrix $\ZM$, which is defined to split into a real
orthogonal matrix $\R(\delta\tmix)$ and a diagonal matrix $\ZM_{{\rm
    diag}}$, \ie $\ZM = \R(\delta\tmix)\,\ZM_{{\rm diag}}$.  The
rotational part can be reinterpreted as a shift in the mixing
angle~\cite{ourdecay,diaz}, given by $\tmixn-\delta\tmix \equiv
\tmix$:
\begin{eqnarray}
\left( \begin{array}{c} \ste   \\ \stz  \end{array}  \right)
= 
\left( \begin{array}{cc} Z^{-1/2}_{{\rm diag},\,11} & 0 \\ 
                         0 & Z^{-1/2}_{{\rm diag},\,22}
       \end{array}  \right)  
\left( \begin{array}{cc} \cos\tmix & \ \sin\tmix\\ 
                        -\sin\tmix & \ \cos\tmix
       \end{array}  \right)  
\left( \begin{array}{c} \stln  \\ \strn \end{array}  \right)
\end{eqnarray}  
This counterterm for the mixing angle allows the diagonalization of the
real part of the inverse stop propagator matrix in any fixed-order
perturbation theory.
\begin{alignat}{9}
{\rm Re}\left[ D_{\rm ren}^{-1}(p^2) \right] 
=&
\big( \ZM \big)^T\,
  \left[ p^2 \:{\rm\bf 1} - {\cal M}^2 + {\rm Re}\,\Sigma(p^2) \right]\,
\big( \ZM \big) \notag \\
=&
\big( \ZM_{\rm diag} \big)^T\, 
  \left[ p^2 \:{\rm\bf 1} 
   - \R(\delta\tmix)^{-1} 
     \left( {\cal M}^2 + {\rm Re}\,\Sigma(p^2) \right) 
     \R(\delta\tmix)
  \right]\,
\big( \ZM_{\rm diag} \big) \notag \\
=&
\big( \ZM_{\rm diag} \big)\, 
{\rm Re}\,D_{\rm diag}^{-1}(p^2)\,
\big( \ZM_{\rm diag} \big) 
\label{eq_susy_diagonalize}
\end{alignat}  
This holds as long as the real part of the unrenormalized stop
self-energy matrix ${\rm Re}\Sigma(p^2)$ and  thereby the whole
next-to-leading order mass matrix is symmetric\footnote{The
  next-to-leading order SUSY-QCD correction to the stop mass matrix is
  \vspace*{-2mm}
  \begin{alignat}{2}
    \Sigma_{12}(p^2) &= - 2 \pi C_F \as \Big[
       \svt A(\msz) - \svt A(\mse) + 8 \mg m_t \czt B(p;\mg,\mt) \Big]
    = \Sigma_{21}(p^2) \notag \\
    \Sigma_{11}(p^2) &= - 4 \pi
        C_F \as \Big[ 
        (1+\czt^2) A(\mse) + \sztq A(\msz) -2 A(\mg) - 2 A(\mt) 
        \notag \\[-2mm]
        &\phantom{= - 4 \pi i C_F \as } 
        -2 (p^2 + \mse^2) B(p;\lambda,\mse)
        +2 (p^2 - \mg^2 - \mt^2 + 2 \mg\mt\szt) B(p;\mg,\mt) \Big]
    = \PX^{-1} \Sigma_{22}(p^2) \notag \\
    \dot{\Sigma}_{ij}(p^2) &\equiv 
        \partial \Sigma_{ij}(p^2)/ \partial p^2 
  \end{alignat}}. 
The mixing angle depends on the scale of the self energy matrix
\begin{equation}
\tan \left( 2\delta \tmix(p^2) \right) = 
\frac{2\,{\rm Re}\,\Sigma_{12}(p^2)}{\mse^2-\msz^2
           +{\rm Re}\,\Sigma_{22}(p^2)-{\rm Re}\,\Sigma_{11}(p^2)}
=\frac{2\,{\rm Re}\,\Sigma_{12}(p^2)}{\mse^2-\msz^2}
+ \order(g^2)
\end{equation}
We fix the renormalization constants by imposing the following two
conditions on the renormalized stop propagator matrix: (i) the
diagonal elements should approach the form $1/D_{{\rm ren},\,jj}(p^2)
\to p^2-\msj^2 + i \msj \Gamma_{\stj}$ for $p^2 \to \msj^2$, with
$\msj$ denoting the pole masses; (ii) the renormalized (real) mixing
angle $\tmix$ is defined by requiring the real part of the
off-diagonal elements $D_{{\rm ren},\,12}(p^2)$ and $D_{{\rm
    ren},\,21}(p^2)$ to vanish. The three relevant counter terms for
external scalar particles are
\begin{equation}
\delta \msj^2 = \real\Sigma_{jj}(\msj^2) \qquad
\delta Z_{jj} = -\real\dot{\Sigma}_{jj}(\msj^2) \qquad
\delta \tmix(p^2) = - \frac{\real\Sigma_{12}(p^2)}{\msz^2-\mse^2}  
\end{equation}
Thus, for the fixed scale $p^2$ the real particles $\ste$ and $\stz$
propagate independently of each other and do not oscillate.\medskip

\begin{figure}[t] \begin{center}
\epsfig{file=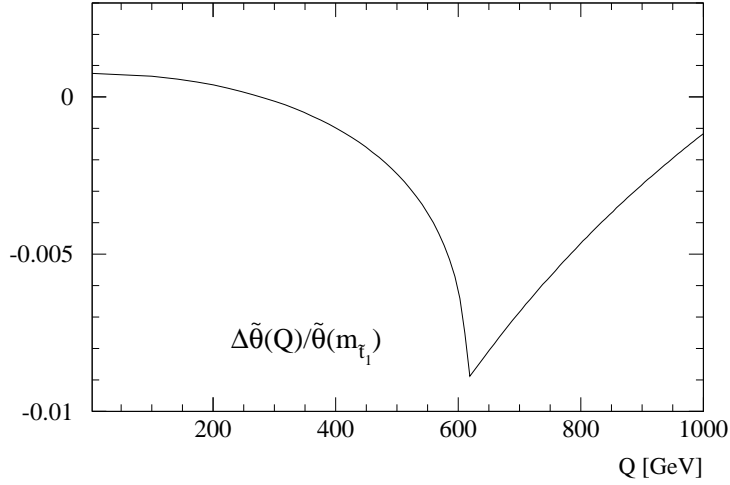,width=10cm}
\end{center}
\vspace*{-1cm}
\caption[]{\it The dependence of $\tmix(Q^2)$ on the renormalization 
  scale $Q$. The input mass values are: $m_{1/2}=150\gev$,
  $m_0=800\gev$, $A_0=200\gev$, $\mu>0$, for which the leading-order
  mixing angle is given by $1.24$ rad.  The minimum of the correction
  corresponds to the threshold $Q=\mg+\mt$ in the scalar integral.
  \label{fig_susy_angle}}
\end{figure}

The so-obtained (running) mixing angle depends on the renormalization
point $Q$, which we will indicate by writing $\tmix(Q^2)$. The
appropriate choice of $Q$ depends on the characteristic scale of the
observable that is analyzed.  The real shift connecting two different
values of the renormalization point is given by the renormalization
group, leading to a finite shift at next-to-leading order SUSY-QCD
\begin{equation}
  \tmix(Q_1^2) - \tmix(Q_2^2) 
      = \frac{C_F \as\mg\mt \cos(2\tmix)}{\pi(\msz^2 - \mse^2)}
      \,{\rm Re}\!\left[ \; B(Q_2;\mg,\mt)\!-\!B(Q_1;\mg,\mt) \; \right]
\end{equation}
This shift is independent  of the regularization. In the limit of
large scales the difference behaves as $\log(Q^2/{Q'}^2)$.  A
numerical example is presented in Fig~\ref{fig_susy_angle}. As a
noteworthy consequence of the running-mixing-angle scheme, we mention
that some LO symmetries of the Lagrangean are retained in the NLO
observables.  For instance, if for only one kind of external stop
particle one chooses $Q=\mst$, the results for the other stop particle
can be derived by the simple operation $\PX$, eq.(\ref{eq_app_perm}),
which then also acts on the argument of the mixing angle.\medskip 

 Considering virtual stop states with arbitrary $p^2$, the
off-diagonal elements of the propagator matrix can be absorbed into a
redefinition of the mixing of the stop fields, described by an
effective (complex) running mixing angle $\tmix_{\rm eff}(p^2) \equiv
\tmixn-\delta\tmix_{\rm eff}(p^2)$. This generalization amounts to a
diagonalization of the complex symmetric stop propagator matrix
$D_{{\rm ren}}$, including the full self-energy $\Sigma(p^2)$, by a
complex orthogonal matrix $\R(\delta\tmix_{\rm eff})$\footnote{Any
  real symmetric matrix can be diagonalized by a real orthogonal
  transformation $O^TAO$ where $O^{-1}=O^T$. One generalization is the
  complex unitary diagonalization of a complex symmetric matrix
  $U^TAU$ with $U^\dagger=U^{-1}$, where the diagonal matrix is real
  and positive. Another one is the complex orthogonal diagonalization
  of a complex symmetric matrix $O^TAO$, $O^{-1}=O^T$ where the
  diagonalized matrix is still complex. Note that a hermitian matrix
  can only be diagonalized by a unitary transformation $U^{-1}AU$.}
exactly in analogy to eq.(\ref{eq_susy_diagonalize}).  The so-defined
effective running mixing angle is given by
\begin{equation}
\tmix_{\rm eff}(p^2) = \tmixn -
   \frac{1}{2}\arctan\!\left[ \frac{2\Sigma_{12}(p^2)}
   {\mse^2-\msz^2+\Sigma_{22}(p^2)-\Sigma_{11}(p^2)} \right]
\stackrel{{\rm NLO}}{\rightarrow}
\tmix(p^2) + \frac{ {\rm Im}\,\Sigma_{12}(p^2)}{\msz^2-\mse^2}
\end{equation}
The complex argument of the trigonometric functions leads to
hyperbolic functions. From this point of view the use of a diagonal
Breit--Wigner propagator matrix is straightforward. For instance, in
the toy process $t \gt \to t \gt$ all NLO stop-mixing contributions to
the virtual stop exchange can be absorbed by introducing the effective
mixing angle in the LO matrix elements. The argument of this effective
mixing angle is given by the virtuality of the stop particles in the
$s$ channel. This procedure also applies to multi-scale processes like
$q\bar{q} \to t\sti\gt\,$ or $e^+e^- \to \ste\stzb$, where the
effective $g\ste\stz/\gamma\ste\stz$ couplings become non-zero due to
the different scales of the redefined stop fields.\medskip

There exist other renormalization schemes for the stop mixing angle,
either fixing the scale of the running mixing angle at some
appropriate scale or absorbing certain diagrams \eg contributing to
the production process $e^+e^- \to \ste \stzb$~\cite{wien}. Any of
these schemes can be regarded as a prescription to measure the mixing
angle, either in the mixed production at $e^+e^-$ linear colliders or
in decay modes or quantum corrections. The mixed production induced
scheme however has the disadvantage of introducing the $Z \ste \stz$
weak coupling constants into the QCD counter terms. The measured
values of the mixing angle can be translated from one scheme into
another by comparing the counter terms.  In Fig.~\ref{fig_susy_angle}
the numerical effect of the finite renormalization can be seen to be
small; the same holds for the different renormalization schemes, which
are numerically almost equivalent.\smallskip

When fixing the counter term for the stop mixing angle $\tmix$, one
can express the angle in terms of the parameters appearing in the mass
matrix eq.(\ref{eq_susy_squarkmass}). The counter term $\delta (\sin
(2\tmix))$ can be linked to the counter terms of these parameters:
\begin{alignat}{9}
\szt =& \; \frac{2 \mt ( A_t + \mu {\rm cot} \beta)}{\mse^2 - \msz^2}
\notag  \\
\frac{\delta \szt}{\szt} =& \; 
- \frac{\delta (\mse^2-\msz^2)}{\mse^2-\msz^2}
+ \frac{\delta \mt}{\mt}
+ \frac{\delta (A_t+\mu {\rm cot} \beta)}{A_t+\mu {\rm cot} \beta}
\end{alignat}
where $\delta x$ denotes the counter term of the parameter $x$. Since
$\mu$ and $\beta$ appear in the scalar potential only in the weakly
interacting sector, they will not be renormalized in next-to-leading
order SUSY-QCD. However, $\delta A_t$ can be calculated from the mass
and mixing angle counter terms. This reflects the fact, that the
system of observables used in the Feynman rules is non-minimal, \ie
the on-shell scheme for the masses and the running mixing angle
determine the renormalization of the couplings $\ste \stz G^0$ and
$\ste \stz A^0$, where $A_t$ appears explicitly~\cite{hhg}.

\section{GUT inspired Mass Spectrum}
\label{sect_susy_gut} 

Next-to-leading order calculations in the framework of light-flavor
SUSY-QCD~\cite{roland} only incorporate a few free parameters: the
Standard Model set and the gluino and the light-flavor squark mass.
Including mixing stops and the mixing neutralinos/charginos the number
of low-energy parameters becomes large. Hence, for a rough
phenomenological analysis we will use a simplifying scenario, which
could be a SUSY-GUT scenario, either supergravity~\cite{sugra} or
gauge mediation~\cite{gauge_med} inspired.

\subsubsection{SUSY-GUT Scenario}

Inspired by the unification of the three Standard Model gauge
couplings in supersymmetric GUT models we will assume a relation
between these couplings and the gaugino masses.  Independent of the
actual form of the simple gauge group and the connected GUT scenario,
the three Standard Model gauge groups are embedded into, and
independent of intermediate scale particles and thresholds, we  
can assume gauge coupling unification.
\begin{equation}
 \frac{M_1(Q)}{\alpha_1(Q)}
=\frac{M_2(Q)}{\alpha_2(Q)}
=\frac{M_3(Q)}{\alpha_3(Q)}
=\frac{m_{1/2}(M_X)}{\alpha_{\rm GUT}(M_X)}
\end{equation}
where $m_{1/2}$ is the mass entry in the scalar potential, defined at
the unification scale $M_X$. There the three gauge couplings unify to
$\alpha_{\rm GUT}\simeq 1/26$. For the masses at the weak scale this
leads to~\cite{drees_martin}
\begin{alignat}{9}
m_{\tilde{B}} \simeq& \; M_1(m_Z) \simeq 0.4 \; m_{1/2} \notag \\ 
m_{\tilde{W}} \simeq& \; M_2(m_Z) \simeq 0.8 \; m_{1/2} \notag \\ 
\mg           \simeq& \; M_3(m_Z) \sim   2.6 \; m_{1/2}
\label{eq_susy_gaugemass}
\end{alignat}
However, the gluino mass is strongly dependent on the scale which can
lead to a difference of $30\%$ between the pole mass $\mg$ and the
running mass $M_3(M_3)$~\cite{drees_martin}. For the derivation of
these mass relations it is only necessary to assume a simple
unification gauge group arising at a scale $M_X\sim 2 \cdot
10^{16}\gev$. The gaugino mass unification can be tested
experimentally at the LHC~\cite{lhc_rep} as well as at a future linear
collider~\cite{lincol_rep}.

\subsubsection{Mass Unification}

In a supergravity inspired MSSM [mSUGRA] the scalar masses and the
trilinear couplings are assumed to be universal at the unification
scale $M_X$\footnote{Several unification scales may arise as the gauge
  coupling unification scale and the string scale only few orders
  below the Planck scale. Numerically the variation of the scale $M_X$
  between these physical scales leads to a small effect only.}. In
simple supergravity models they depend on the gravitino mass scale
$m_{3/2}$ and on the cosmological constant~\cite{sugra}.  The
universal parameters at the unification scale $M_X$ will be refered to
as $m_0$ and $A_0$.  The parameter $\mu B$ occuring in the Higgs
sector of the scalar potential [section \ref{sect_susy_soft}] will be
fixed by the choice of $m_{1/2},m_0,A_0,\tan\beta$ and the Standard
Model parameters, and by the requirement of electroweak symmetry
breaking, up to its sign. The light-flavor squark masses can be
expressed in terms of the universal scalar and gaugino masses, the
other parameters only enter the off-diagonal elements of the mass
matrix eq.(\ref{eq_susy_squarkmass}) and can be neglected
\begin{equation}
m_{\tilde{q}_{\SP L}}^2 \simeq m_0^2 + 6.3 m_{1/2}^2 + 0.35 D 
\qquad \qquad
m_{\tilde{q}_{\SP R}}^2 \simeq m_0^2 + 5.8 m_{1/2}^2 + 0.16 D 
\end{equation}
where $D=m_Z^2 \cos(2\beta)<0$. For mSUGRA scenarios a general
prediction for the light-flavor squark mass can be
given~\cite{drees_martin}
\begin{equation}
\ms \gtrsim 0.85 \mg
\end{equation}

\subsubsection{Approximate Solution}

\begin{figure}[t] \begin{center}
\vspace*{-0.7cm}
\epsfig{file=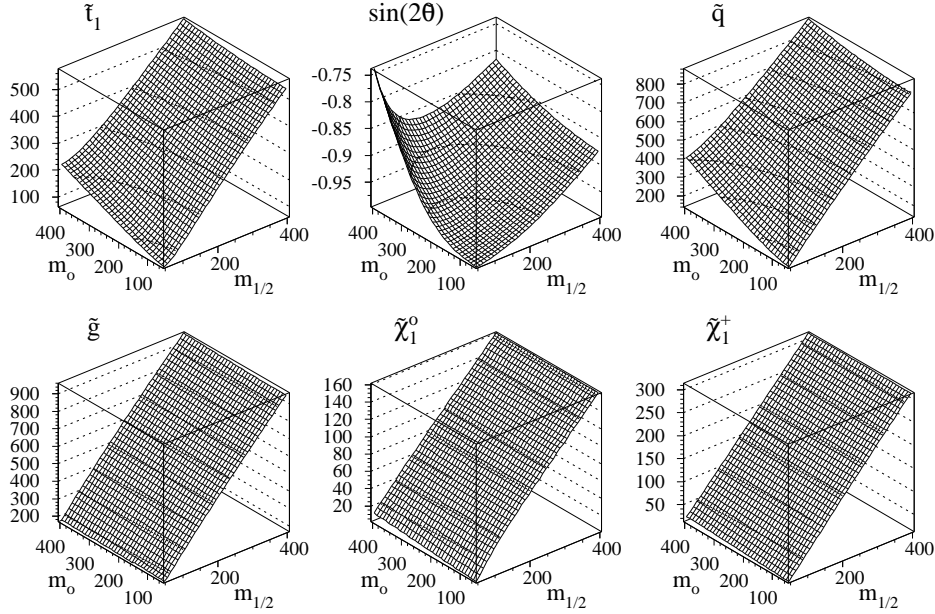,width=12.5cm}
\end{center}
\vspace*{-1.1cm}
\caption[]{\it Some relevant masses in the approximate mSUGRA scenario 
  for $A_0=300\gev, \tan\beta=4, \mu>0$; $m_0$ and $m_{1/2}$
  are varied between 50 and $400\gev$.
  \label{fig_susy_sugra}}
\end{figure}

The stop masses can be expressed in terms of the top Yukawa coupling
$Y_t=h_t^2/(4\pi)$. For small $\tan \beta$ they approximately read 
\begin{alignat}{9}
m_{\tilde{t}_{\SP L}}^2 \simeq& \;
     m_0^2 \left( 1 - \frac{Y_t}{2 Y_t^{\rm IR}} \right) 
   + m_{1/2}^2 \left( 6.3 - \frac{7Y_t}{3Y_t^{\rm IR}}
                          - \left( \frac{Y_t}{Y_t^{\rm IR}} \right)^2
               \right) + 0.35 D \notag \\
m_{\tilde{t}_{\SP R}}^2 \simeq& \;
     m_0^2 \left( 1 - \frac{Y_t}{Y_t^{\rm IR}} \right) 
   + m_{1/2}^2 \left( 5.8 - \frac{14Y_t}{3Y_t^{\rm IR}}
                          - \left( \frac{2Y_t}{Y_t^{\rm IR}} \right)^2
               \right) + 0.16 D \notag \\
A_t \simeq& \; \left( 1 - \frac{Y_t}{Y_t^{\rm IR}} \right)
    - 2 m_{1/2} \notag \\
\mt =& \; \frac{v s_\beta}{\sqrt{2}} h_t
\end{alignat}
The IR fixed point of the top mass is $Y_t^{\rm IR} \simeq 8
\alpha_3/9$ and $Y_t/Y_t^{\rm IR}$ varies from 0.75 to 1 dependent on
$\tan \beta$, becoming unity for $\tan \beta=1$. In this limit the
universal scalar mass does not influence the lighter right handed stop
mass. If the doublet soft breaking mass is larger than the right
handed soft breaking mass, the $\ste$, defined as the light stop, will
be mostly right-handed and the angle will prefer values around
$\pi/2$.

The higgsino mass parameter in this limit will be given as 
\begin{equation}
\mu^2 + \frac{m_Z^2}{2} = - m_0^2 - \frac{1}{2} m_{1/2}^2
+ \; {\rm terms \; including} \; \frac{Y_t}{Y_t^{\rm IR}} 
\end{equation}
The analyses in the following chapters are carried out using this
approximate mSUGRA renormalization group solution\footnote{This is
  implemented in the initialization routine of {\sc
    SPYHTIA}~\cite{spythia}. Some comments concerning the 5.7 version
  can be found in the bibliography.}. If not explicitly stated
otherwise we will vary the high-scale parameters around one central
point:
\begin{alignat}{20}
 m_{1/2}    =& 150 \gev \quad
&m_0        =& 100 \gev \quad
&A_0        =& 300 \gev \quad
&\tan \beta =& 4        \quad
&\mu        >& 0        \notag \\
 \mu        =& 277 \gev \quad
&M_2        =& 122 \gev \quad
&A_t        =& 355 \gev &&&& \notag \\
 \mne       =& 55  \gev \quad
&\mnz       =& 103 \gev \quad
&\mnd       =& 283 \gev \quad
&\mnv       =& 309 \gev && \notag \\
 \mce       =& 100 \gev \quad
&\mcz       =& 307 \gev &&&&&& \notag \\
 \mg        =& 401 \gev \quad
&\ms        =& 352 \gev \quad
&\mse       =& 198 \gev \quad
&\msz       =& 427 \gev \quad
&\sin(2 \tmix) =&-0.97
\label{eq_susy_scen}
\end{alignat}
In Fig.~\ref{fig_susy_sugra} some relevant low energy mass parameters
are given as a function of $m_0$ and $m_{1/2}$ to illustrate the
qualitative behavior described above. Typical features are the large
mass difference between the stop mass eigenstates, nearly independent
of the value $A_0$, and the clustered neutralino masses, where the two
light states are gaugino-type and the two heavy states are
higgsino-type. The latter results from the large value for $\mu$ in
the mSUGRA scenario. The lightest Higgs mass in this scenario in the
given approximation is larger than 100$\gev$ and will not be excluded
by LEP2.

\section{Mass Spectrum and Experimental Limits}

\subsubsection{Neutralinos and Charginos}

\begin{figure}[t] \begin{center}
\epsfig{file=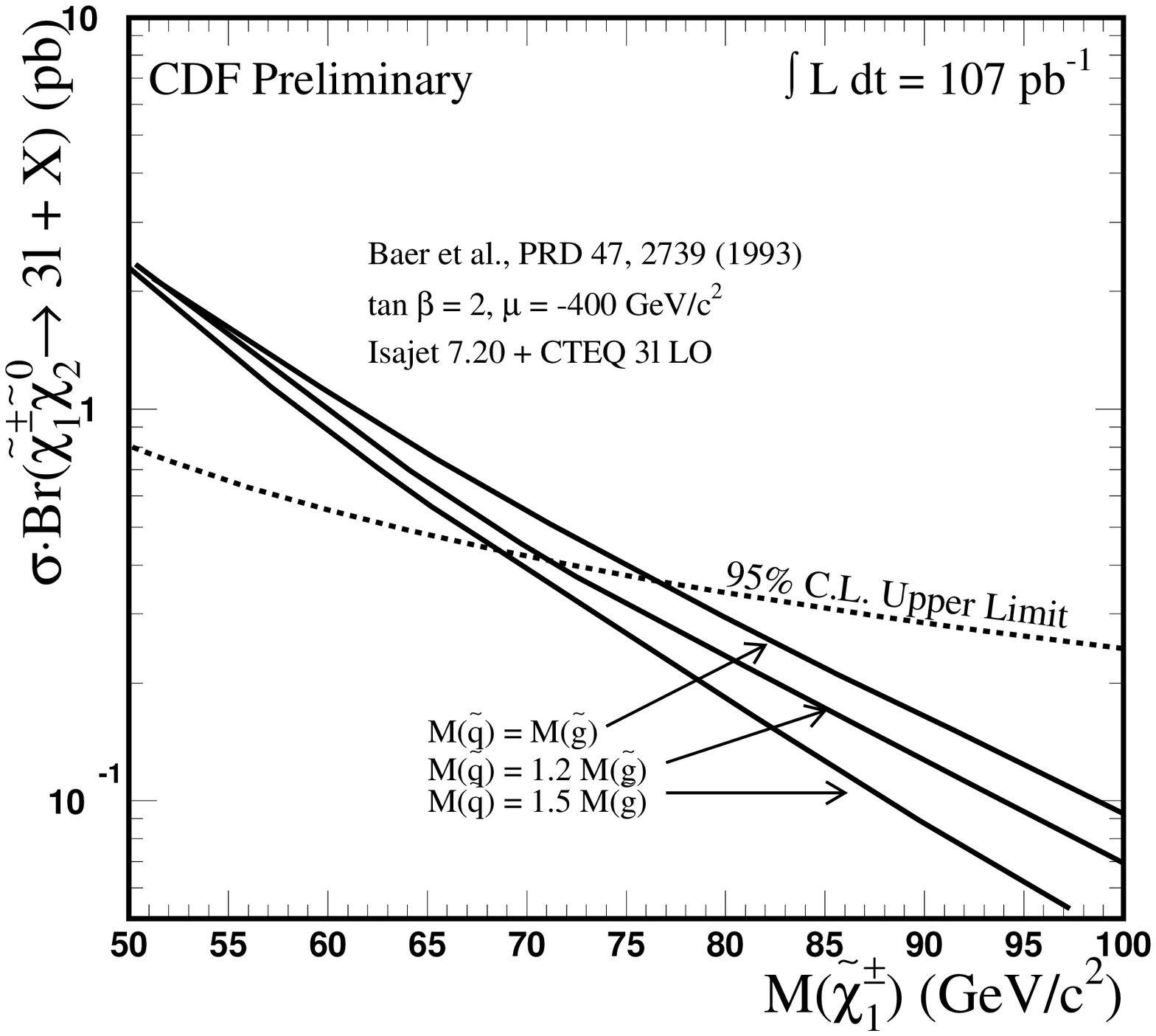,width=7.6cm}
\raisebox{-8mm}{\epsfig{file=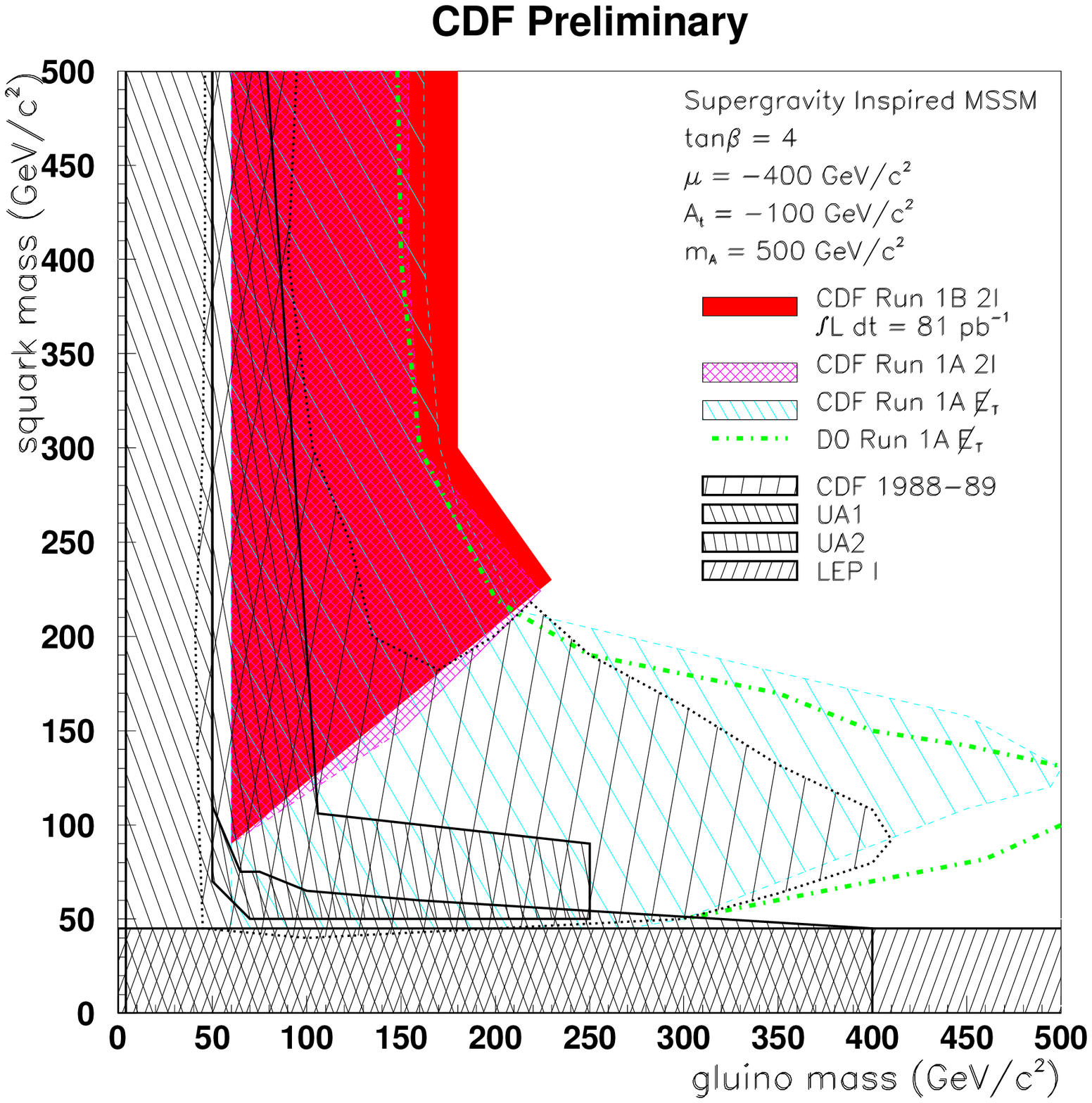,width=7.4cm}}
\end{center}
\vspace*{-0.9cm}
\caption[]{\it Left: The CDF limits on $\sigma\cdot \br$ for the 
  $\nnz \cpe$ production from the search for trilepton events, 
  eq.(\ref{eq_susy_tril}); Right: The CDF limits on the squark 
  and gluino mass including the (jets+$\Slash{E_T}$) and the 
  like-sign lepton signal for the gluino pair 
  production~\cite{tev_search}. In part of the parameter space 
  the NLO cross sections have been used~\cite{roland}. 
  \label{fig_susy_tril}}
\end{figure}

Searches for neutralinos and charginos have been carried out at the
Tevatron~\cite{tev_search} as well as at LEP~\cite{lep_search}. Due to
low energy $R$ parity conservation they can only be produced in pairs
$\nni \nnj$, $\cpi \cpj$, and $\cpi \nnj$. If the lightest neutralino
is the LSP, then the heavier particles have to decay via a cascade
into the LSP.  However the two and three parton decay channels are
strongly dependent on the mass spectrum:
\begin{alignat}{9}
\nnj &\longrightarrow  \; Z^* \nne, \; h^* \nne
     &\longrightarrow& \; \ell^+\ell^-\nne, \; q\bar{q} \nne 
 \notag \\
\nnj &\longrightarrow  \; \ell \tilde{\ell}, \; \tilde{\nu} \nu &\cdots&
 \notag \\
\cpj &\longrightarrow  \; {W^+}^* \nne, \; {H^+}^* \nne &\cdots& 
 \notag \\
\cpj &\longrightarrow  \; \tilde{\ell} \nu, \; \tilde{\nu} \ell &\cdots&
\end{alignat}
The decay $\nnj \to \sq q$ will be dominant if kinematically allowed,
but in a SUGRA inspired mass scenario this will be only the case for
the two heavy neutralinos. Besides, the chargino can enter the
neutralino decay chain via $\nnj \to \cpi H^-, \cpi W^-$. One very
promising final state for the mixed neutralino/chargino production is
the trilepton event
\begin{equation}
pp / p\bar{p} \longrightarrow \cpe \; \nnz 
 \longrightarrow \ell \nu \nne \; \ell \ell \nne 
 \longrightarrow \ell \ell \ell + \Slash{E_T}
\label{eq_susy_tril}
\end{equation}
where three charged leptons are present in the final state and the
missing transverse energy $\Slash{E_T}$ is based on three invisible
particles.  The exclusion plot is given in Fig.~\ref{fig_susy_tril}.
The cross section for chargino/neutralino production times the
branching ratio into the trilepton channel is given for different
squark masses, the gluino mass is fixed by the neutralino/chargino
mass and the gaugino mass unification. The mass limits for $\cpe$ can
be read off the axis, they vary between 60 and $80\gev$.

\subsubsection{Squarks and Gluinos}

The gluino will in general be assumed heavy, as suggested by SUSY-GUT
scenarios. The experimental exclusion limits from the direct search
for squarks and gluinos are given in the mass plane in
Fig.~\ref{fig_susy_tril}. The absolute lower limit on the gluino mass
is $\mg > 180\gev$~\cite{tev_search}. The decay channels considered
for the light-flavor squarks and for the gluinos are
\begin{alignat}{9}
\sq &\longrightarrow \; q \nnj, \;  q' \cpj 
    &\longrightarrow& \; {\rm jets} + \Slash{E_T} + \cdots 
  \notag \\ 
\gt &\longrightarrow \; q \bar{q} \nnj, \; q' \bar{q} \cpj 
    &\longrightarrow& \; {\rm jets} + \Slash{E_T} + \cdots 
  \notag \\
\gt &\longrightarrow \; q' \bar{q} \cpj 
    &\longrightarrow& \; {\rm jets} + \Slash{E_T} + \ell \ell \cdots 
  \notag \\
\gt &\longrightarrow \; \ste \bar{t}
    &\longrightarrow& \; b \cpj \bar{t} \cdots
\end{alignat}
The final state neutralino/chargino decays via a cascade to the
lightest neutralino, which is assumed to be the LSP. Products in this
decay chain are denoted by the dots. If it is not kinematically
forbidden, the gluino can first decay into a squark and a quark, and
{\it vice versa}. This leads to one more jet in the final state. The
stop decay channel of the gluino leads to a higher multiplicity of
Standard model particles and bottom jets. A typical signature for the
Majorana gluinos arises from the decay via a chargino. Since the
gluino is a singlet under the electro-weak gauge group, it decays to
$\cpj$ and $\cmj$ with the same probability, leading to like-sign
leptons in the final state of gluino pair production. A considerable
Standard Model background is not present for this signature.

Since supergravity inspired SUSY-GUT relations are used for the
experimental search at hadron colliders, there are no strong limits on
the squark mass if the gluino mass exceeds $550\gev$, see
Fig.~\ref{fig_susy_tril}. The supergravity inspired GUT scenarios as
described in section \ref{sect_susy_gut} do not allow for a gluino
mass being much larger than the light-flavor squark mass. In this
region of the ($\ms-\mg$) plane only the general unification of the
gaugino masses can be kept. The mass of the lightest neutralino,
assumed to be the LSP, grows with the gluino mass and becomes large
enough for the squark to decay into an LSP almost at rest.  The
missing transverse momentum would then become too small to be
measured.

The limits on the neutralino/chargino mass from the search at LEP
could be translated into limits on the gluino mass, using the gauge
coupling unification eq.(\ref{eq_susy_gaugemass}). Those are much
stricter than the Tevatron limits but model dependent. 

\subsubsection{Stops}

\begin{figure}[t] \begin{center}
\epsfig{file=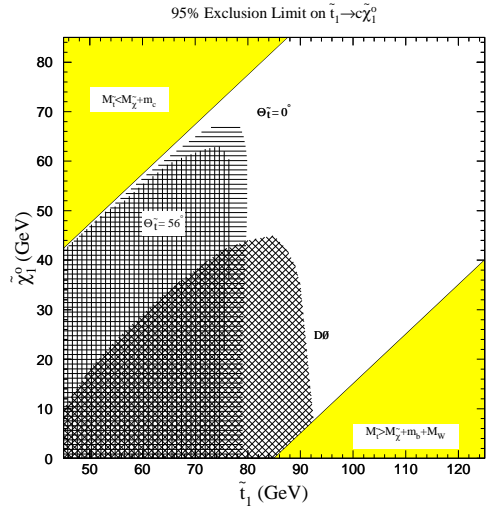,width=7cm}
\end{center}
\vspace*{-0.5cm}
\caption[]{\it Mass limits from the 
  $\ste$ pair production at D0 and LEP followed by the decay 
  $\ste \to c \nne$. The 
  dependence on the mixing angle enters through the coupling
  $\ste\ste Z$ at LEP.  
  \label{fig_susy_ste}}
\end{figure}

The limits on the stop mass arise from a search for stop pairs
decaying into $\ste \to c \nne$ and are therefore strongly dependent
on the mass of the lightest neutralino.  For a light stop mass this
decay mode will be dominant. In this mass regime the light stop can be
produced at LEP $e^+e^-\to\ste\steb$,
Fig.~\ref{fig_susy_ste}~\cite{lep_search}. The production cross
section depends on the mixing angle, arising from the $\ste \ste Z$
coupling, and thereby also the mass bound. As will be shown in chapter
\ref{chap_stops}, the hadroproduction cross section is independent of
the mixing angle, and both analyses, at LEP and at the Tevatron, yield
a mass bound on the lightest stop $\mse$ around
$79\gev$~\cite{lep_search,tev_search}.  However, these limits are only
valid as long as the $\nne$ is light enough and the decay channel
$\ste \to c \nne$ is dominant.  Additional limits arising from the
search for the decay $t \to \ste \nne$ are strongly dependent on the
branching ratio of this decay mode and therefore weaker than those
from the direct search. The different stop decay modes are described
in chapter~\ref{chap_decay}, and the direct search at hadron colliders
is investigated in chapter~\ref{chap_stops}.

\section{Regularization and Supersymmetric Ward Identities}
\label{sect_susy_ward}

\subsubsection{Dimensional Regularization and Reduction}

The $\msbar$ renormalization scheme is by definition related to the
regularization of infrared and ultraviolet divergences in dimensional
regularization (DREG)~\cite{dim_reg}. This regularization scheme
respects gauge symmetry and therefore the gauge symmetry Ward
identities\footnote{We will not focus on the problem of the chiral
  projector matrix $\gamma_5$, since consistent schemes have been
  developed~\cite{dim_reg,gamma_five} to deal with $\gamma$ traces in
  $n$ dimensions. In one-loop order a naive scheme can be used,
  however for neutralino/chargino production it has explicitly been
  checked that the ambiguous scheme dependent terms do not
  contribute~\cite{dim_reg}.}. It is less well-suited for
supersymmetric theories, since all Lorentz indices are evaluated in
$n$ dimensions, whereas the spinors are still four dimensional.  This
leads to a mismatch between the degrees of freedom carried \eg by a
physical gluon ($n-2$) and a gluino (2). A modified dimensional
reduction scheme (DRED) has been introduced to cope with this
problem~\cite{dim_red}.  The number of space-time dimensions is
compactified from four to $n$ dimensions, leaving the number of gauge
fields invariant \ie the gauge fields carry the $n$ dimensional
Lorentz indices. The remaining ($4-n$) dimensions form the $\epsilon$
scalars.  These particles render the $\gamma$ algebra four
dimensional. The gauge bosons and the gauginos carry the same number
of 4 degrees of freedom. The DRED scheme will be used to illustrate
the modified $\msbar$ scheme. Except for the unsolved problem of mass
factorization in DRED~\cite{bible,roland} it can be shown that both
dimension based schemes are consistent for calculations in the
framework of supersymmetric gauge theories.\medskip

Starting with a Lagrangean ${\cal L}[W^a_\mu,\lambda^a,D]$ for a
non-abelian supersymmetric gauge theory in the Wess-Zumino gauge one
can show that the supersymmetric variation $\delta_S$ of the
Lagrangean only vanishes in the limit of ($n \to 4$) dimensions, 
up to a total derivative~\cite{jack_jones}
\begin{equation}
\delta_S {\cal L}[W^a_\mu,\lambda^a,D]
\; \stackrel{n \to 4}{\longrightarrow} \; 0
\end{equation}
The component fields $W$ indicate the gauge fields, and $\lambda$ the
[Majorana] gauginos.  This leads to the Ward identity including the
ghost and gauge fixing term ${\cal L}_G$, where in $n$ dimensions the
variation $\delta_S {\cal L}$ has to be kept.
\begin{alignat}{9}
0 =& \; \langle \; 
\int d^nx \left[ J^\mu \; \delta_S W_\mu 
               + \bar{j} \; \delta_S \lambda
               + \bar{j}_D \; \delta_S D
               + \delta_S {\cal L}_G
               + \delta_S {\cal L}   \right] \; \rangle \notag \\
& 
\langle X \rangle \equiv \; \int d\{W_\mu\} d\{\lambda\} d\{D\} X \; 
e^{i \int d^nx [{\cal L}+{\cal L}_G
                +J^\mu W_\mu+\bar{j}\lambda+\bar{j}_DD]}
\end{alignat}\smallskip 

Although DREG and thereby the $\msbar$ scheme cannot be shown being
inconsistent with supersymmetry, they do not respect supersymmetry on
the level of  naively used Feynman rules. The problem is similar
to applying DRED to gauge theories: Evanescent couplings renormalize
in a manner different from the physical couplings . In NLO-DREG
this results in a finite renormalization of Feynman diagrams which
restores supersymmetry explicitly. At higher orders these additional
counter terms even include poles in $\epsilon$.

\subsubsection{Finite Renormalization}

Explicit calculations show that Green's functions calculated from the
MSSM Lagrangean using dimensional regularization may not respect
supersymmetry. The supersymmetry transformation mirrors \eg the gauge
coupling $g(qqg)$ to the gauge coupling $g(\sq \sq g)$ and the Yukawa
coupling $\hat{g}(q \sq \gt)$. In regularization schemes which respect
supersymmetry, like dimensional reduction\footnote{The difference
  between DREG and DRED are $\epsilon$ terms arising from DREG Dirac
  traces including gauge fields. They combine with a pole $1/\epsilon$
  in a scalar integral, leading to a finite contribution.  These terms
  are exactly those leading to the difference \eg in
  eq.(\ref{eq_susy_yukshift}).}, the supersymmetric limits of these
couplings are identical in any order perturbation theory. In DREG the
supersymmetric limit of the Yukawa coupling differs from the gauge
couplings at one loop level~\cite{martin_vaughn}
\begin{equation}
\hat{g} = g 
\left[ 1 + \frac{g^2}{32\pi^2} \left( \frac{4}{3}C(G) - C(r) \right)
\right]
\label{eq_susy_coupshift}
\end{equation}
The Casimir invariants $C$ are defined for the Dirac fermions in the
fundamental $(r)$, and for the gauge boson and the Majorana gauge
fermions in the adjoint ($G$) representation\footnote{The SU(3)
  coupling $qqg$ yields $C(r)=C_F$ and $C(G)=C_A=N$.}. This difference
has to be compensated to render the calculation supersymmetric. Since
the Standard Model quark-gluon coupling $g(qqg)$ is by definition the
measured quantity, the Yukawa coupling will be shifted $\hat{g} \to g$
in the expression for the final observable.  This finite shift is not
a finite field theoretical renormalization of any measured parameter
and it is not only present for gauge vs. Yukawa couplings. It is an
artifact arising from the supersymmetry violation of naive dimensional
regularization.  \smallskip

Supersymmetry relates the weak Higgs Yukawa coupling $Y(qqh)$ to the
vertices $Y(\sq\sq h)$ and $Y(g\sq\tilde{h})$. The three couplings $Y$
in the supersymmetric limit and calculated in DREG are not identical
in NLO
\begin{equation}
Y(qqh)
= Y(\sq \sq h) \left[ 1 + \frac{g^2}{16\pi^2} C(r) \right]
= Y(q \sq \tilde{h}) \left[ 1 + \frac{3g^2}{32\pi^2} C(r) \right]
\label{eq_susy_yukshift}
\end{equation}
where in the case of weak coupling only $C(r)$ occurs. These two
finite differences $\propto \alpha_s$ in couplings mediated by $G_F$
have to be compensated to make dimensional regularization compatible
with supersymmetry. \medskip 

The usual parameterization of the Yukawa coupling constant is $Y=mg$,
where $g$ is defined in the $\msbar$ scheme and $m$ is the pole mass,
\ie renormalized in the on-shell scheme. However, the pole mass has to
be calculated in the DREG scheme, and the mass appearing in the
different couplings eq.(\ref{eq_susy_yukshift}) is --- in the
supersymmetric limit --- only numerically the same. In fact, the scalar
mass set to $m$ and the fermion mass set to $m$ behave differently in
next-to-leading order, since the counter term for the scalar and the
fermion on-shell mass in DREG is not the same.
\begin{equation}
m_q = \left[  1 + \frac{g^2}{16\pi^2} C(r) \right] \, m_{\sq}
\end{equation}
This behavior breaks supersymmetry explicitly and has therefore be
removed. The mass shift is responsible for the difference between
$Y(qqh)$ and $Y(\sq\sq h)$, and it renders the difference to $Y(q \sq
\tilde{h})$ compatible with the general difference between the gauge
and Yukawa coupling as given in eq.(\ref{eq_susy_coupshift}):
\begin{equation}
g(qqh)
= g(\sq \sq h) 
= g(q \sq \tilde{h}) \left[ 1 + \frac{g^2}{32\pi^2} C(r) \right]
\end{equation}
The observable coupling is again defined as in the Standard Model
value $Y(qqh)$.

\chapter{Production of Neutralinos and Charginos}
\label{chap_neut}

\section{Born Cross Sections}

\subsubsection{Partonic Cross Sections}
Neutralinos and Charginos can be produced at hadron colliders in
several combinations, all starting from a pure quark incoming state
\begin{alignat}{9}
q \, \bar{q} \; &\longrightarrow \; \nni \nnj \notag \\
q \, \bar{q} \; &\longrightarrow \; \cpi \cmj \notag \\
u \, \bar{d} \; &\longrightarrow \; \cpi \nnj \notag \\
d \, \bar{u} \; &\longrightarrow \; \cmi \nnj 
\end{alignat}
The first two processes are possible for a general quark-antiquark
pair. For the latter, charge conservation requires $u$ and $d$ type
quarks in the initial state~\cite{neut_lo}.

\begin{figure}[b] \begin{center}
\epsfig{file=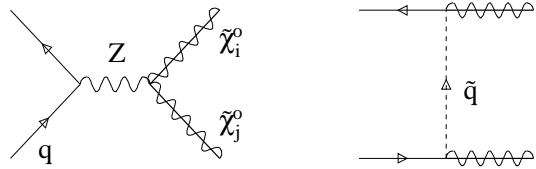,width=8cm}
\end{center}
\caption[]{\it Generic Born diagrams for neutralino/chargino 
  production \label{fig_neut_feyn1}}
\end{figure}

Two generic Born Feynman diagrams contribute
[Fig.~\ref{fig_neut_feyn1}]: an $s$ channel gauge boson ($\gamma,W,Z$)
Drell-Yan like and ($t,u$) channel squark exchange diagrams. Two final
state neutralinos are produced by the first diagram purely as
higgsino-type.  Final state charginos can couple to the $s$ channel
gauge boson as gauginos and as higgsinos. For mixed
neutralino/chargino production the $s$ channel diagram contributes to
all current eigenstates as well. In the given approximation of a
trivial squark CKM matrix, the $t,u$ channel squark couples flavor
conserving to the incoming quark and will therefore be regarded as
light-flavored; the incoming quark originates in the parton density of
the proton, and will consistently be assumed massless. This makes the
higgsino Yukawa coupling vanish for all possible final states. For the
gaugino-like charginos this coupling also vanishes in case of $\sqr$,
since the $q \sq \tilde{\chi}$ coupling respects the helicity
eigenstates.\smallskip

The LO partonic cross section $\hat{\sigma}$, which is proportional to
the matrix element squared in the limit of $(n\to4)$ dimensions can
for all possible final states be written as [The $n$ dimensional Born
cross section is required for the NLO contribution.]
\begin{alignat}{9}
\frac{d \hat{\sigma}_{ij}}{dt} = \; 
\frac{2 \pi \alpha^2}{N s^2} &\Bigg[ \; \;
 \frac{2 m_i m_j}{t_{\sq}} 
  \left( A_{Li} A_{Lj}^* C_{T2} + A_{Ri} A_{Rj}^* C_{T4} \right) 
+\frac{2 t_i t_j}{st_{\sq}}
  \left( A_{Li} A_{Lj}^* C_{T1} + A_{Ri} A_{Rj}^* C_{T3} \right)
\notag \\ &  
-\frac{2 m_i m_j}{u_{\sq}} 
  \left( \Ac_{Lj} \Ac_{Li}^* C_{T1} 
       + \Ac_{Rj} \Ac_{Ri}^* C_{T3} \right) 
-\frac{2 u_i u_j}{su_{\sq}}
  \left( \Ac_{Lj} \Ac_{Li}^* C_{T2} 
       + \Ac_{Rj} \Ac_{Ri}^* C_{T4} \right)
\notag \\ &  
+\frac{t_i t_j + u_i u_j}{s^2} C_{S1}
+\frac{m_i m_j}{s} C_{S2}
+\frac{2( u_1 u_2 - t_1 t_2)}{s(s-M^2)} C_{S3}
\notag \\ &  
-\frac{8 s m_i m_j}{t_{\sq}u_{\sq}}
   \left( A_{Lj} A_{Li}^* \Ac_{Lj} \Ac_{Li}^* 
        + A_{Rj} A_{Ri}^* \Ac_{Rj} \Ac_{Ri}^* \right)
\notag \\ &  
+\frac{8 t_i t_j}{t_{\sq}^2}
   \left( A_{Li} A_{Li}^* A_{Lj} A_{Lj}^* 
        + A_{Ri} A_{Ri}^* A_{Rj} A_{Rj}^* \right)
\notag \\ &  
+\frac{2 m_i m_j}{t_{\sq}}
   \left( A_{Lj} A_{Li}^* C_{T2}^* + A_{Rj} A_{Ri}^* C_{T4}^* \right)
+\frac{2 t_i t_j}{st_{\sq}}
   \left( A_{Lj} A_{Li}^* C_{T1}^* + A_{Rj} A_{Ri}^* C_{T3}^* \right)  
\notag \\ &  
-\frac{8 s m_i m_j}{t_{\sq}u_{\sq}}
   \left( A_{Li} A_{Lj}^* \Ac_{Li} \Ac_{Lj}^* 
        + A_{Ri} A_{Rj}^* \Ac_{Ri} \Ac_{Rj}^* \right)
\notag \\ &  
+\frac{8 t_i t_j}{u_{\sq}^2}
   \left( \Ac_{Li} \Ac_{Li}^* \Ac_{Lj} \Ac_{Lj}^* 
        + \Ac_{Ri} \Ac_{Ri}^* \Ac_{Rj} \Ac_{Rj}^* \right)
\notag \\ &  
+\frac{2 m_i m_j}{u_{\sq}}
   \left( \Ac_{Li} \Ac_{Lj}^* C_{T1}^* 
        + \Ac_{Ri} \Ac_{Rj}^* C_{T3}^* \right)
+\frac{2 t_i t_j}{su_{\sq}}
   \left( \Ac_{Li} \Ac_{Lj}^* C_{T2}^* 
        + \Ac_{Ri} \Ac_{Rj}^* C_{T4}^* \right)  
\Bigg] \notag \\
& \qquad \text{where} \qquad t_{\sq}=t-\ms^2 \qquad u_{\sq}=u-\ms^2
\label{eq_neut_born}
\end{alignat} 
$M$ is the $W$ or $Z$ mass of the $s$ channel gauge boson. The
coupling parameters $A$ correspond to the $t,u$ channel couplings for
the outgoing particles $i,j$ and are defined in
Tab.~\ref{tab_app_feynneut2}. The charge conjugate coupling $A^c$ is
identical to $A$ for the neutralinos. In the chargino case $A_j$ is
the coupling for outgoing $\cmj$ containing the mixing matrix $U$, and
$\Ac_j$ for an outgoing $\cpj$ containing $V$. The typical couplings
$C_S,C_T$ follow from the Feynman rules Fig.~\ref{fig_neut_feyn1}:
\begin{alignat}{9}
C_{S1} =& \; 
  X_c^2 - \frac{X_c s}{2(s-M^2)} \real \left[ (\ell+r)(L+R) \right]
  + \frac{s^2}{4(s-M^2)^2} \, (|\ell|^2+|r|^2) \, 
    (|L|^2+|R|^2) \notag \\
C_{S2} =& \; 
  X_c^2 - \frac{X_c s}{2(s-M^2)} \real \left[ (\ell+r)(L+R) \right]
  + \frac{s^2}{2(s-M^2)^2} \, (|\ell|^2+|r|^2) \, 
    \real \left[ |L||R| \right] \notag \\
C_{S3} =& \; 
  \frac{X_c}{4} \real \left[ (\ell-r)(L-R) \right]
  + \frac{s}{8(s-M^2)} \, (|\ell|^2-|r|^2) \, (|L|^2-|R|^2) \notag \\
C_{T1} =& \; X_c + \frac{s}{s-M^2} \; \ell R \qquad \qquad \qquad
C_{T2} =  \; X_c + \frac{s}{s-M^2} \; \ell L \notag \\
C_{T3} =& \; X_c + \frac{s}{s-M^2} \; r L \qquad \qquad \qquad
C_{T4} =  \; X_c + \frac{s}{s-M^2} \; r R \notag \\
X_c    =& \; - Q \qquad \text{only for} \quad \cmj \cpj
\label{eq_neut_coup}
\end{alignat}
The gauge boson-quark couplings $r,\ell$ are given in
Tab.~\ref{tab_app_feynqcd}, the neutralino-chargino couplings in
Tab.~\ref{tab_app_feynneut1}. Final state charginos require one
subtlety in the matrix elements: either the $t$ or the $u$ channel
diagrams contribute to the amplitude with a fixed quark flavor, except
for the pure neutralino case. For two final state charginos, $A$ only
couples to $u$ type, $\Ac$ to $d$ type quarks. In the mixed production
processes the couplings $A$ and $\Ac$ have to be arranged making use
of charge conservation.

The factor of $C_{S3}$ in the Born cross section
eq.(\ref{eq_neut_born}) originates from the contraction of two CP odd
Dirac traces ${\rm Tr}(\gamma_5 \gamma^\mu \slash{k}_1 \gamma^\nu
\slash{k}_2) {\rm Tr}(\gamma_5 \gamma^\mu \slash{p}_1 \gamma^\nu
\slash{p}_2)$, where the definition of the momenta is given in
appendix~\ref{chap_app_phase}.  Using a naive $\gamma_5$ scheme, this
term cannot be fixed consistently. We therefore keep this kind of
structure in the Born, real gluon and virtual gluon contributions. The
different choices for the $\gamma_5$ scheme result in
$\order(\epsilon)$ corrections and do not contribute to the final
expression, since the corresponding diagrams are finite. The
calculation performed in the consistent 't~Hooft-Veltman
scheme~\cite{dim_reg} agrees with the naive calculation.

\subsubsection{Hadronic Cross Section}

The hadronic cross section for $pp/p\bar{p}$ collisions is given by a
convolution of the partonic cross section with the parton densities
for the quarks in the proton, \eg for two hadrons $H_1 H_2$
\begin{equation}
\sigma(S,Q^2) = \sum_{{\rm partons \,} ij} \;
  \int_{\tau_0}^1 dx_i \int_{\frac{\tau_0}{x_i}}^1 dx_j \; 
   \left[ f^{H_1}_i(x_i,Q^2) \, f^{H_2}_j(x_j,Q^2) 
     + (H_1 \leftrightarrow H_2) \right] \;
   \hat{\sigma}_{ij}(x_ix_jS,Q^2)
\end{equation}
where $k_1$ and $k_2$ are the incoming parton momenta, $S=(k_1+k_2)^2$
is the hadronic cm energy; $m_j$ are the masses of the final state
particles, and $\tau_0=(m_1+m_2)^2/S$ is the kinematical limit.
$f^{H_j}_i$ are the parton densities, forming the convoluted hadronic
luminosity
\begin{alignat}{9}
& \sigma(S,Q^2) \, = \sum_{{\rm partons \,} ij} \;
  \int_{\tau_0}^1 d\tau \; 
  \frac{d{\cal L}_{ij}}{d\tau}(\tau,Q^2) \;
  \hat{\sigma}_{ij}(\tau S,Q^2) \notag \\
& \frac{d{\cal L}_{ij}}{d\tau}(\tau,Q^2) \, =
  \left[ f_i \otimes f_j \right](\tau,Q^2)
  + \left[ f_j \otimes f_i \right](\tau,Q^2) \notag \\
&\left[ f \otimes g \right](\tau,Q^2) \equiv
 \int_\tau^1 \frac{dx}{x} \, f(x,Q^2) \, 
                          g\left(\frac{\tau}{x},Q^2\right)
\label{eq_neut_lumi}
\end{alignat}  
where the hadrons $H_1,H_2$ are implicitly fixed by the order of the
convolution of the parton densities. For identical incoming gluons a
factor 1/2 has to be incorporated.

\section{Next-to-leading Order Cross Sections}

\subsection{Virtual and Real Gluon Emission}

\begin{figure}[t] \begin{center}
\epsfig{file=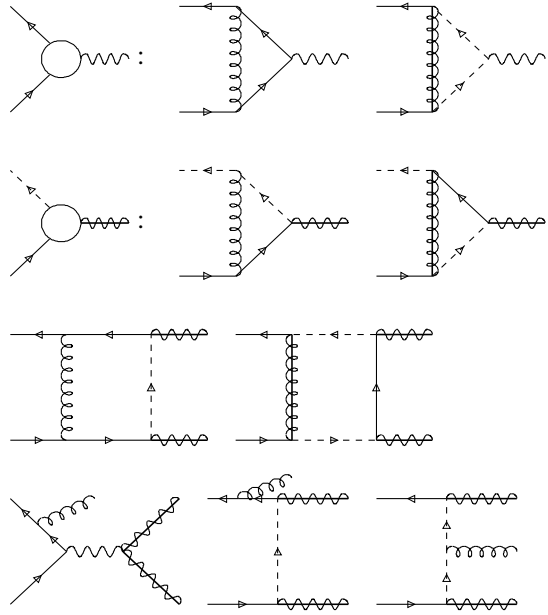,width=8cm}
\end{center}
\caption[]{\it Generic NLO diagrams for neutralino/chargino 
  production, the self energy contributions 
  are not shown. \label{fig_neut_feyn2}}
\end{figure}

The NLO cross section includes the radiation of real quarks and gluons
and virtual gluons and gluinos. The generic diagrams are given in
Fig.~\ref{fig_neut_feyn2} for the $q\bar{q}$ incoming state. The
additional $qg$ and $g\bar{q}$ diagrams are obtained by crossing one
quark to the final and the gluon to the initial state. The virtual
contributions are regularized by dimensional regularization. Therefore
a finite shift of the couplings eq.(\ref{eq_susy_yukshift}) has to be
applied to restore supersymmetry.  The divergences appear as poles in
$\epsilon$, as shown in appendix~\ref{chap_app_virt}. The UV poles
require renormalization; the only parameter in the Born term
eq.(\ref{eq_neut_born}) which undergo the renormalization procedure is
the squark mass, defined as the pole mass, \ie in the on-shell scheme.
The soft gluon poles cancel with the real gluon emission. The phase
space integration for the real gluon emission is given in
appendix~\ref{chap_app_phase}. These matrix elements have been
computed using phase space subtraction, \ie the additional gluon phase
space is integrated numerically. After subtracting the dipole terms
the remaining divergences are of collinear type and removed by mass
factorization, appearing in the subtraction term, see
appendix~\ref{chap_app_subtract}.

\subsection{Mass Factorization}
\label{sect_neut_massfac}

The parton densities eq.(\ref{eq_neut_lumi}) form observable structure
functions [\eg $F_2$], which contain divergences in next-to-leading
order QCD~\cite{aem}.  These divergences arise from the collinear
radiation of gluons and have a universal structure which is fixed by
the $Q^2$ evolution.  They have to be absorbed into the definition of
the parton densities to render the physical structure function finite.
In analogy to a UV renormalization procedure it is possible to absorb
additional finite parts into the re-definition. The minimal set is the
$\msbar$ scheme, and it leaves the next-to-leading order contribution
to the measured structure function with a non-zero finite term.  This
minimal choice respects the required sum rules naively.\medskip

Due to the factorization theorem, the universal form of the partonic
cross section in the collinear limit is independent of the order of
perturbation theory.
\begin{alignat}{9}
s^2 \frac{d^2\hat{\sigma}_{ij}}{dt_2ds_4} =& \; 
 \int_0^1 \frac{dx_i}{x_i} \int_0^1 \frac{dx_j}{x_j} \;
 \Gamma_{li}(x_i,Q^2) \, \Gamma_{mj}(x_j,Q^2)
 \left( s^2 \frac{d^2\hat{\sigma}_{lm}^{\rm red}}{dt_2ds_4}
 \right)_{x_ik_i, x_jk_j} \notag \\
&\Gamma_{ij}(x,Q^2) = \;
 \delta_{ij} \, \delta(1-x) - 
 \frac{\alpha_s}{2\pi\epsilon} \, 
 \frac{\Gamma(1-\epsilon)}{\Gamma(1-2\epsilon)}
 \left( \frac{4\pi Q^2}{Q_F^2} \right)^\epsilon
    P_{ij}(x) 
\label{eq_neut_split}
\end{alignat}
$\Gamma_{ij}$ is called splitting function and describes the splitting
of a parton $i$ to a parton $j$ in the collinear limit. It is
evaluated perturbatively and consists of the trivial LO term and a
divergent NLO contribution. The appearance of the Altarelli-Parisi
kernels $P_{ij}$ fixes the $Q^2$ evolution, they are given in
eq.(\ref{eq_app_ap}). Other non-minimal schemes lead to a finite
renormalization $\Gamma_{ij} \to \Gamma_{ij}+f_{ij}$.  The reduced
cross section $\hat{\sigma}^{\rm red}$ is finite and, as well as the
splitting function, depends on the factorization scale $Q_F$. This
scale dependence should flatten after adding higher order perturbative
contributions, since it is a perturbative artifact.

The renormalization of the parton densities has to cancel the
remaining collinear poles in the matrix elements and leave the final
expression finite. The counter term which has to be added to the bare
cross section to obtain the reduced one in the $\msbar$ scheme can be
read off eq.(\ref{eq_neut_split})
\begin{alignat}{9}
s^2 \frac{d^2 \hat{\sigma}^{MF}_{ij}}{dt_2ds_4} =
 \frac{\alpha_s}{2\pi\epsilon}
  \frac{\Gamma(1\!-\!\epsilon)}{\Gamma(1\!-\!2\epsilon)} 
  \left( \frac{4\pi Q^2}{Q_F^2} \right)^\epsilon 
  \int_0^1 \frac{dx}{x} 
\left[ 
  P_{li}(x) \!
\left( s^2 \frac{d^2 \hat{\sigma}^B_{lj}}{dt_2 ds_4} \right)_{x k_l} 
+ P_{mj}(x) \!
\left( s^2 \frac{d^2 \hat{\sigma}^B_{im}}{dt_2 ds_4} \right)_{x k_m} 
\right] \notag \\
\end{alignat}

\subsection{On-Shell Subtraction}
\label{sect_neut_os}

\begin{figure}[t] \begin{center}
\epsfig{file=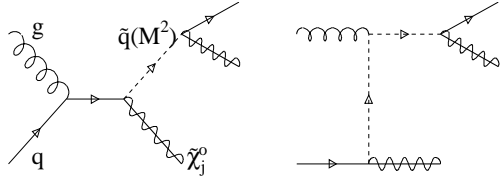,width=7cm}
\end{center}
\caption[]{\it Feynman diagrams for crossed channel production 
  of neutralinos/charginos including on-shell 
  intermediate states, which have to be subtracted. 
  \label{fig_neut_feyn3}}
\end{figure}

Apart from the UV and IR divergences another kind of divergences can
occur, due to on-shell intermediate particles. After crossing the NLO
production matrix elements, different incoming states may contribute to
the ($\chi \chi$+jet) inclusive final state  
\begin{alignat}{9}
q g       &\longrightarrow \chi_i \chi_j q \notag \\
g \bar{q} &\longrightarrow \chi_i \chi_j \bar{q} 
\end{alignat}

As depicted in Fig.~\ref{fig_neut_feyn3}, these can proceed via an
on-shell squark. A natural way of solving the problem would be
introducing finite widths for all particles under consideration.
However, a finite squark width would spoil gauge invariance. In
addition, it would yield a strong dependence of the next-to-leading
order production cross section on the physical widths of intermediate
states. This dependence would only vanish after including the decays
into the calculation. Therefore we instead differentiate between
off-shell and on-shell particle contributions, the latter regarded as
final states in the set of supersymmetric production cross
sections.\smallskip

Considering an analysis of all production processes for two MSSM
particles at hadron colliders this differentiation removes a double
counting of the on-shell contributions of the squark, as it would
occur in the case of general finite widths:
\begin{alignat}{4}
&gq \to \sq^* \chi_i \to q \chi_j \chi_i 
\quad && \text{neutralino pair production} \notag \\
&gq \to \sq \chi_i \cdot \br (\sq \to q\chi_j) 
\quad && \text{squark neutralino production}
\label{eq_neut_double}
\end{alignat}
The on-shell squark contribution is subtracted from the crossed
$\chi_i \chi_j$ production matrix element, leaving it as a
contribution to direct $\chi \sq$ production,
eq.(\ref{eq_neut_double}). The off-shell contribution is kept for the
first of the processes under consideration. To distinguish these
contributions numerically, one regularizes the possibly divergent
propagator by introducing the Breit-Wigner propagator $(p^2-m^2) \to
(p^2-m^2+im\Gamma)$. Since this width can be regarded not as a
physical property of the final state particle, but as a mathematical
cut-off, the matrix element can be evaluated in the narrow width
approximation, regarding the final state particles as
quasi-stable.\medskip

Assuming an on-shell divergence in the variable $M^2$, the hard
production cross section in the narrow width approximation reads
\begin{alignat}{9}
\frac{d \sigma}{d M^2} &= \;
 \sigma\left( gq \to \sq\chi_i \right) \,
 \frac{\ms \Gamma_{\sq}/\pi}{(M^2-\ms^2)^2 + \ms^2\Gamma_{\sq}^2} \,
  \br \left( \sq \to q\chi_j \right) 
 + \order \left( \frac{1}{M^2-\ms^2} \right) 
\notag \\
&\longrightarrow \; 
 \sigma\left( gq \to \sq\chi_i \right) \,
  \br \left(\sq \to q\chi_j \right) \,
 \delta(M^2-\ms^2) + \order \left( \frac{1}{M^2-\ms^2} \right) 
\end{alignat}

\begin{figure}[b] \begin{center}
\epsfig{file=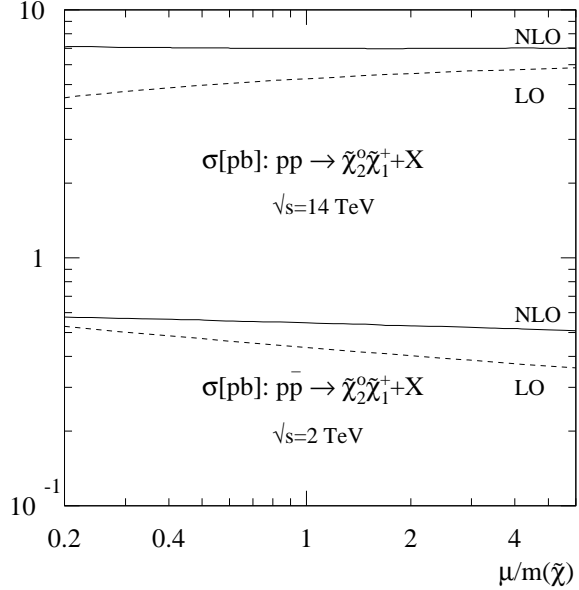,width=8cm}
\end{center}
\vspace*{-0.5cm}
\caption[]{\it The renormalization/factorization scale dependence 
  of the total cross section for $\nnz \cpe$ production at the 
  upgraded Tevatron and the LHC. There is no maximum in either 
  of the next-to-leading order curves, and the LO and NLO do not 
  meet for a scale around the average mass. 
  The SUSY scenario determining the masses
  is given in eq.(\ref{eq_susy_scen}).
  \label{fig_neut_scale}}
\end{figure}

In case of the neutralino/chargino production $M^2=s_3+m_1^2$ and
$M^2=s_4+m_2^2$ are relevant for the on-shell squarks, the extended
set of Mandelstam variables is defined in appendix~\ref{chap_app_rad}.
The leading divergence is subtracted from the crossed channel matrix
element, as described before. The complete crossed channel matrix
element can be written as $|\M|^2=f(M^2)/[(M^2-\ms^2)^2 +
\ms^2\Gamma_{\sq}^2]$; then the subtraction for an intermediate squark
is defined as
\begin{equation}
\frac{f(M^2)}{(M^2-\ms^2)^2+\ms^2\Gamma_{\sq}^2} 
\, - \, \frac{f(\ms^2)}{(M^2-\ms^2)^2+\ms^2\Gamma_{\sq}^2}
 \, \Theta(\hat{s}-(\ms+m_i)^2)
 \, \Theta(\ms-m_j)
\end{equation}
Since an over-all factor $\delta(M^2-\ms^2)$ is absent in the
subtracted term, the Breit-Wigner propagator has to be integrated over
the phase space variable $M^2$. The matrix
element, including the remaining phase space integration is evaluated
for $M^2=\ms^2$. 

The remaining non-leading divergences, arising from interference
between finite and divergent Feynman diagrams, are integrable and
well-defined using a principal-value integration. Numerically this
principal value can be implemented by introducing a small imaginary
part ($M^2 \to M^2-i \varepsilon$). Since the matrix element squared
may contain subtractions in more than one variable this imaginary part
may lead to finite contributions and has therefore to be taken into
account. 

\section{Results}

\begin{table}[b] \begin{center} \begin{small} 
\begin{tabular}{|c|c||c|c||c|c||c|c|}
\hline
      $\nni \nnj$
    & $K_{\rm LHC}$ \rule[0mm]{0mm}{5mm}
    & $\nni \cpj$
    & $K_{\rm LHC}$ 
    & $\nni \cmj$
    & $K_{\rm LHC}$ 
    & $\cpi \cmj$ 
    & $K_{\rm LHC}$ \\[2mm] \hline
      $\nne \nne$  \rule[0mm]{0mm}{5mm}
    & 1.51
    & $\nne \cpe$
    & 1.35
    & $\nne \cme$
    & 1.37
    & $\cpe \cme$
    & 1.33 \\
      $\nne \nnz$  \rule[0mm]{0mm}{5mm}
    & 1.50
    & $\nnz \cpe$
    & 1.33
    & $\nnz \cme$
    & 1.34
    & $\cpe \cmz$
    & 1.44 \\
      $\nne \nnd$  \rule[0mm]{0mm}{5mm}
    & 1.35
    & $\nnd \cpe$
    & 1.35
    & $\nnd \cme$
    & 1.33
    & $\cpz \cme$
    & 1.41 \\
      $\nne \nnv$  \rule[0mm]{0mm}{5mm}
    & 1.39
    & $\nnv \cpe$
    & 1.90
    & $\nnv \cme$
    & 1.98
    & $\cpz \cmz$
    & 1.32 \\
      $\nnz \nnz$  \rule[0mm]{0mm}{5mm}
    & 1.44
    & $\nne \cpz$
    & 1.38
    & $\nne \cmz$
    & 1.40
    & 
    & \\
      $\nnz \nnd$  \rule[0mm]{0mm}{5mm}
    & 1.35
    & $\nnz \cpz$
    & 2.51
    & $\nnz \cmz$
    & 2.65
    & 
    & \\
      $\nnz \nnv$  \rule[0mm]{0mm}{5mm}
    & 1.45
    & $\nnd \cpz$
    & 1.35
    & $\nnd \cmz$
    & 1.34
    & 
    & \\
      $\nnd \nnd$  \rule[0mm]{0mm}{5mm}
    & 1.30
    & $\nnv \cpz$
    & 1.31
    & $\nnv \cmz$
    & 1.32
    & 
    & \\
      $\nnd \nnv$  \rule[0mm]{0mm}{5mm}
    & 1.33
    & 
    & 
    & 
    & 
    & 
    & \\
      $\nnv \nnv$  \rule[0mm]{0mm}{5mm}
    & 1.38
    & 
    & 
    & 
    & 
    & 
    & \\[1mm]  \hline
\end{tabular} \end{small} \end{center}
\caption[]{\it A complete set of $K$ factors for neutralino 
  and chargino production at the LHC. The masses are chosen 
  according to the default SUGRA inspired scenario,
  eq.(\ref{eq_susy_scen}). The renormalization and 
  factorization scales are set to the average final state mass.
  Although the $K$ factors are of a similar size $1.3 \cdots 1.5$ for 
  each diagram contributing, large cancelations lead to huge
  corrections for the scenario under consideration. 
  \label{tab_neut_kfac}}  
\end{table}

\subsubsection{Scale Dependence}

Since the leading order hadro-production cross section for neutralinos
and charginos does not contain the QCD coupling constant, it only
depends on the factorization scale through the parton densities. This
renders the leading order scale dependence smaller than $\sim 30\%$.
The variation of the cross section with the scale is therefore not a
good measure for the theoretical uncertainty. In next-to-leading
order, this factorization scale dependence becomes weaker; however, an
additional dependence on the renormalization scale arises.  For
$\mu=\mu_F=\mu_R$ this yields a generally weak scale dependence of
$\lesssim 20\%$ at the upgraded Tevatron and $\lesssim 5\%$ at the
LHC. As can be seen from the leading order curves in
Fig.~\ref{fig_neut_scale}, the combination of factorization and
renormalization scale dependence leads to a different behavior at the
Tevatron and at the LHC, due to different momentum fractions $x$
contributing; in contrast to the strong coupling induced processes a
maximum cross section for some small scale does not
occur,~Fig.~\ref{fig_neut_scale}.

\subsubsection{Numerical Results}

\begin{figure}[p] \begin{center}
    \epsfig{file=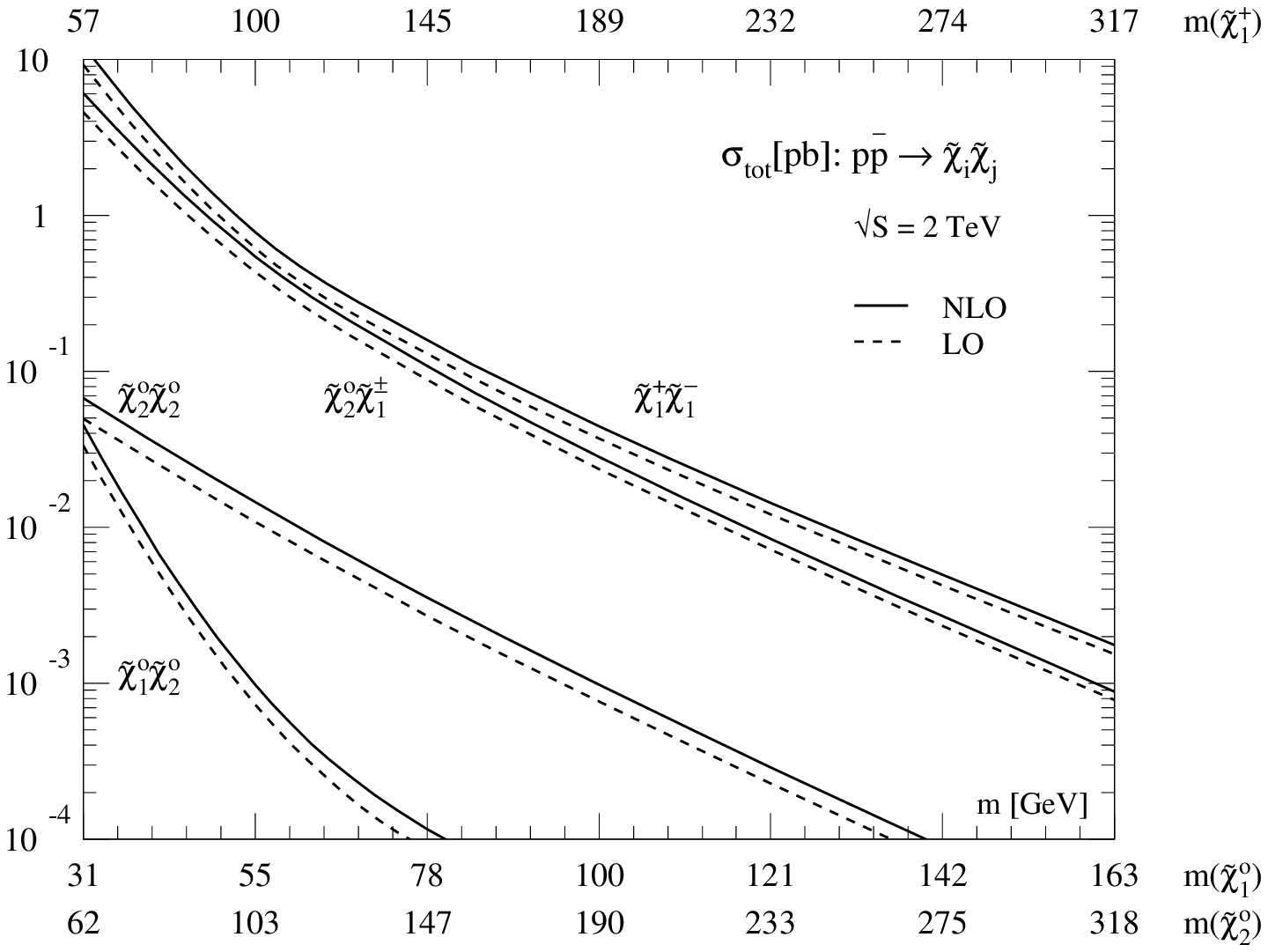,width=12cm}
    \epsfig{file=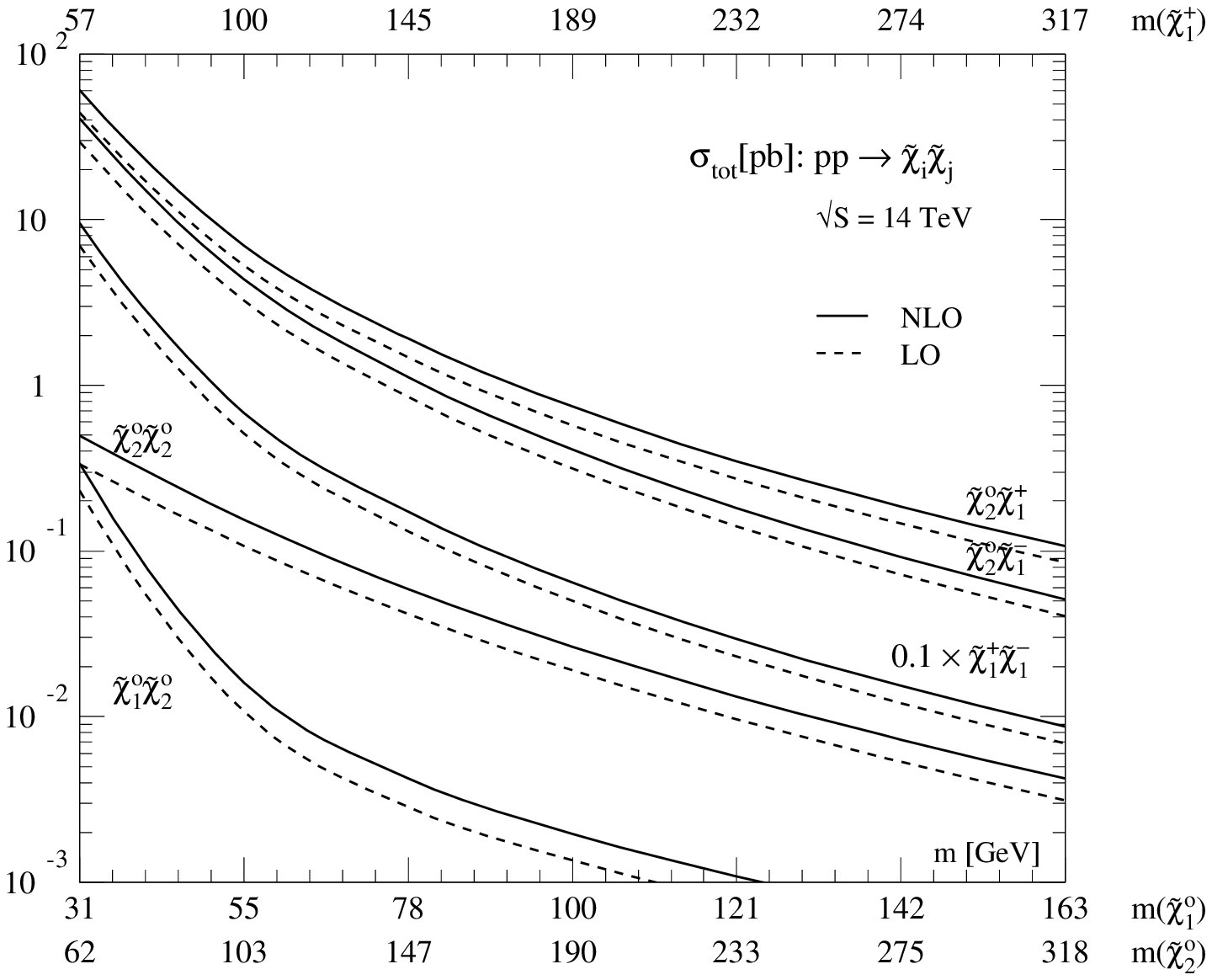,width=12cm}
\end{center}
\vspace*{-0.5cm}
\caption[]{\it Some total cross sections for pairs of neutralinos and 
  charginos at the Tevatron and at the LHC as a function of $m_{1/2}$. 
  The different masses
  of the particles involved are given on secondary axis. The 
  strongly suppressed heavy higgsino cross sections are 
  not given. The $\nnz \cpe$ and $\nnz \cme$ cross sections 
  are different due to the non-symmetric parton luminosities.
  \label{fig_neut_resu}}
\end{figure}

The production of neutralinos and charginos can be probed at the
upgraded Tevatron, a $p\bar{p}$ collider with a center-of-mass energy
of 2$\tev$, and at the future LHC, a $pp$ collider with an energy of
14$\tev$. The cross section for several combinations of light
neutralinos and charginos, which turn out to be gaugino-like in the
considered scenario, are given in Fig.~\ref{fig_neut_resu}. The size
of the cross sections strongly depends on the mixing matrix elements
associated with the different couplings. This yields \eg a larger
cross section for $\nnz \cpe$ pairs compared to $\nne \cpe$
production. In general, the processes containing no final state
chargino are suppressed, independent of the masses, which are almost
the same for $\nnz$ and $\cpe$. Whereas the cross section for the
production of positively and negatively charged mixed pairs are
identical at the Tevatron, they differ significantly at the LHC, due
to non-symmetric parton luminosities. The dependence on SUSY masses
and parameters, which are not contained in the leading order cross
section, like the gluino mass, is weak in next-to-leading order.  The
virtual corrections are generically small [$\lesssim 10\%$] compared to
the real gluon emission; however, they are not universal and even do
not have a unique sign for the different gaugino and higgsino-type
outgoing particles.\smallskip

The next-to-leading order $K$ factor is consistently defined as
$K=\sigma_{NLO}/\sigma_{LO}$. It is dominated by the gluon emission
off the incoming partons and therefore similar for all considered
processes and a constant function of the masses,
Tab.~\ref{tab_neut_kfac}. Although the real gluon corrections to any
diagram contributing to the production process are of the order $1.3
\cdots 1.5$, large cancelations give rise to huge $K$ factors. The
same effect occurs for the virtual corrections, which grow up to
50$\%$ \eg for the $\nnz \tilde{\chi}_2^\pm$ or the $\nnv
\tilde{\chi}_1^\pm$ channel. Varying the common gaugino mass $m_{1/2}$
reduces the $K$ factor to values expected by regarding the other
channels.

With an integrated luminosity of $\int {\cal L}=20\fb^{-1}$ in run~II,
the upgraded Tevatron will have a maximal reach for the mass of the
produced particles when probing the $\nnz \cpe$ channel. For masses
smaller than 150$\gev$, $10^3$ to $10^5$ events could there be
accumulated.  Although the $\cpe \cme$ cross section is compatible
with the mixed neutralino/chargino channel for a fixed value of the
common gaugino mass, the particle masses, which can be probed, stay
below 80$\gev$ in the considered SUGRA inspired scenario.  The same
holds for the LHC, where for typical masses of the $\cpe$ and $\nnz$
below 300$\gev$ and an integrated luminosity of $\int {\cal
  L}=300\fb^{-1}$ a sample of $10^4$ to $10^6$ events can be
accumulated. In the given scenario the higgsino type neutralinos and
charginos are strongly suppressed compared with the lighter gauginos.

\chapter{Scalar Top Quark Decays}
\label{chap_decay}

Scalar top quarks can decay into two or three on-shell particles via
the strong or electroweak coupling~\cite{porod}. The possible two body
decays are --- kinematically allowed for an increasing stop mass in
typical mass scenarios:
\begin{alignat}{9}
\stj &\longrightarrow \; c \nne, \quad b \cpi, \quad t \nni, \quad 
 [\; Z \ste, \quad
     h \ste, \quad 
     W^+ \sbx, \quad 
     H^+ \sbx, \quad 
     t \gt         \;]
 \notag \\
\stj &\longrightarrow \; W^+ b \nne, \quad H^+ b \nne \quad \cdots 
\label{eq_dec_channels}
\end{alignat}
The channels in brackets are possible only for the heavier stop, since
the $\ste$ is assumed to be the lightest scalar quark. The decay into
a charm jet is induced by a one-loop amplitude, and will therefore be
suppressed, if any other tree-level two or three body decay channel is
open. In the intermediate mass range, when the $b \cpi$ channel is
still closed, the three particle decay into $W b \nne$ is dominant.
For a heavy $\stz$ the strong decay mode including a final state
gluino will be the leading one, as will be shown later in this chapter.

\section{Strong Decays}
\label{sect_dec_strong}

\subsection{Born Decay Widths}

Since the Yukawa $q \sq \gt$ couplings are flavor diagonal, any decay
involving a scalar top quark
\begin{alignat}{9}
\stj \longrightarrow& \; t + \gt 
 \qquad \qquad 
 &&\left[ \msj > \mt + \mg \right] \notag \\ 
\gt \longrightarrow& \; \tb + \stj \quad \text{and c.c.} 
 \qquad \qquad 
 &&\left[ \mg > \mt + \msj \right]
\label{eq_dec_strong}
\end{alignat}
includes a top quark in the final state, \ie the strong decays will
only be possible for large mass scenarios. For the light stop $\ste$
 the weak decays in eq.(\ref{eq_dec_channels}) will be the only
kinematically allowed.

The calculation including the stop mixing and a massive top quark is a
generalization of the light-flavor decay width~\cite{roland_dec}. To
lowest order the partial widths for the stop
and gluino decay, eq.(\ref{eq_dec_strong}), are given by
by\footnote{$\Lambda(x,y,z)=x^2+y^2+z^2-2(xy+xz+yz)$}
\begin{alignat}{2}
\Gamma (\stez \rightarrow t\,\gt) &= 
  \frac{2 \as}{N \msez^3} \Lambda^{1/2}(\msez^2,\mt^2,\mg^2)
  \left[ \msez^2 - m_t^2 - \mg^2 \pm 2 m_t \mg \sin(2 \tmix) \right]  
  \notag \\
\Gamma (\gt \rightarrow \tb\,\stez) &=
  - \frac{\as}{(N^2-1) \mg^3} \Lambda^{1/2}(\msez^2,\mt^2,\mg^2) 
  \left[ \msez^2 - m_t^2 - \mg^2 \pm 2 m_t \mg \sin(2 \tmix) \right] 
\label{eq_dec_born_s}  
\end{alignat}
 The different factors in front are due to the color and spin
averaging of the decaying particle, and the crossing of a fermion
line. Interchanging $\ste$ and $\stz$ in the two leading-order decay
widths corresponds to the symmetry operation $\PX$ in the Lagrangean,
as described in section~\ref{sect_susy_stop}. 

\subsection{Next-to-leading Order SUSY-QCD Corrections}
\label{sect_dec_nlo}

\subsubsection{Massive Gluon Emission}

\begin{figure}[t] \begin{center}
\epsfig{file=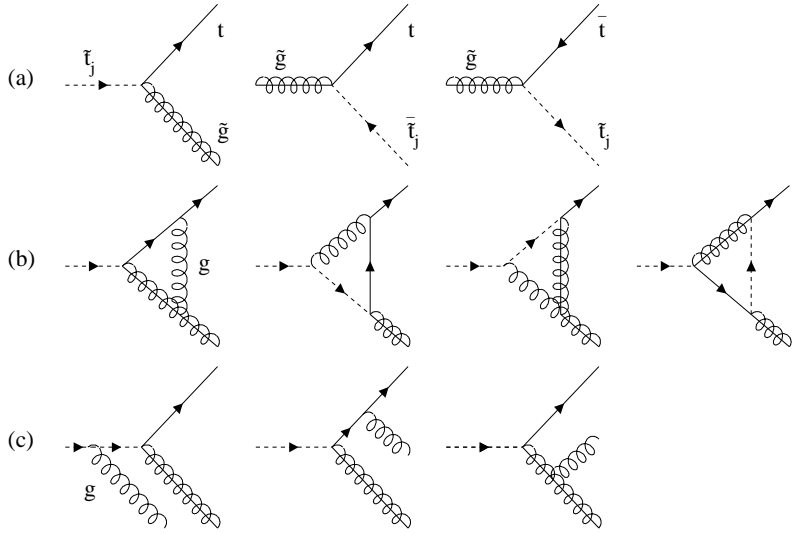,width=11cm}
\end{center}
\caption[]{\it (a) Born diagrams for stop and gluino decays; (b)
  vertex corrections; (c) real gluon emission. The correction to 
  the gluino decay can be obtained by crossing the diagrams
  \label{fig_dec_feyn}}
\end{figure}

The NLO corrections~\cite{ourdecay} include the emission of an
on-shell gluon, Fig.~\ref{fig_dec_feyn}c. This gluon leads to IR
singularities which are regularized using a small gluon mass
$\lambda$, subsequently appearing in logarithms $\log \lambda^2$.  The
massive gluon scheme breaks gauge invariance for the non-abelian SU(3)
symmetry. Hence the scheme has to be extended by new counter terms if
a non-abelian contribution arises from a three or four gluon vertex,
otherwise the SU(3) Ward identities would not be satisfied anymore.
This is not the case for the stop decays Fig.~\ref{fig_dec_feyn}. The
gluon behaves like a photon and its mass can be regarded as a
mathematical cut-off parameter. After integration over the whole phase
space the small mass parameter drops out and yields a finite sum of
virtual and real gluon matrix elements.  However, these massive gluon
matrix elements must not be interpreted as exclusive cross sections,
since gauge invariance is only restored for inclusive observables, \ie
the gluon integrated out.

In the considered process the logarithms of the gluon mass arise from
the integration over the soft and collinear divergent three particle
phase space, eq.(\ref{eq_app_dennerint}). The same kind of logarithms
enter through the virtual gluon contributions, \eg the scalar three
point function eq.(\ref{eq_app_massgluon}) and cancel
analytically.

\subsubsection{Virtual Corrections}

\begin{figure}[t] \begin{center}
\epsfig{file=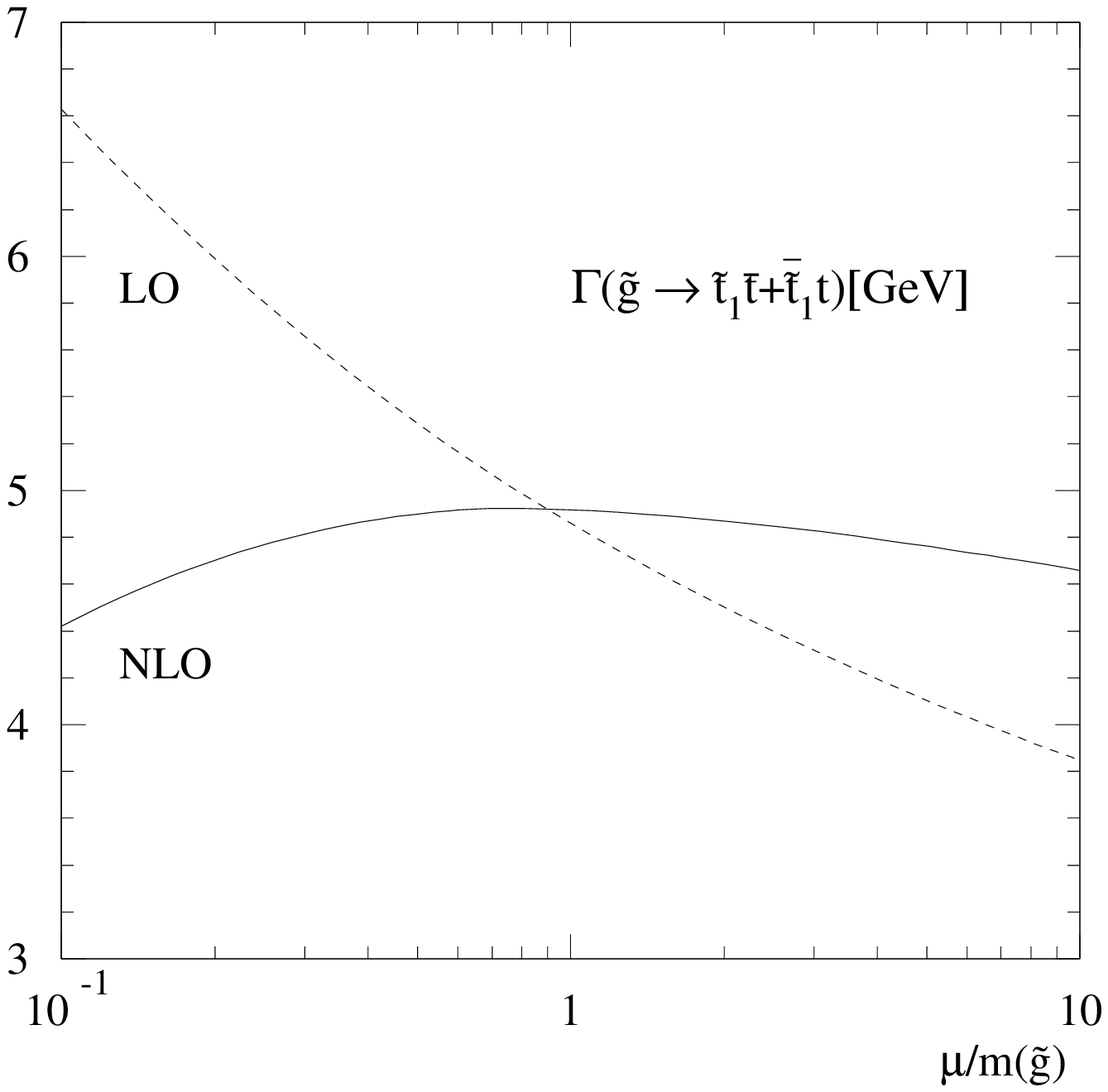,width=7cm}
\hspace*{0.5cm}
\epsfig{file=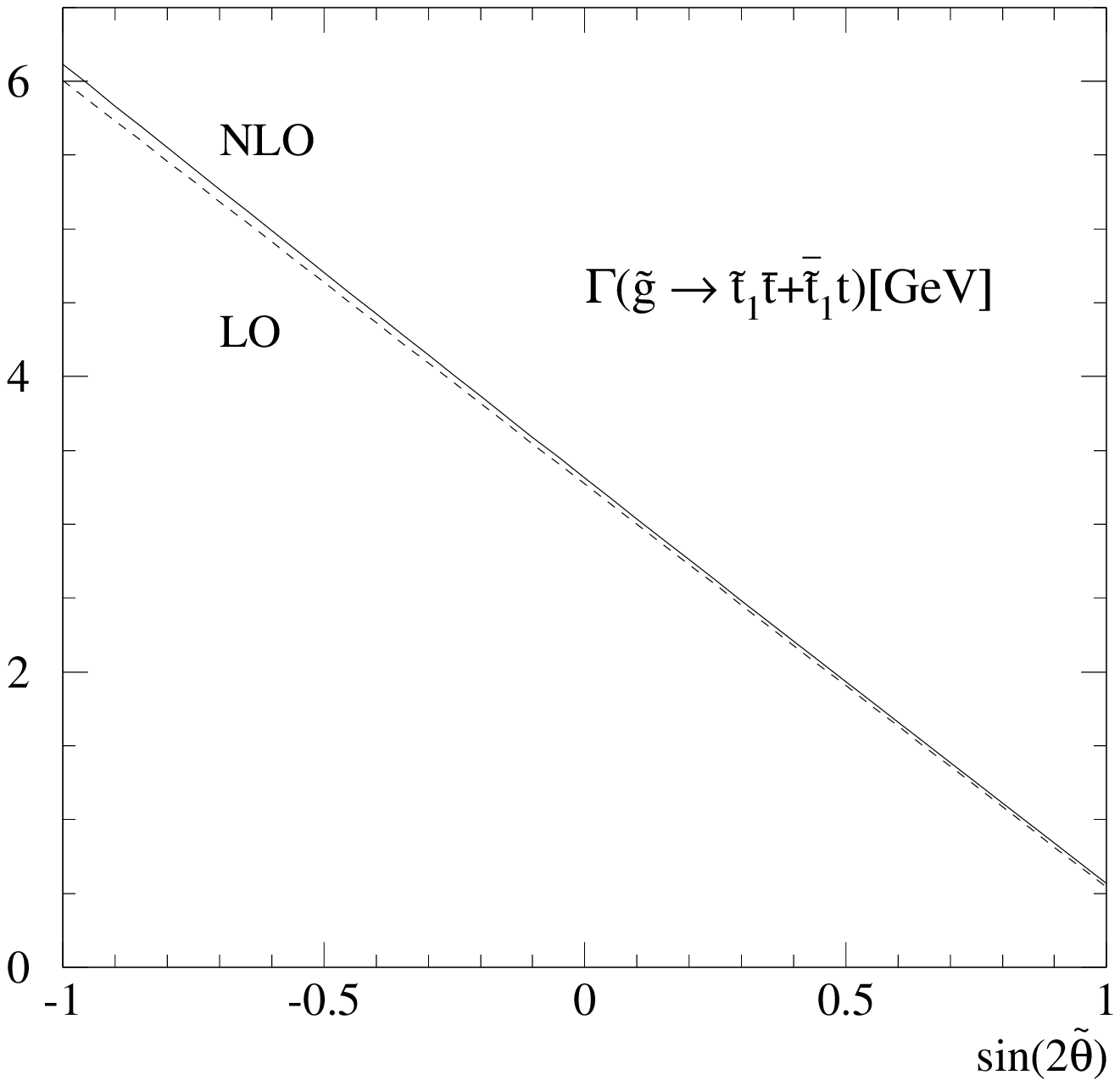,width=7cm}
\end{center}
\vspace*{-1cm}
\caption[]{\it Left: Renormalization scale dependence
  of the gluino decay width for the same SUGRA inspired scenario 
  as chosen in Fig.~\ref{fig_dec_resu}: $\mse=449\gev$,$\msz=847\gev$,
  $\sin(2\tmix)=-0.59$, and $\mg=637\gev$. The renormalization scale 
  is varied as a fraction of the mass of the decaying gluino;
  Right: mixing angle dependence of this gluino decay, where the 
  stop mixing has been varied over the whole range, independent
  of the other low energy parameters.
  \label{fig_dec_small}}
\end{figure}

The virtual gluon corrections, including self energy diagrams for all
external particles and vertex corrections Fig.~\ref{fig_dec_feyn}c,
are also regularized using the massive gluon scheme. The additional UV
divergences have to be regularized dimensionally. The poles
$1/\epsilon$ are absorbed into the renormalization of the masses, the
strong coupling, and the mixing angle, which are the parameters
appearing in the Born decay width eq.(\ref{eq_dec_born_s}). The
counter terms for mass renormalization in the on-shell scheme and the
renormalization of $\alpha_s$ in $\msbar$ can be found in
appendix~\ref{chap_app_counter}.  The mixing angle is renormalized by
introducing the running mixing angle and absorbing the mixing stop
self energy contributions. This scheme restores the ($\ste
\leftrightarrow \stz$) symmetry $\PX$ in NLO. The dependence on the
mixing angle in NLO can be described by a constant $K$ factor,
Fig.~\ref{fig_dec_small}, possibly large contributions from the
gluino-top loop are absorbed into the definition of the mixing angle.
Renormalizing the strong coupling in the $\msbar$ scheme breaks
supersymmetry; adding a finite counter term, derived in
eq.(\ref{eq_susy_coupshift}), restores supersymmetry.\smallskip

The Born decay widths are proportional to $\Lambda^{1/2}$, \ie the
relative momentum of the produced particles. One of the vertex
correction diagrams is constructed by exchanging a virtual gluon
between outgoing color charged particles. Near threshold the exchange
of a gluon between two slowly moving particles picks up a factor
$\Lambda^{-1/2}$, the Coulomb singularity, which cancels against the
phase space suppression factor in the virtual correction matrix
element. The NLO decay width therefore does not vanish at threshold.
The narrow divergence can be removed by resummation of the
contributions near threshold. Moreover, the screening due to a
non-zero life time of the final state particles reduces the Coulomb
effect considerably.\medskip

\begin{figure}[t] \begin{center}
\epsfig{file=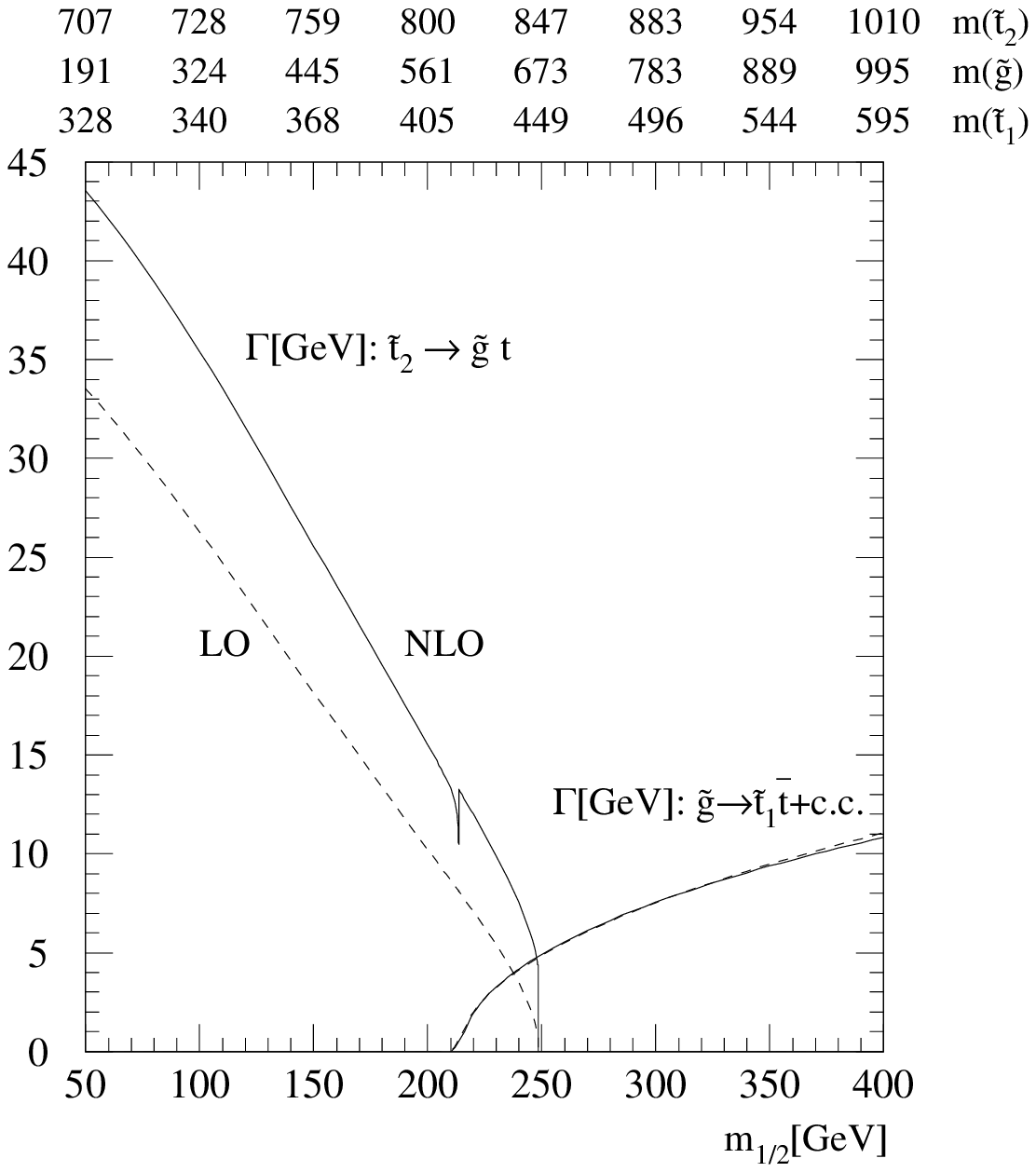,width=8.2cm}
\hspace*{-.6cm}
\epsfig{file=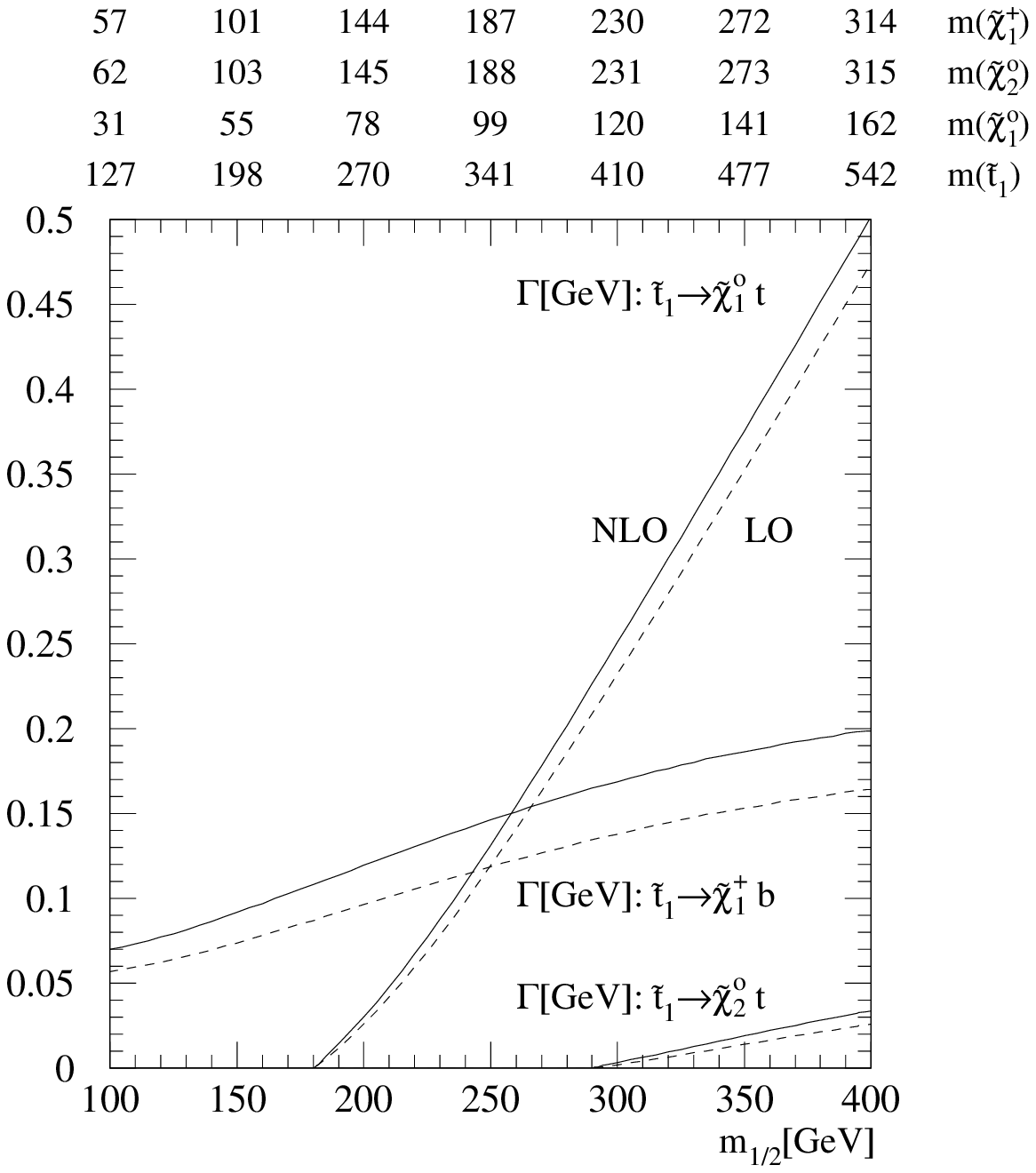,width=8.2cm}
\end{center}
\vspace*{-1cm}
\caption[]{\it Left: SUSY-QCD corrections to the strong decays 
  $\stz \to \gt t$ and $\gt \to t \steb + \tb \ste$ as a function of 
  the common gaugino mass. The masses of the particles involved 
  are labeled in the additional axis. The input parameters have 
  been chosen as $m_0=900\gev, A_0=900\gev, \tan \beta=2.5, \mu>0$,
  in order to see the possible structures of the curves. The kink in 
  the NLO stop decay width results from the gluino self energy and 
  could be smoothed by inserting a finite gluino width; 
  Right: SUSY-QCD corrections to weak decays of the light $\ste$. The 
  input parameters are the usual $m_0=100\gev, A_0=300\gev,
  \tan \beta=4, \mu>0$. Only the decay into the two lightest 
  neutralinos and into the light chargino is possible in the 
  $m_{1/2}$ range considered. Note that the mass of the $\cpe$ and 
  the $\nnz$ are almost identical in the SUGRA inspired scenario.
  The dashed line denotes the LO, the solid one the NLO results 
  in both figures.
  \label{fig_dec_resu}}
\end{figure}

The complete analytical expression for the stop decay width is given
in appendix~\ref{chap_app_dec}. The numerical results are shown
together with the weak decays in Fig.~\ref{fig_dec_resu}.

\section{Weak Decays}
\label{sect_dec_weak}

The possible weak decay modes including a stop will be dominant once
the strong channels  are kinematically forbidden. Although this
region is not preferred by the mSUGRA scenario even the crossed top
decay could be possible, which leads to experimental limits on the
branching ratio of this decay mode and thereby on the masses
involved~\cite{tev_search}
\begin{alignat}{9}
\stj \longrightarrow& \; t + \nnj 
 \qquad \qquad 
 &&\left[ \msj > \mt + \mnj \right] \notag \\ 
\stj \longrightarrow& \; b + \cpj 
 \qquad \qquad 
 &&\left[ \msj > m_b + \mnj \right] \notag \\ 
t \longrightarrow& \; \ste + \nnj
 \qquad \qquad 
 &&\left[ \mt > \mse + \mnj \right]
\label{eq_dec_weak}
\end{alignat}
The Born decay width for the $\ste$ decay to a neutralino
reads\footnote{This decay width has also been calculated in NLO by
  other groups~\cite{wien}; we have analyzed it for the sake of
  comparison and to illustrate the running mixing angle. The three
  calculations are in agreement.}
\begin{alignat}{9}
\Gamma (\ste \rightarrow t\,\nnj) =& \; 
  \frac{2\alpha}{N \mse^3} \Lambda^{1/2}(\mse^2,\mt^2,\mnj^2)
  \left[ \left( \mse^2 - m_t^2 - \mnj^2 \right) (C_L^2+C_R^2) 
         + 4 m_t \mnj C_L C_R \right]  
  \notag \\
  C_L =& \; A_L \cos \tmix + B_L \sin \tmix \notag \\ 
  C_L =& \; B_R \cos \tmix - A_R \sin \tmix  
\label{eq_dec_weaklo}
\end{alignat}
The couplings $A$ and $B$ are given in Tab.~\ref{tab_app_feynneut2}
for the neutralino involved. The $\stz$ decay can be derived using the
$\PX$ operation. The decay channel producing a bottom quark and a
chargino can be read off using Tab.~\ref{tab_app_feynneut2} by setting
the mixing angle to zero, as long as sbottom mixing is neglected. The
NLO calculation is performed exactly as for the strong decay channel,
whereby some virtual and real correction diagrams in
Fig.~\ref{fig_dec_feyn} vanish for a Majorana particle without color
charge. Again the finite shift eq.(\ref{eq_susy_yukshift}) has to be
added to the weak coupling vertex, no matter if a gaugino or a
higgsino is involved.

\section{Results}
\label{sect_dec_results}

In the calculation the renormalization scale of the process is fixed
to the mass of the decaying particle. Since the scale dependence should
vanish after adding all orders of perturbation theory one expects the
variation of the width with the scale to be weaker  in NLO than in
LO. This is shown in Fig.~\ref{fig_dec_small} for the strong coupling
gluino decay.\smallskip 

The numerical results for the strong decay channels can be seen in
Fig.~\ref{fig_dec_resu}. Assuming for illustration a SUGRA inspired
mass spectrum the light stop $\ste$ can decay only via the weakly
interacting channels. The strong decays are possible for the gluino
and for the heavy stop. With increasing $m_{1/2}$ the gluino becomes
heavier compared with the stop masses, \ie the decay into the gluino
vanishes and the gluino decay channel $\gt \to \steb t+\ste \tb$
opens. A kink in the NLO $\stz$ decay widths occurs at the production
threshold $\gt\to t \ste$, where the gluino self energy exhibits a
large discontinuity. It can be smoothed out by introducing a finite
width of the gluino.  The Coulomb singularity is present in both of
the strong decay channels.  However, it can be seen only in the stop
decay, since the kink in the gluino self energy and the Coulombic
vertex contribution to the NLO decay width cancel each other
numerically near threshold.  Since each of the large contributions is
narrow, the phenomenological consequences are negligible.

The large difference in the size of the virtual corrections between
the stop and the gluino decay is due to $\pi^2$ terms which are
determined by the sign of $(\mst-\mg)$ and arise through the
analytical continuation of the matrix element squared into the
different parameter regions, Fig.~\ref{fig_dec_resu}. For the gluino
decay they give rise to destructive interference effects of the
different color structures, and render the over-all NLO corrections
small.  The size and the sign of the NLO correction to the gluino
decay depends on the masses involved. The $K$ factor for the stop
decay is always large and positive $K=1.35\cdots 1.9$ decreasing far
above threshold, the $K$ factor for the gluino decay is in general
modest and tends to be smaller than one, $K=0.8\cdots 1$.\bigskip

The weak decay widths of the light stop $\ste$ are shown in
Fig.~\ref{fig_dec_resu}.  They are generically suppressed compared to
the strong decay widths, due to the coupling constant. This yields
about one order of magnitude between the different contributions.
Moreover the typical weak coupling factor includes mixing matrices of
the neutralinos and charginos, which may lead to a further
suppression.  Given that the masses of the four neutralinos cluster
for the higgsino type and for the lighter gaugino type mass
eigenstates, even the decay width into the heavier
neutralinos/charginos can exceed the width to the lighter
one~\cite{ourdecay}. Since the top quark is heavy, the $b \cpe$ decay
mode is typically the first tree level two particle decay
kinematically allowed. The neutralino channels open only for higher
stop masses, but will then be of a comparable size. The NLO
corrections exceed $15 \%$ for special choices of masses and
parameters only~\cite{wien}.

\section{Heavy Neutralino Decay to Stops}

\begin{figure}[hb] \begin{center}
\epsfig{file=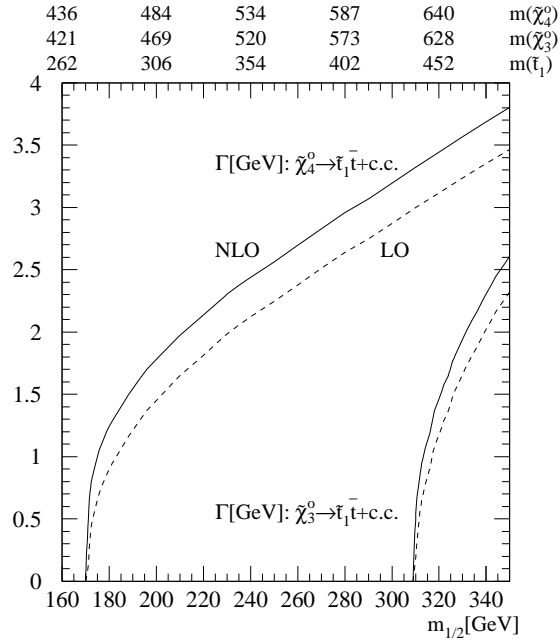,width=8.2cm}
\end{center}
\vspace*{-0.5cm}
\caption[]{\it SUSY-QCD corrections to the weak decays 
  of heavy neutralinos to a stop $\nnj \to t \steb + \tb \ste$ 
  as a function of the common gaugino mass. The masses of the 
  particles involved are labeled in the additional axis. 
  The input parameters have 
  been chosen as $m_0=450\gev, A_0=600\gev, \tan \beta=4, \mu>0$,
  in order to see the Coulomb singularity at threshold.
  The dashed line denotes the LO, the solid one the NLO results. 
  \label{fig_dec_neut}}
\end{figure}

Heavy neutralinos will be produced at a future $e^+e^-$ linear
collider~\cite{lincol_rep}. In most supergravity inspired scenarios
they are higgsino-like, and will therefore not decay into light-flavor
quark jets. However, the large top Yukawa coupling may open the decay
channel $\nnj \to \ste \bar{t} + \steb t \; [j=3,4]$ for a light stop.
The analytical expression for this decay can be obtained from the stop
decay width, eq.(\ref{eq_dec_weaklo}), by crossing the stop and the
neutralino.
\begin{equation}
\Gamma (\nnj \rightarrow t\,\ste) = 
  -\frac{\alpha}{\mnj^3} \Lambda^{1/2}(\mse^2,\mt^2,\mnj^2)
  \left[ \left( \mse^2 - m_t^2 - \mnj^2 \right) (C_L^2+C_R^2) 
         + 4 m_t \mnj C_L C_R \right]  
\end{equation}
The couplings are defined as for the stop decay. The numerical result
is shown in Fig.~\ref{fig_dec_neut}. Similar to the strong decay width
the neutralino decay to two strongly interacting particles exhibits a
Coulomb singularity, due to the exchange of a slowly moving gluon
between the decay products. Especially for the steeply rising decay of
the second heaviest neutralino, this Coulomb singularity is very
narrow, as can be seen in Fig.~\ref{fig_dec_neut}. Unlike the gluino
decay in section \ref{sect_dec_strong} the next-to-leading order
corrections can be of the order $10\%$ and are positive over the whole
mass range. The clustering of the masses of the heavy higgsino-like
neutralinos is a typical behavior in supergravity inspired scenarios,
since the off-diagonal entries of the neutralino mass matrix are small
compared to $\mu$.

\chapter{Production of Scalar Top Quarks}
\label{chap_stops}

\section{Diagonal Stop Pair Production}

\subsection{Born Cross Section}

Diagonal pairs of stop particles can be produced at lowest order QCD
in quark-antiquark annihilation and gluon-gluon
fusion:
\begin{alignat}{9}
q \bar{q} &\longrightarrow \ste \steb 
\quad \text{and} \quad \stz \stzb \notag \\
g g       &\longrightarrow \ste \steb  
\quad \text{and} \quad \stz \stzb 
\label{eq_stop_diagonal}
\end{alignat}
Mixed pairs $\ste \stzb$ and $\stz \steb$ cannot be produced in lowest
order since the $g\st\st$ and $gg\st\st$ vertices are diagonal in the
chiral as well as in the mass basis of the squarks involved. The
relevant diagrams for the reactions (\ref{eq_stop_diagonal}) are shown
in Fig.~\ref{fig_stop_feyn}~a. The corresponding cross sections for
these partonic subprocesses may be written as~\cite{stop_lo,my_stop}:
\begin{alignat}{9}
\hat{\sigma}_{LO}[q\bar{q}\to\st\stb] =& \;  
    \frac{\alpha_s^2 \pi}{s} && \,\frac{2}{27}\,\beta^3 \notag \\
\hat{\sigma}_{\rm LO}[gg\to\st\stb]       =& \;
    \frac{\alpha_s^2 \pi}{s} && \,\left\{ 
        \beta \left( \frac{5}{48} + \frac{31\mst^2}{24s} \right)
      + \left( \frac{2\mst^2}{3s} 
             + \frac{\mst^4}{6s^2} \right)
        \log\left(\frac{1-\beta}{1+\beta}\right) 
                                       \right\}
\label{eq_stop_lo}
\end{alignat}
The invariant energy of the subprocess is denoted by $\sqrt{s}$, the
velocity by $\beta=\sqrt{1-4 \mst^2/s}$. The cross sections coincide
with the corresponding expressions for light-flavor
squarks~\cite{roland}
\begin{alignat}{9}
\hat{\sigma}_{\rm LO}[q\bar{q'}\to\sq\sqb'] =& \;  
    \frac{\alpha_s^2 \pi}{s} && \,\Bigg\{
     \beta \left( \frac{4\mg^2s}{9(\mu^4 + \mg^2s)} 
                - \frac{8}{9}
           \right) 
     + \left( \frac{4}{9} + \frac{8\mu^2}{9s} \right) L \notag \\
&&& \quad + \delta_{q q'} \left[
        \frac{4n_{\sq}\beta^3}{27}
      + \beta \left( \frac{4}{27} + \frac{8\mu^2}{27s} \right)
      + \left( \frac{8\mu^2}{27s} 
             + \frac{8\mu^4}{27s^2} \right) L
                            \right]
                                       \Bigg\} \notag \\
\hat{\sigma}_{\rm LO}[gg\to\sq\sqb]       =& \;
    \frac{2 n_{\sq} \alpha_s^2 \pi}{s} && \Bigg\{ 
        \beta \left( \frac{5}{48} + \frac{31\ms^2}{24s} \right)
      + \left( \frac{2\ms^2}{3s} 
             + \frac{\ms^4}{6s^2} \right)
        \log\left(\frac{1-\beta}{1+\beta}\right) 
                                       \Bigg\} \notag \\
&&&\phantom{hhhaaalllllooo} \mu^2=\mg^2-\ms^2 \notag \\
&&&\phantom{hhhaaalllllooo} 
 L=\log((1-\beta+2\mu^2/s)/(1+\beta+2\mu^2/s))
\label{eq_stop_losq}
\end{alignat}
in the limit of large gluino masses and for $2n_{\sq}=1$. The
main difference between these two cross sections results from the
flavor diagonal $q\sq\gt$ coupling, which makes the $t$ channel gluino
contribution for the squark production eq.(\ref{eq_stop_losq}) vanish
in case of stops.  This yields a $\beta^3$ dependence of the $q
\bar{q} \to \st \stb$ cross section.  As described in
section~\ref{sect_susy_susyqcd} an additional factor $2n_{\sq}$ arises
for the mass degenerate light-flavor squark production.

Internal gluon propagators in the LO and the NLO calculation are
evaluated in the Feynman gauge. External gluons are restricted to
their physical degrees of freedom.  The sum over the physical
polarizations in the axial gauge is
\begin{alignat}{9}
P^{\mu\nu}
= \sum_{\rm transverse \, DOF} {\varepsilon^\mu_T}^*(k)
                          \, \varepsilon^\nu_T(k) 
&= - g^{\mu\nu}
  + \frac{n^\mu k^\nu + k^\mu n^\nu}{(nk)}
  - \frac{n^2 k^\mu k^\nu}{(nk)^2} \notag \\
&k_\mu P^{\mu\nu} = 0 = k_\nu P^{\mu\nu} 
\end{alignat}
with an arbitrary four vector $n$. In the final result this vector $n$
drops out according to gauge invariance. Using the Slavnov-Taylor
identity (\ref{eq_app_ward}) this is equivalent to using the
polarization sum $(-g^{\mu\nu})$ and removing the momenta $k_1^\mu$
and $k_2^\nu$ from the tensor matrix element $\M^{\mu\nu}(k_1,k_2)$.

\subsection{Next-to-leading Order Cross Section}

The incoming gluons in the virtual and real correction matrix elements
are treated the same way as in the Born matrix elements.  The
Feynman diagrams for the virtual gluon correction are shown in
Fig.~\ref{fig_stop_feyn}~b,c. The masses are renormalized in the
on-shell scheme, the coupling constant  $\alpha_s$ in the $\msbar$
scheme. The renormalization is performed in such a way, that the heavy
particles (top quarks, gluinos, squarks) decouple smoothly for scales
smaller than their masses, as described in eq.(\ref{eq_app_coupren}).
Note that no vertex requiring a finite renormalization according to
section~\ref{sect_susy_ward} occurs in the Born term, again in
contrast to the light-flavor squark production.\medskip 

The calculation of the gluon bremsstrahlung matrix element has been
performed in the cut-off scheme, appendix~\ref{chap_app_cutoff}. The
Feynman diagrams for the different incoming states
$gg,q\bar{q},g\bar{q},qg$ are given in Fig.~\ref{fig_stop_feyn}~d. The
angular integrals have been calculated analytically, which leads to an
analytic cancelation of the IR poles in $\epsilon$ between the virtual
correction, the real correction, and the mass factorization matrix
elements squared. The latter one is described in
section~\ref{sect_neut_massfac}.\medskip 

At lowest order, the cross sections for $\ste\steb$ and $\stz\stzb$
production are given by the same analytical expression, since the
mixing angle does not occur. At next-to-leading order the $t\st\gt$
and four squark coupling introduce an explicit dependence on the
mixing angle. The $\stz\stzb$ cross section can be obtained using the
operation $\PX$ described in eq.(\ref{eq_app_perm}). However, the
dependence on the mixing angle turns out to be very weak.\medskip

To perform a more detailed analysis the partonic cross section is
expressed in form of scaling functions
\begin{equation}
\hat{\sigma}_{ij} = 
\frac{\alpha_s^2(\mu^2)}{\mst^2}
\left\{
f_{ij}^B(\eta) + 4\pi\alpha_s(\mu^2)
\left[ f_{ij}^{V+S}(\eta,m_j,\tmix)
     + f_{ij}^H(\eta)
     + \bar{f}_{ij}(\eta) \log \left( \frac{\mu^2}{\mst^2} \right)
\right] \right\}
\end{equation}
where $ij$ are the incoming partons, ($\eta=s/(4\mst^2)-1$) with the
partonic cm energy $s$, $m_j$ generically denote the set of masses
entering the virtual corrections, and $\tmix$ is the stop mixing
angle. For the sake of simplicity we have identified the
renormalization and the factorization scale $\mu_f=\mu_R=\mu$. The
scaling function $f^B$ contains the Born term, $f^{V+S}$ the virtual
and soft-gluon contributions\footnote{Dividing the real gluon
  contribution into soft and hard gluons leads of course to some
  ambiguity in the definitions of $f^{V+S}$ and $f^H$. This is in
  detail described in appendix~\ref{chap_app_rad}.}, $f^H$ the
hard-gluon contribution, and $\bar{f}$ the scale dependence. The
function $\bar{f}$ combined with the running of the strong coupling
constant and the scale dependence of the parton densities should yield
a decreased dependence on $\mu$.

\begin{figure}[t] \begin{center}
\epsfig{file=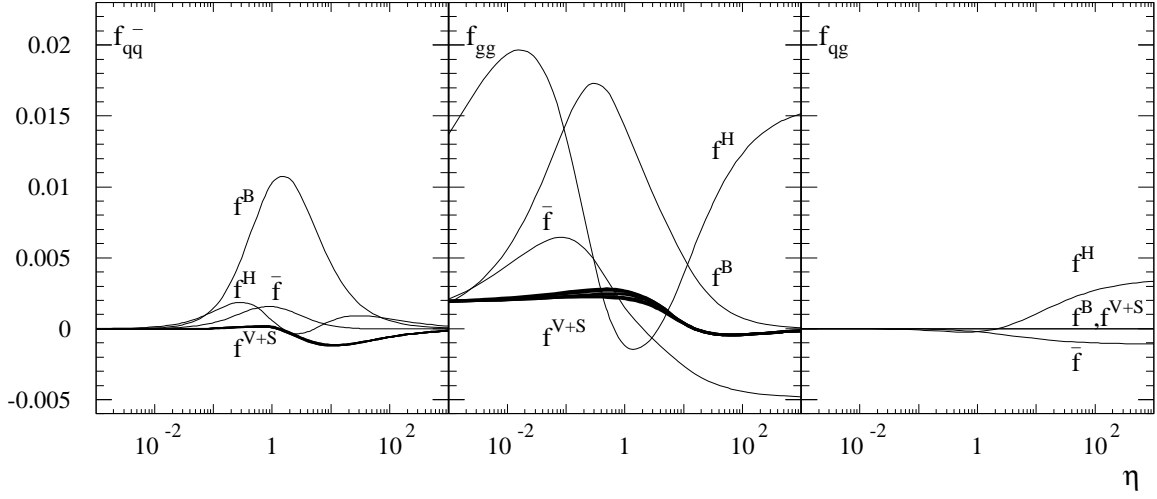,width=16cm}
\end{center}
\vspace*{-0.5cm}
\caption[]{\it The scaling functions for the production of $\ste\steb$ 
  pairs as a function of $\eta=s/(4\mst^2)-1$. The variation of 
  $f^{V+S}$ for all possible values of the mixing angle $\tmix$ is 
  indicated by the line-thickness of the curves.
  \label{fig_stop_scalingfct}}
\end{figure}

The scaling functions are shown in Fig.~\ref{fig_stop_scalingfct}. The
scaling functions $f_{g\bar{q}}$ are identical to $f_{qg}$. Only the
scaling function $f^{V+S}$ depends on the mixing angle $\tmix$ and on
the additional squark and gluino masses. The contribution of $f^{V+S}$
compared to $f^H$ is small in general, and the dependence on the
mixing angle is suppressed. In contrast to the light-flavor squark
production only the gluonic stop production cross section is
proportional to $\beta$, whereas the $q\bar{q}$ collision leading to
an $s$ channel gluon exhibits an over-all factor $\beta^3$, as can be
seen in eq.(\ref{eq_stop_lo}). A Coulomb singularity similar to the
one for the stop decays in section~\ref{sect_dec_strong} appears: The
scaling function $f^{V+S}_{gg}$ approaches a non-zero limit near
threshold $\eta \to 0$.\medskip

The emission of soft gluons from the incoming partons leads to an
energy dependence $\beta \log^i \beta$ near threshold. The leading
$\log^2\beta$ terms are universal and could be exponentiated. All
scaling functions approach a simple form in the limit $\beta \ll 1$:
\begin{alignat}{9}
  &f_{gg}^B =  \frac{7 \pi \beta}{384} &\qquad
  &f_{q\bar{q}}^B = \frac{\pi \beta^3}{54} 
\notag \\
  &f_{gg}^{V+S}  =  f_{gg}^{B} \,\frac{11}{336 \beta} &\qquad
  &f_{q\bar{q}}^{V+S} =  -f_{q\bar{q}}^{B} \,\frac{1}{48 \beta} 
\notag \\
  &f_{gg}^{H} =  f_{gg}^{B} 
  \left[ 
   \frac{3}{2\pi^2}\log^2(8\beta^2)
  -\frac{183}{28\pi^2} \log(8\beta^2) 
  \right] &\qquad
  &f_{q\bar{q}}^{H}  =  f_{q\bar{q}}^{B} 
  \left[ 
   \frac{2}{3\pi^2}\log^2(8\beta^2) 
  - \frac{107}{36\pi^2}\log(8\beta^2) 
  \right] 
\notag \\
  &\bar{f}_{gg} = -f_{gg}^B \,\frac{3}{2\pi^2}\log(8\beta^2)  &\qquad
  &\bar{f}_{q\bar{q}} = -f_{q\bar{q}}^B \,\frac{2}{3\pi^2}\log(8\beta^2)
\end{alignat} \medskip 

In the high energy limit $\eta \gg 1$ the LO cross section scales
$\propto 1/s$, eq.(\ref{eq_stop_lo}). The NLO cross sections involving
at least one gluon in the initial state approach a finite value in
this limit. This is caused by the exchange of soft gluons in the $t$
or $u$ channel. Exploiting $k_T$ factorization~\cite{kt_fac} the
non-zero limits of the scaling functions can be determined:
\begin{alignat}{3}
  f_{gg}^H  &=&  \frac{2159}{43200\pi} &\qquad 
  f_{qg}^H  &=&  \frac{2159}{194400\pi} \notag \\
 \bar{f}_{gg} &=&  -\frac{11}{720\pi} &\qquad
 \bar{f}_{qg} &=&  -\frac{11}{3240\pi}
\end{alignat}
The ratio of $f_{gg}$ and $f_{qg}$ is given by $2N:C_F$ which are the
color factors for the exchange of a gluon between an incoming quark
or one of the two incoming gluons, and the Born diagram, which in both
cases is the $gg$ one.

\subsection{Results}

\begin{figure}[b] \begin{center}
\epsfig{file=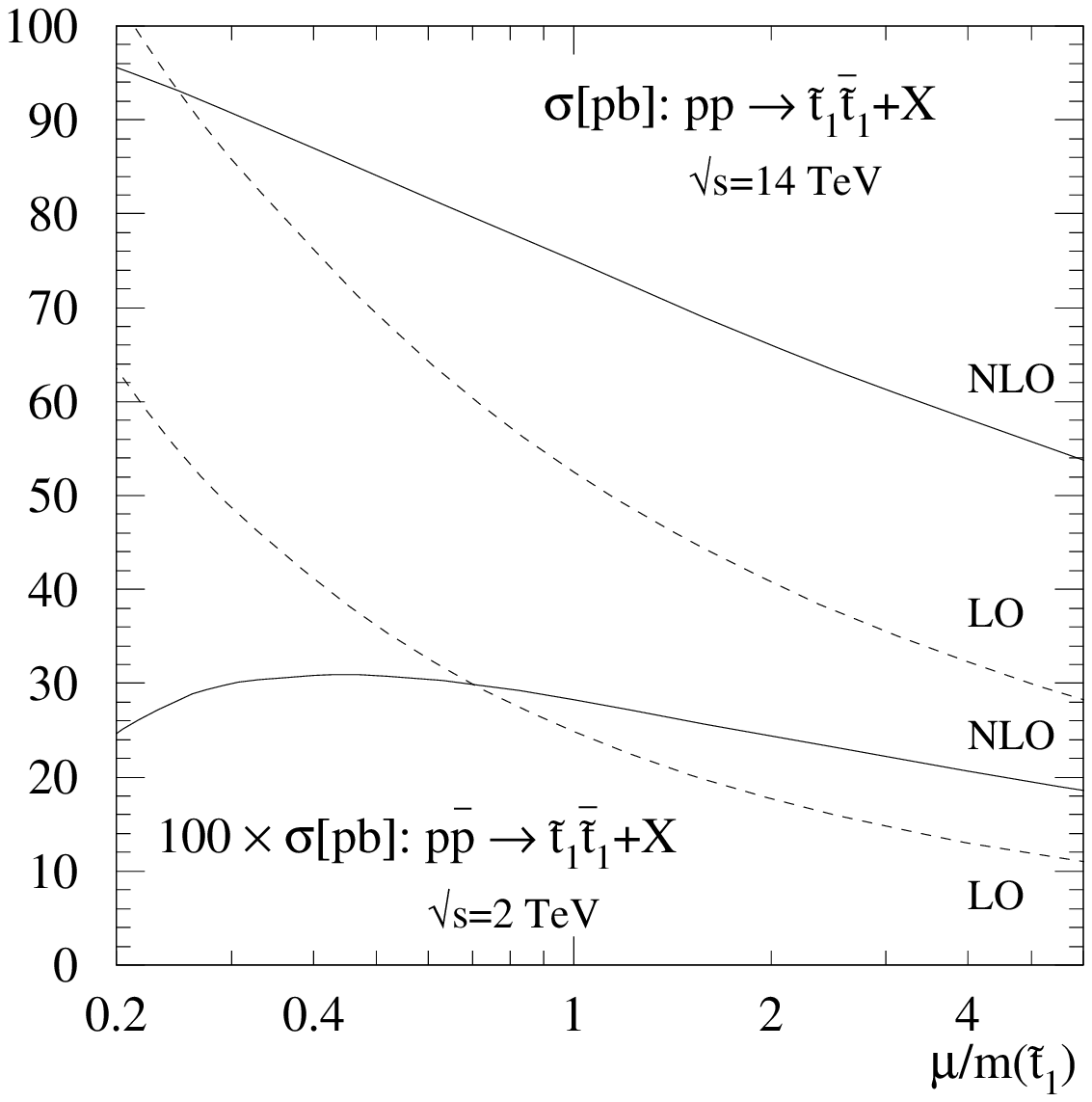,width=7cm}
\hspace*{0.5cm}
\epsfig{file=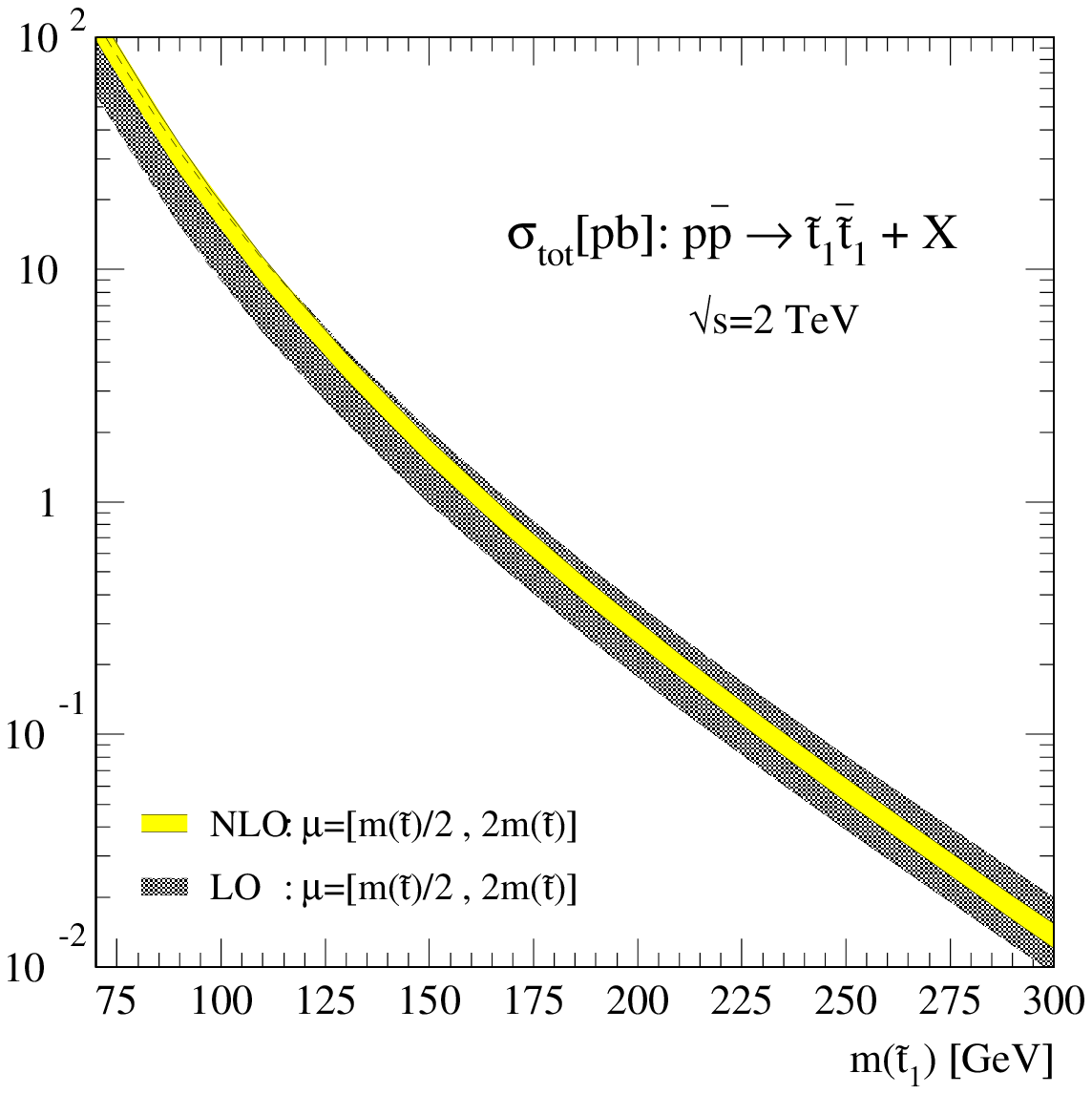,width=7cm}
\end{center}
\vspace*{-0.5cm}
\caption[]{\it Left: The renormalization/factorization scale dependence 
  of the total cross section for $\ste$ pair production at the 
  Tevatron and the LHC. the maximum for the NLO cross section at the 
  LHC is reached only for very small scales; Right: Effect 
  of the variation of the 
  scale on the upgraded Tevatron production cross section, as a 
  function of the stop mass. The LO and NLO bands show the 
  improvement of the theoretical uncertainties in the 
  derivation of mass bounds. The SUSY scenario determining the masses
  is given in eq.(\ref{eq_susy_scen}).
  \label{fig_stop_scale}}
\end{figure}

The hadronic cross section is obtained from the partonic by
convolution with the parton densities, eq.(\ref{eq_neut_lumi}). The
phase space integration for the Born cross section as well as for the
real gluon emission is given in appendix~\ref{chap_app_phase}.\medskip

\subsubsection{Scale Dependence}

The dependence on the scale $\mu=\mu_F=\mu_R$ has been analyzed for
the production of $\ste$ pairs both at the upgraded Tevatron and at
the LHC. The hadronic cross sections include $\alpha_s$ and the
CTEQ4~\cite{cteq} parton densities consistently in LO or NLO, which
also enters the definition of the $K$ factor $K=\sigma_{\rm
  NLO}/\sigma_{\rm LO}$. The scale dependence is shown in
Fig.~\ref{fig_stop_scale}. The leading order dependence is monotonic
and varies by about $\sim 100\%$ for scales between $\mse/2$ and
$2\mse$.  The increase for small scales results from the large value
of the running QCD coupling in leading order. At next-to-leading order
the variation with the scale is reduced to $\sim 30\%$. The monotonic
behavior of the leading order curve is corrected in next-to-leading
order, yielding a maximum value at some small scale,
Fig.~\ref{fig_stop_scale}.

\subsubsection{Supersymmetric Parameter Dependence}

\begin{figure}[t] \begin{center}
\pspicture(0,0)(16,6.5)
\rput[cl]{0}(0,3.9){
\begin{small} \begin{tabular}{|c||c|c||c|c|}
\hline
      $\mse$ \rule[0mm]{0mm}{5mm}
    & $K_{\rm Tev}$
    & $gg_{\rm in}$ : $q\bar{q}_{\rm in}$
    & $K_{\rm LHC}$
    & $gg_{\rm in}$ : $q\bar{q}_{\rm in}$ \\[2mm] \hline
      70  \rule[0mm]{0mm}{5mm}
    & 1.43
    & 0.69 : 0.31
    & 1.27
    & 0.96 : 0.04 \\
      110
    & 1.33
    & 0.46 : 0.54
    & 1.33
    & 0.95 : 0.05 \\
      150
    & 1.23
    & 0.29 : 0.71
    & 1.38
    & 0.94 : 0.06 \\
      190
    & 1.15
    & 0.19 : 0.81
    & 1.42
    & 0.92 : 0.08 \\
      230
    & 1.10
    & 0.12 : 0.88
    & 1.45
    & 0.91 : 0.09 \\
      270
    & 1.06
    & 0.08 : 0.92
    & 1.48
    & 0.89 : 0.11 \\
      310
    & 1.03
    & 0.06 : 0.94
    & 1.50
    & 0.88 : 0.12 \\
      $[\gev]$ &&&& \\ \hline
\end{tabular} \end{small}
}
\rput[cl]{0}(8.5,2.8){
\epsfig{file=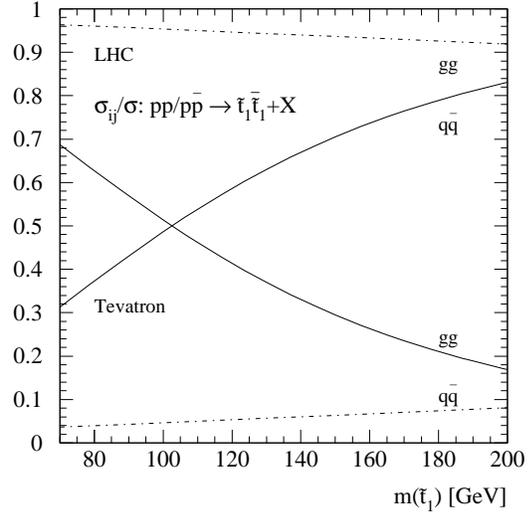,width=7cm}
}
\endpspicture
\end{center}
\caption[]{\it $K$ factors for diagonal stop-pair production at 
  the upgraded Tevatron  and the LHC for a sample of stop masses. 
  Scale choice: $\mu=\mse$. For a comparison of Tevatron and LHC 
  also the LO 
  initial-state $gg$ and $q\bar{q}$  fractions are given.
  \label{fig_stop_kfac}}  
\end{figure}

\begin{figure}[p] \begin{center}
\epsfig{file=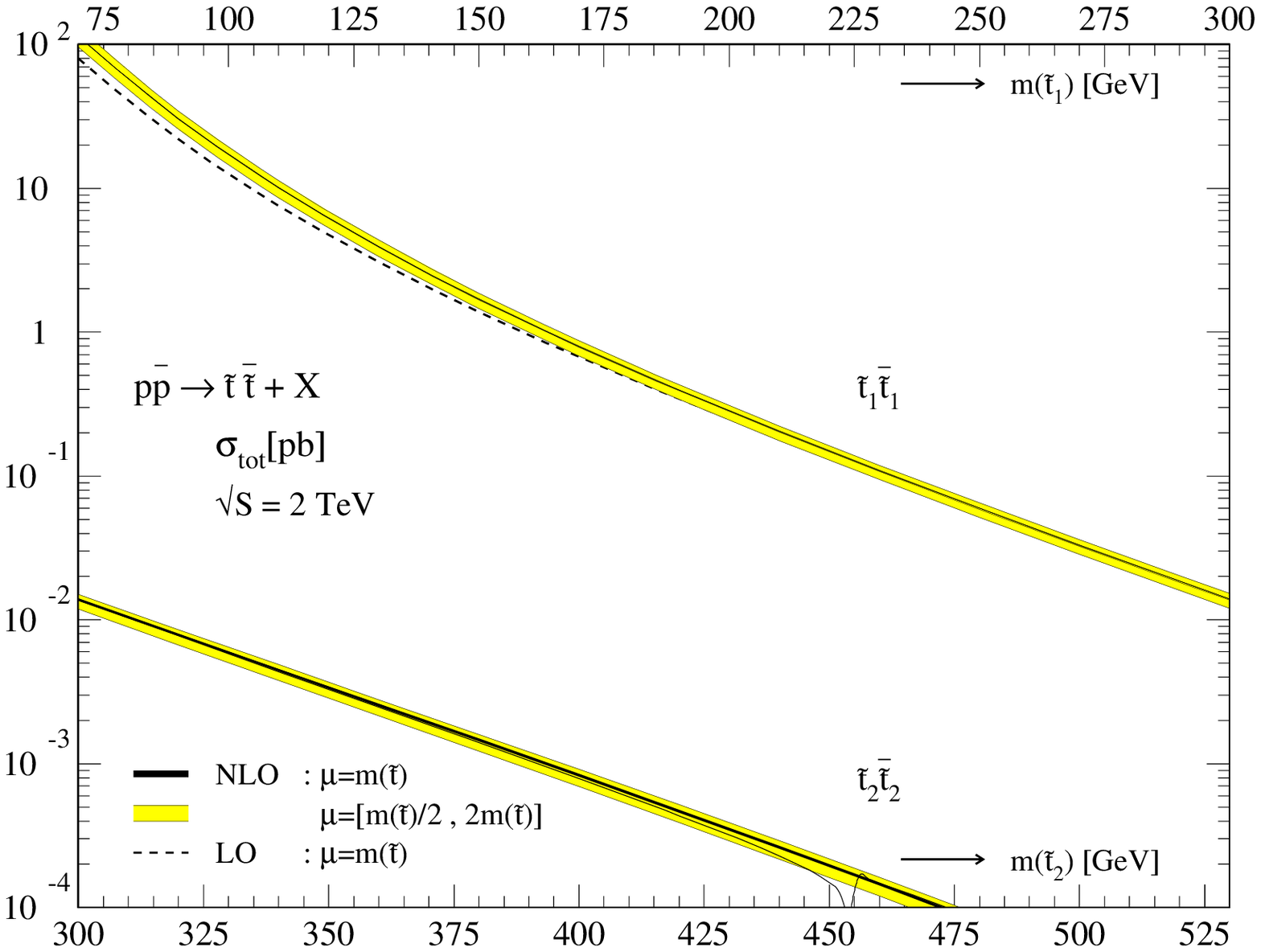,width=12cm}
\epsfig{file=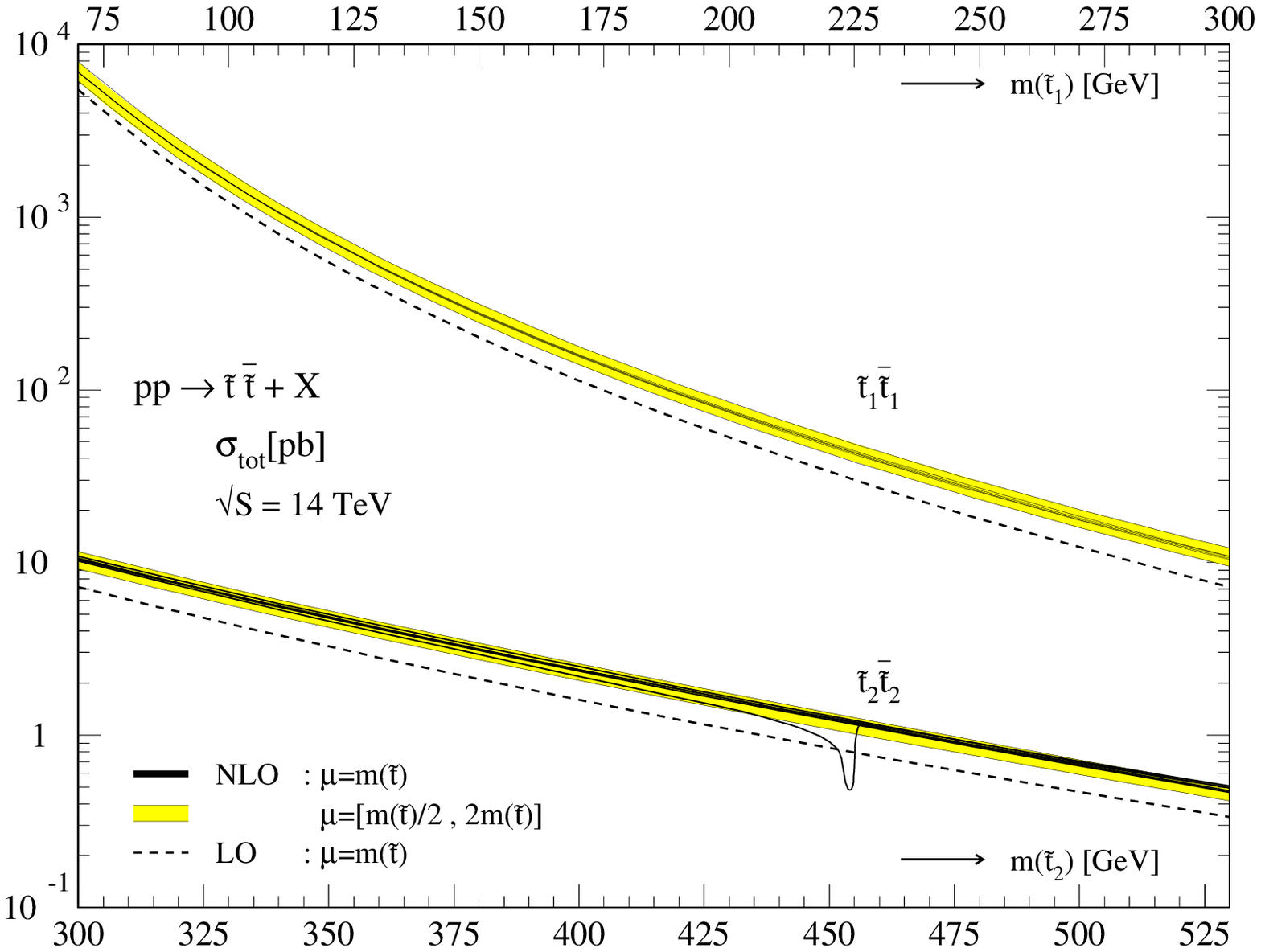,width=12cm}
\end{center}
\vspace*{-0.7cm}
\caption[]{\it The total cross section for $\ste$ and $\stz$ pair 
  production at the Tevatron and at the LHC. The band for the 
  NLO results indicates the uncertainty due to the scale dependence.
  The mass scenario is given in eq.(\ref{eq_susy_scen}). The line 
  thickness of the NLO curves represents the variation of the 
  gluino mass between 280 and $900\gev$ and of the mixing angle 
  over its full range. The kink in both of the cross section 
  results from the on-shell decay of the heavy stop to a very light 
  gluino and a top, and can be regularized by introducing a finite stop 
  width. 
  \label{fig_stop_cxn}}
\end{figure}

The total hadronic cross sections at the Tevatron and at the LHC are
given in Fig.~\ref{fig_stop_cxn}. The masses involved are fixed by a
SUGRA inspired scenario~eq.(\ref{eq_susy_scen}). All squarks except
the top squark are assumed to be mass degenerate. The mass range of
the outgoing stop is varied independently of the other mass
parameters.  The same is done for the mixing angle and for the gluino
mass. The dependence of the cross section on these internal mass
parameters is small, as can be seen from the finite width of the
central lines in the figures, \ie the cross sections depend
essentially on the outgoing masses and do not distinguish between
$\ste$ and $\stz$. The light-flavor squark and the gluino
contributions appearing in the loops are decoupled even for
numerically small masses. The search for stops therefore yields
limits on their masses independent of any other
parameter, unlike the squark/gluino or the neutralino/chargino
case.\medskip

One exception of this behavior is the kink in the next-to-leading
order cross sections for the heavy stops. Similarly to the decay width
of a heavy stop to a gluino, Fig.~\ref{fig_dec_resu}, threshold
contributions occur. In the stop production case the heavy stop can
decay into an on-shell gluino and a top quark. The kink will be
regularized by introducing a finite width for the stop, and for an
analysis it has to be removed by resummation. However, for the search
for stops at the Tevatron this parameter region is not of
interest.\medskip

The strong mass dependence of the $K$ factor is due to different $K$
factors for the quark and the gluon channel, both of which are only
weakly mass dependent. However the contribution of the two channels
varies strongly with the mass of the external particle, as can be read
off Fig.~\ref{fig_stop_kfac}. Whereas the gluonic $K$ factor is large
(about 1.3), the quark $K$ factor tends to be smaller than and close
to one. Weighted with the fraction of the incoming state these
combine to the $K$ factor given in Tab.~\ref{fig_stop_kfac}. Since at
the Tevatron the contribution of the incoming quarks decreases as the
stops become light, the $K$ factor grows from 1.03 to 1.43 far from
threshold. At the LHC the gluons dominate over the whole considered
mass region and the $K$ factor varies between 1.27 and 1.50.\smallskip

With cross sections between 0.1 and 100~pb the integrated luminosity
$\int {\cal L}=20\fb^{-1}$ should be sufficient for collecting a
sample of $10^3$ and $10^6$ stop events, provided the particle exists
and with a mass less than $450\gev$. The LHC with an integrated
luminosity of $\int {\cal L}=300\fb^{-1}$ could collect $10^5$
to $10^8$ stop events in the mass range of 200 to $500\gev$.

\begin{figure}[hb] \begin{center}
\epsfig{file=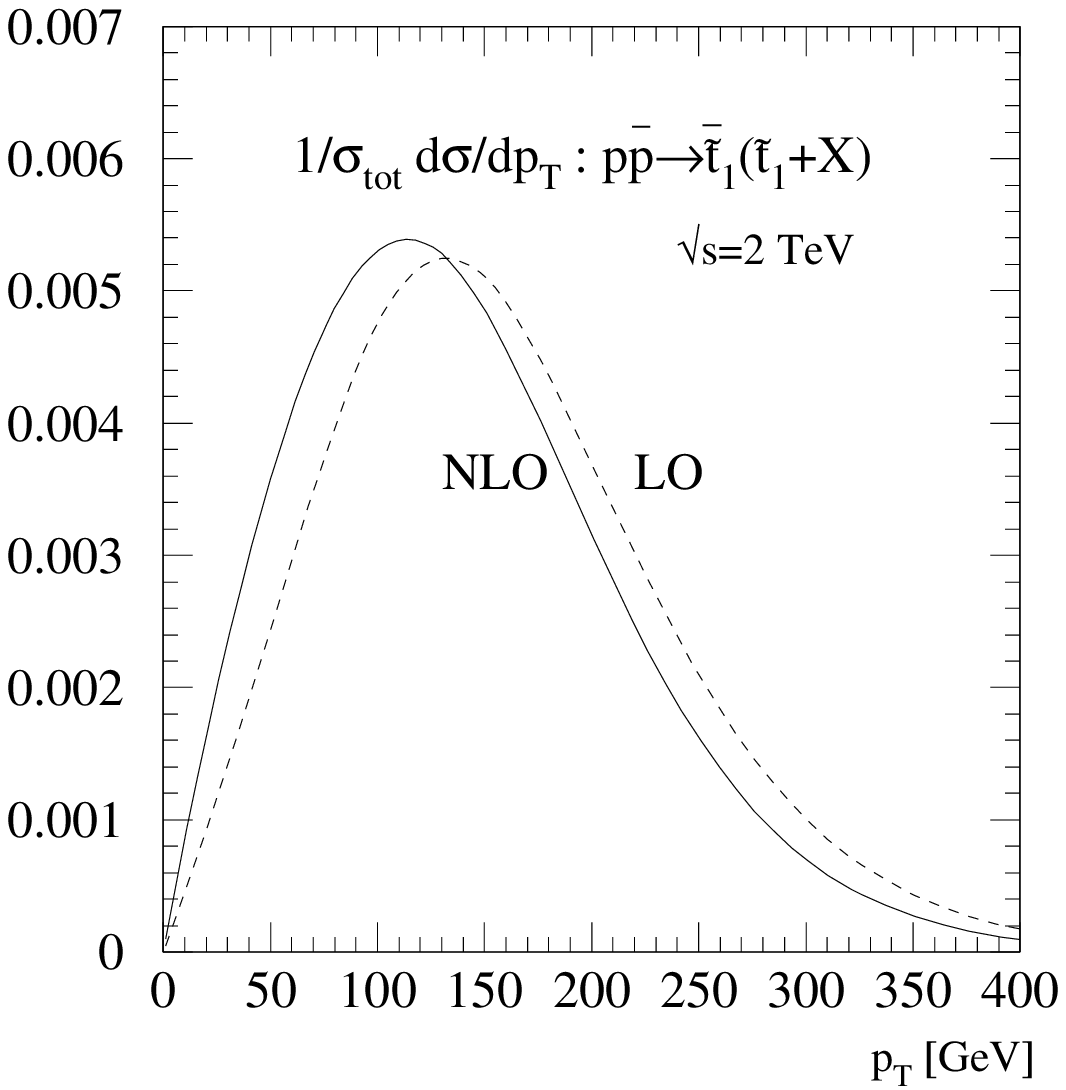,width=7cm}
\hspace*{1cm}
\epsfig{file=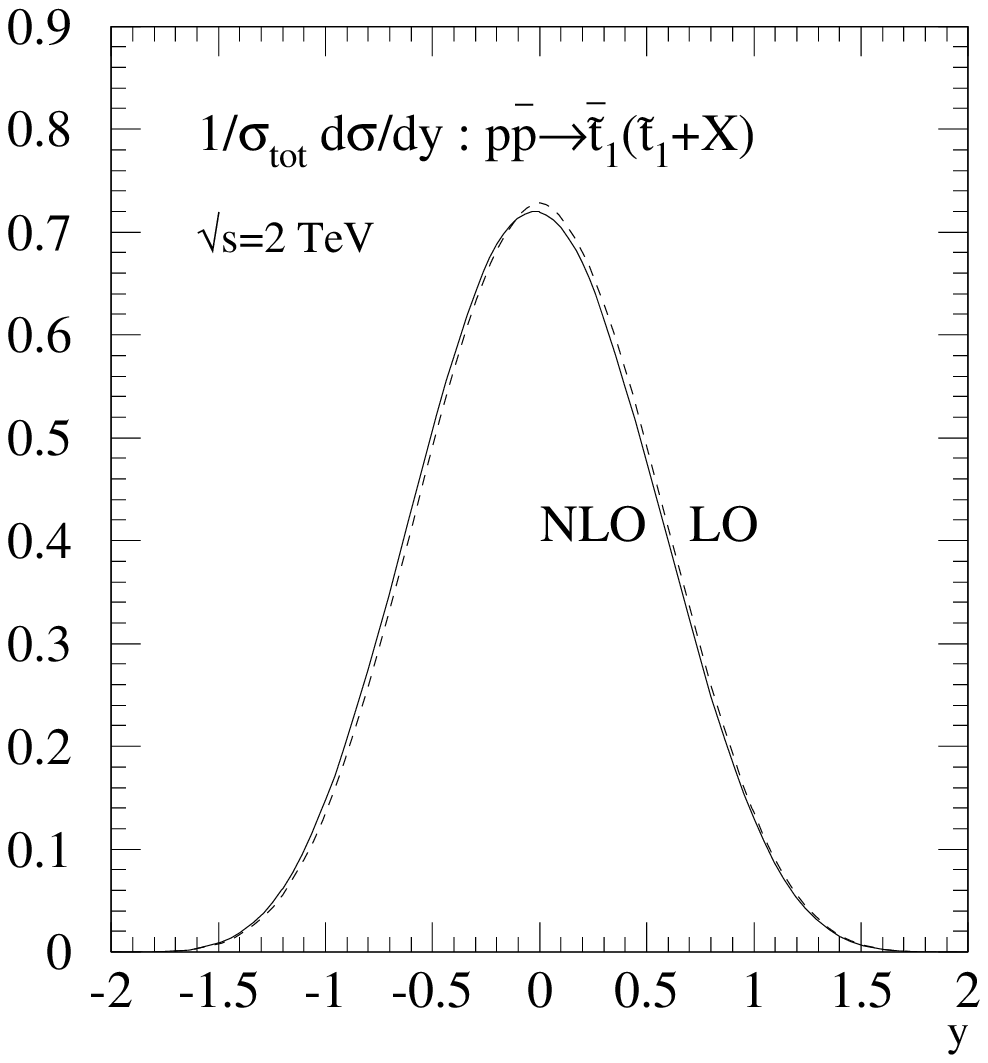,width=7cm}
\end{center}
\vspace*{-0.7cm}
\caption[]{\it The normalized differential cross section for 
  the production of $\ste$ pairs at the Tevatron. The mass 
  scenario using the central scale is defined in 
  eq.(\ref{eq_susy_scen}).
  \label{fig_stop_diff}}
\end{figure}

The normalized differential cross sections with respect to the
transverse momentum and the rapidity is shown in
Fig.~\ref{fig_stop_diff}. The transverse momentum of the outgoing stops
is shifted to a softer regime by the momentum carried by the
additional jet in the final state. A naive description using the $K$
factor would not take into account this shift and therefore lead to
large errors in the size of the cross section for a certain value of
$p_T$. The rapidity distribution keeps almost the same shape in NLO as
in LO. However, it is not symmetric in NLO anymore.

\section{Non-diagonal Stop Production}

\begin{table}[b]
\begin{center} \begin{small}
\begin{tabular}{|c|l||cc||cc|}
\hline \rule[0mm]{0mm}{5mm}
    & $\sigma$[fb]
    & $\sigma_{q\bar{q}}$
    & $\sigma_{q\bar{q}}^{\rm limit}$
    & $\sigma_{gg}$
    & $\sigma_{gg}^{\rm limit}$ \\[1mm] \hline
      Tevatron
    & $\ste \steb$ \rule[0mm]{0mm}{5mm}
    & 0.201 $\cdot 10^{3}$
    & 0.202 $\cdot 10^{3}$
    & 0.087 $\cdot 10^{3}$
    & 0.087 $\cdot 10^{3}$   \\
    & $\stz \stzb$ \rule[0mm]{0mm}{5mm}
    & 0.333 
    & 0.337
    & 0.016 
    & 0.016   \\
    & $\ste \stzb + \stz \steb$ \rule[0mm]{0mm}{5mm}
    & --
    & 0
    & --
    & 0.131 $\cdot 10^{-4}$ \\[1mm] \hline
      LHC
    & $\ste \steb$ \rule[0mm]{0mm}{5mm}
    & 4.137 $\cdot 10^{3}$
    & 4.150 $\cdot 10^{3}$
    & 70.13 $\cdot 10^{3}$
    & 75.00 $\cdot 10^{3}$    \\
    & $\stz \stzb$ \rule[0mm]{0mm}{5mm}
    & 0.169 $\cdot 10^{3}$
    & 0.172 $\cdot 10^{3}$
    & 1.422 $\cdot 10^{3}$
    & 1.458 $\cdot 10^{3}$    \\
    & $\ste \stzb + \stz \steb$ \rule[0mm]{0mm}{5mm}
    & --
    & 0
    & --
    & 0.149   \\[1mm] \hline
\end{tabular}
\end{small} \end{center}
\caption[]{\it Cross sections for diagonal and non-diagonal pair 
  production at the Tevatron and the LHC, using the default 
  SUGRA-inspired scenario eq.(\ref{eq_susy_scen}).
  The non-diagonal results are given 
  without the mixing factor $\sin^2(4 \tmix)$.The superscript 
  'limit' denotes the asymptotic value of the cross section 
  for large gluino masses. 
  \label{tab_stop_mix}}   
\end{table}

\begin{figure}[p] \begin{center}
\epsfig{file=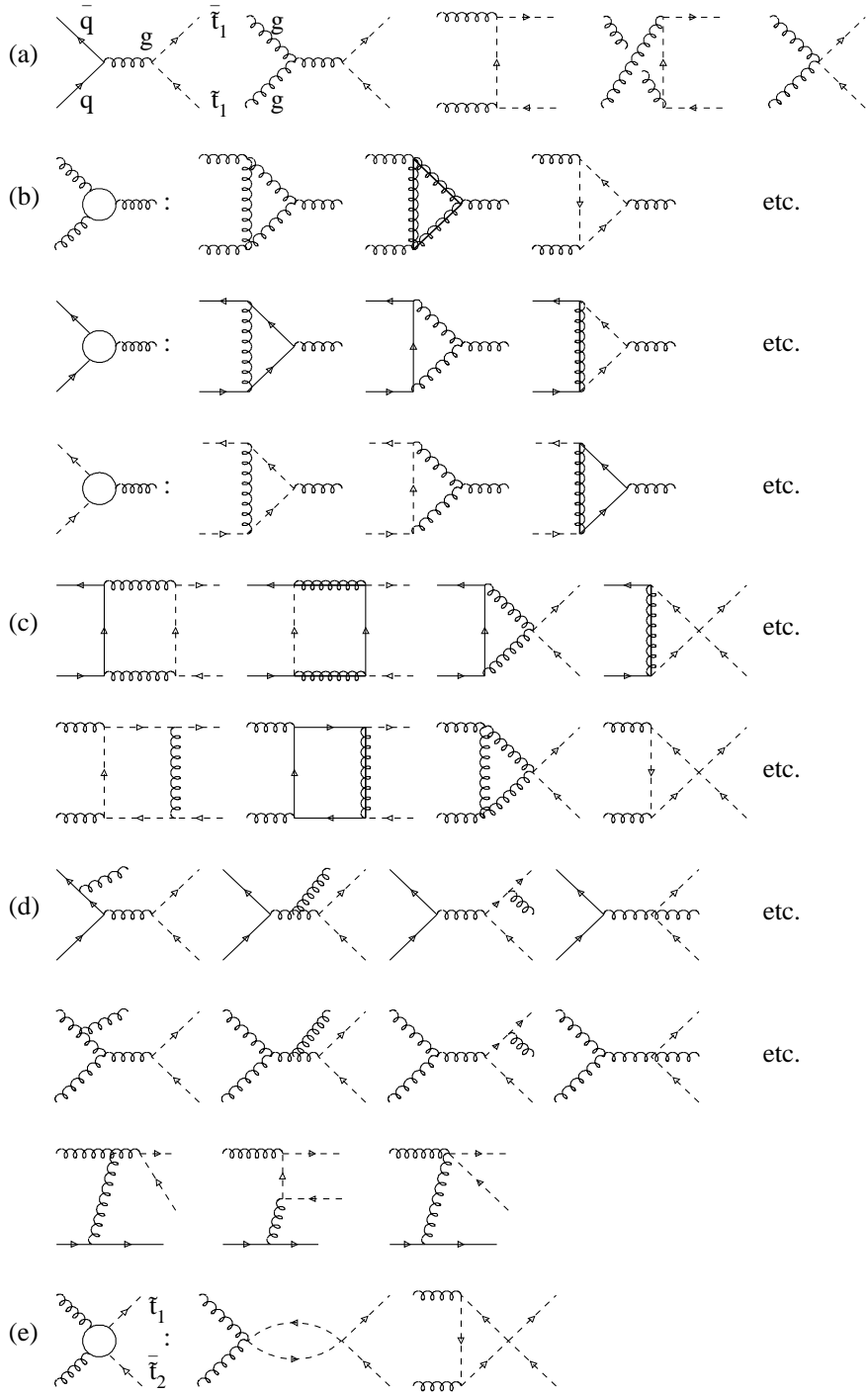,width=12cm}
\end{center}
\caption[]{\it Generic Feynman diagrams for the diagonal stop pair 
  production: (a) Born diagrams for quark and gluon incoming state;
  (b) vertex corrections; (c) box contributions; (d) real gluon/quark
  emission for different incoming states; (e) mixed stop pair 
  production in the limit of decoupled gluinos. The self energy
  contributions are not shown.
  \label{fig_stop_feyn}}
\end{figure}

In hadron collisions mixed pairs $\ste\stzb$ and $\stz\steb$ cannot be
produced in lowest order, unlike in $e^+e^-$ collisions, since the
involved coupling conserve the chirality eigenstate, which does not
hold for the coupling to a $Z$. The mixed production cross section is
therefore $\order(\alpha_s^4)$. For a general mass scenario it is
small but difficult to calculate. In the diagonal production we
observe that the limit of a decoupled gluino gives a good
approximation for the size of the cross sections. Therefore we
calculate the production cross section for $\ste\stzb+\stz\steb$ in
this limit. Only two one-loop diagrams contribute to the amplitude in
the limit, they are given in Fig.~\ref{fig_stop_feyn}~e. They involve
the production of diagonal stop pairs in $gg$ fusion, which are
rescattered to mixed pairs by the four squark vertex. The incoming
quarks are suppressed.

The evaluation of the loops yields
\begin{equation}
\hat{\sigma}_\infty [gg\to\ste\stzb+\steb\stz] = 
\sin^2(4\tmix)\,
\frac{37}{13824}\,
\frac{\alpha_s^4\,\Lambda^{1/2}}{2 \pi s^3}\, 
\left| \mse^2 \log^2(x_1) - \msz^2 \log^2(x_2)
\right|^2
\end{equation}
where the subscript in the cross section $\hat{\sigma}_\infty$
indicates the limit $\mg \to \infty$. The coefficient $\Lambda^{1/2}$
is the usual 2-particle phase-space factor, {\it i.e.}
$\Lambda=[s-(\mse+\msz)^2][s-(\mse-\msz)^2]$, and $x_k =
(\beta_k-1)/(\beta_k+1)$; the logarithmic discontinuities are defined
properly by the infinitesimal shift $s \to s + i\varepsilon$ in
$\beta_k$. The fraction $37/13824$ originates from the color factor
$(N-1)[5N^2-2(N-1)^2]/[256 N^3(N+1)]$.

The cross section depends strongly on the mixing angle $\tmix$ through
the overall factor $\sin^2(4 \tmix)$. Numerical values for the
diagonal and non-diagonal pair cross sections are compared in
Tab.~\ref{tab_stop_mix}. Note that the mixed-pair cross section is
given in this table without the mixing factor $\sin^2(4 \tmix)$. The
values for the cross section for producing mixed stop pairs in the
large $\mg$ limit are very small at the Tevatron as well as at the
LHC.

\chapter{$\mathbf{R}$ Parity Violating Squarks}
\label{chap_lepto}

\section{Production in $\mathbf{ep}$ Collisions}

Limits on the mass and the coupling $\lambda'$ as defined in
eq.(\ref{eq_susy_rviol}) can be derived from the direct search at
different colliders. HERA $ep$ scattering could produce squarks via
the $R$ parity violating coupling $\lambda'$ to quarks and electrons,
where the flavor of the squark has to be chosen consistent with the
whole set of current bounds~\cite{leptoquark_intro,r_theo,r_exp}:
\begin{equation}
e q \longrightarrow \sq
\end{equation}
This resonant $s$ channel production process can be described in terms
of general scalar leptoquarks. The Yukawa matrix $\lambda'$ does not
have to be diagonal in flavor or generations.

For the leading order hadronic cross section the convolution with the
parton density, as defined in eq.(\ref{eq_neut_lumi}) becomes trivial,
since the energy-momentum conservation yields a factor
$\delta(1-m^2/(xS))$
\begin{alignat}{9}
\hat{\sigma}_{\rm LO} =& \; \frac{\pi {\lambda'}^2}{4 m^2} \notag \\
\sigma_{\rm LO} =& \; \frac{m^2}{S} \,
                      f^P_q\left( \frac{m^2}{S},\mu^2 \right) \,
                      \hat{\sigma}_{\rm LO} 
\end{alignat}
The parton densities $f^P_q$ are taken at the factorization scale
$\mu$, and possible flavors of incoming quark are fixed by charge
conservation and the charge of the outgoing squark.

The NLO contributions consist of virtual gluons, real gluon emission,
and the crossed $eg$ incoming state. All other supersymmetric
particles in this scenario are assumed to be decoupled, see
appendix~\ref{chap_app_rpar}. Some generic Feynman diagrams for the
matrix element can be derived from Fig.~\ref{fig_dec_feyn} by
replacing the external gluino by a positron and removing all internal
gluino contributions\footnote{The next-to-leading order calculation
  was performed in parallel to~\cite{kunszt}, and the results are in
  agreement}. After renormalization and mass factorization the
dimensionally regularized NLO cross section is finite and can be
written as
\begin{alignat}{9}
\sigma_{\rm NLO} =& \; \frac{m^2}{S} \,
                    f^P_q(x) \,
                    \hat{\sigma}_{\rm LO} 
   \left[ 1 - \frac{C_F \alpha_s}{\pi} \zeta_2 \right]
   + \Delta_q + \Delta_g \notag \\
\Delta_q =& \; \frac{\alpha_s}{\pi} \, \hat{\sigma}_{\rm LO}
 \int_x^1 dy \, f^P_q(y) \,  
 \Bigg\{ -\frac{z}{2} P_{qq}(z) \log z
         + C_F \left(1+z\right) \notag \\ & \qquad \qquad \qquad
         + C_F \left[ 2 \left( \frac{\log (1-z)}{1-z} \right)_+ 
                     - \left( \frac{1}{1-z} \right)_+ 
                     - (2+z+z^2) \log (1-z) \right]
 \Bigg\} \notag \\
\Delta_g =& \; \frac{\alpha_s}{\pi} \, \hat{\sigma}_{\rm LO}
 \int_x^1 dy \, f^P_q(y) \, \frac{z}{2}
 \Bigg\{ - P_{qg}(z) \left[ \log \frac{z}{(1-z)^2} + 2 \right]
         + z(1-z) \log z + 1 \Bigg\} 
\label{eq_lepto_hera}
\end{alignat}
where $x=m^2/S$ and $z=x/y$, and the renormalization and factorization
scales are set $\mu_F=\mu_R=m$. The splitting functions $P_{ij}$ are
defined in eq.(\ref{eq_app_ap}), and the + distributions are given in
eq.(\ref{eq_app_plus}). The NLO cross section is proportional to
$\hat{\sigma}_{\rm LO}$, \ie the anomalous coupling drops out of the
$K$ factor. In the formulae the factorization and renormalization
scales have been identified with the squark mass.

\begin{figure}[t] \begin{center}
\epsfig{file=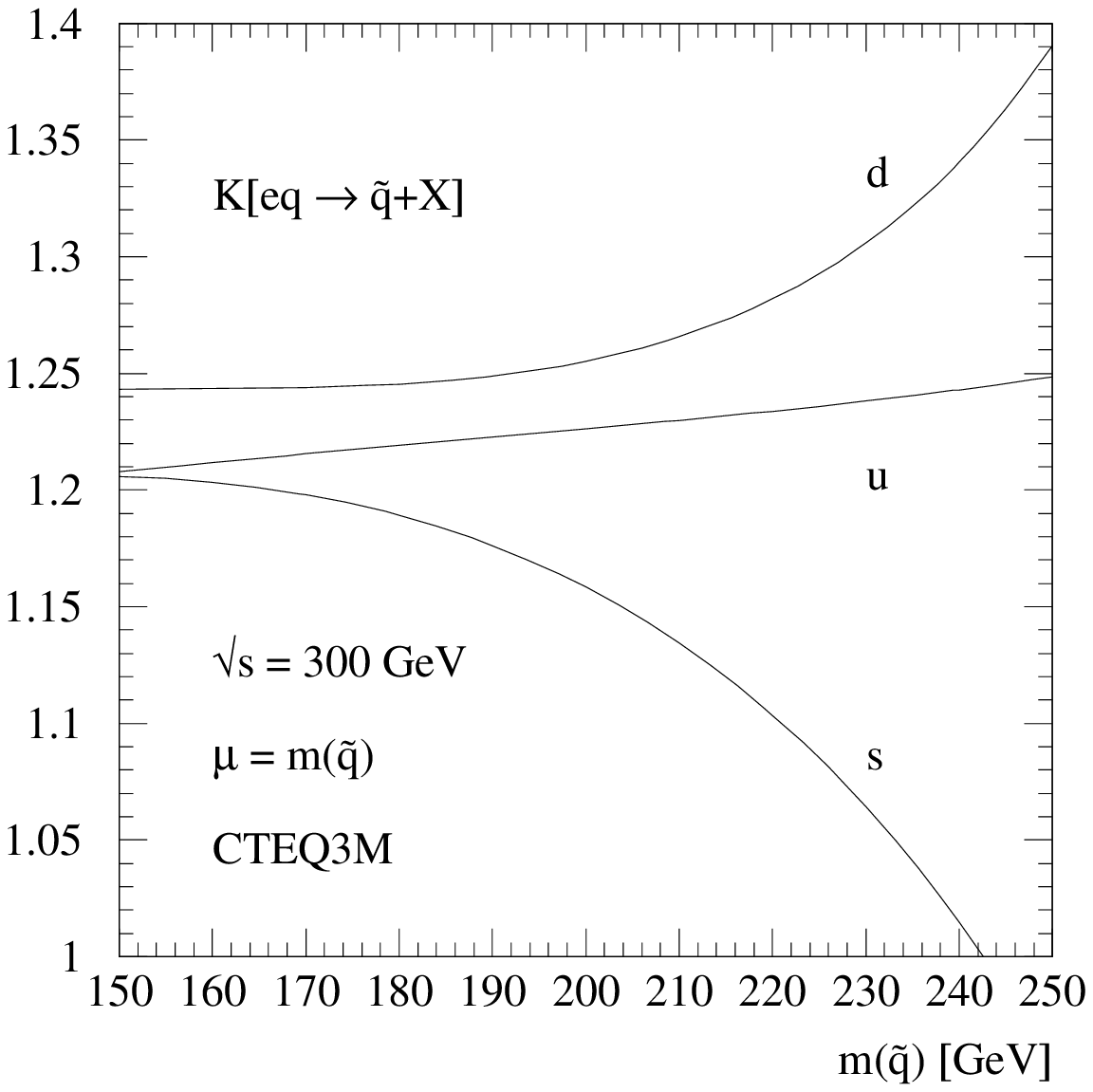,width=7.5cm}
\hspace*{-0cm}
\epsfig{file=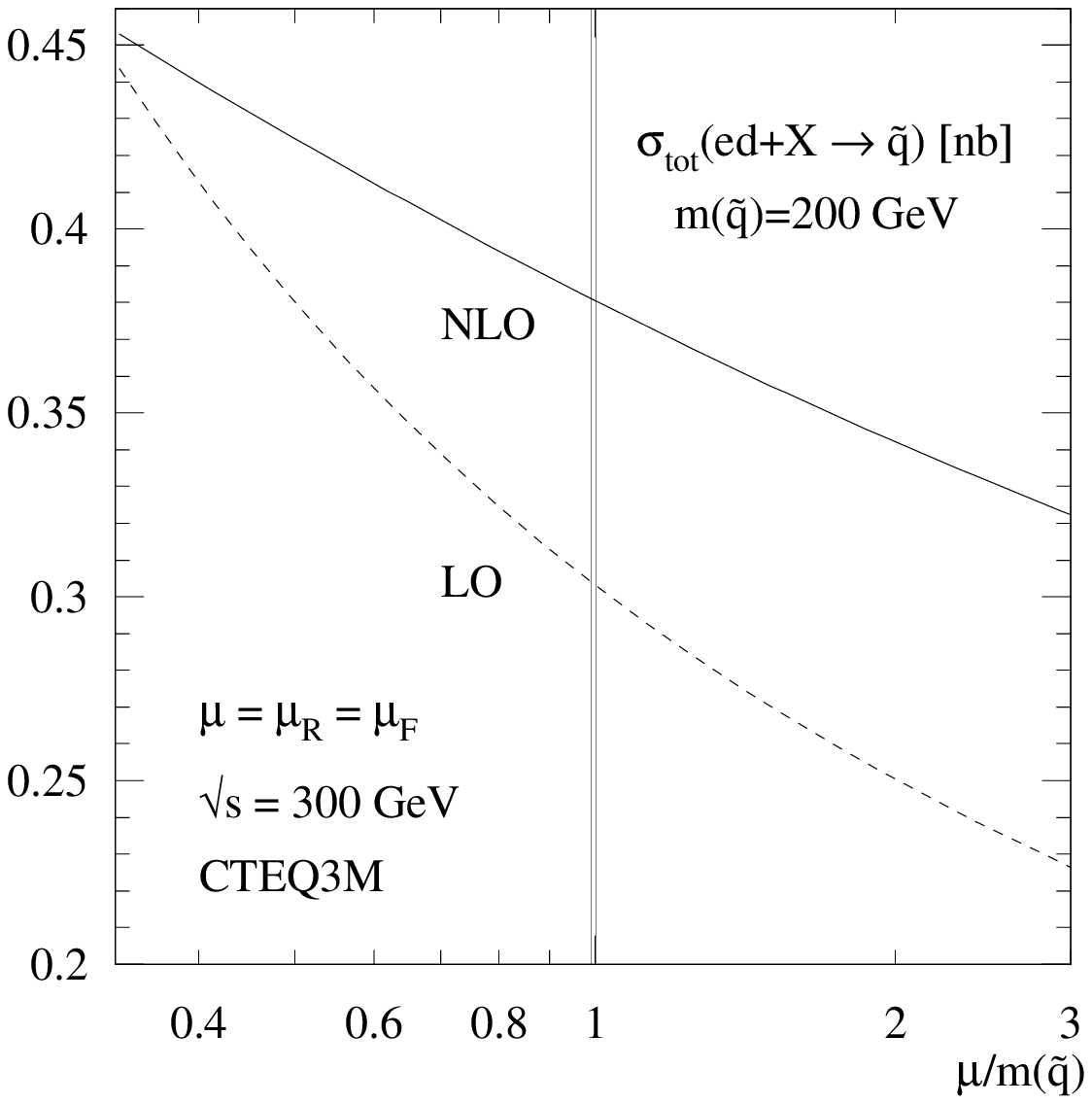,width=7.5cm}
\end{center}
\vspace*{-0.8cm}
\caption[]{\it Left: $K$ factors for $ed,eu \to \sq$ as a function 
  of the mass of the produced squark; Right: $K$ factor for the 
  $d$ type quark vs. the renormalization and factorization scale.
  \label{fig_lepto_hera}}
\end{figure}

The $K$ factors for the production of an up-type and a down-type
squark are given in Fig.~\ref{fig_lepto_hera}. The next-to-leading
order correction for the HERA process is dominated by the gluon
emission from an incoming quark. They are positive in the whole mass
region considered. The virtual correction as well as the crossed
channel are suppressed. Since the parton distributions enter the $K$
factor, the different up and down type valence and sea quarks receive
different corrections.

The dependence on the factorization and renormalization scale can be
made explicit in eq.(\ref{eq_lepto_hera}) by adding $\log
\mu_{FR}^2/m^2$ to the $\zeta_2$ term in the virtual correction and to
the logarithms multiplied by $P_{ij}$. Additionally the mass $m$ in
the argument of the parton densities and the running coupling
$\lambda'$ has to be replaced by $\mu$. The decrease of the scale
dependence of the cross section is shown in Fig.~\ref{fig_lepto_hera}.

The partial width for the decay of a $R$ parity violating squark can
be computed using the results for the scalar top squark. The width for
the decay channel
\begin{equation}
\sq \longrightarrow e q
\end{equation}
in NLO reads 
\begin{equation}
\Gamma_{\rm NLO}(\sq \to eq) = \frac{{\lambda'}^2 m}{16 \pi}
\left[ 1 + \left( \frac{27}{8} - 2 \zeta_2 \right) 
       \frac{C_F \alpha_s}{\pi} \right]
\end{equation}
This correction is small ($\sim 10 \%$). If the $R$ parity breaking
coupling $\lambda'$ is small, then the squark will be long-lived, and
the NLO correction will not change this effect. A typical scenario
could be a squark with a mass of $200\gev$ and a Yukawa coupling
$\lambda' \sim e/10$, which leads to a decay width of $\Gamma \sim
3\gev$. Bound states however will only occur, if $R$ parity conserving
decay modes are kinematically forbidden.

\section{Production in Hadron Collisions}

In contrast to the HERA production process the production of squarks
at hadron colliders, $R$ parity conserving and $R$ parity violating,
is fixed by the QCD coupling, as long as the production process does
not involve any weakly interacting  particles\footnote{Other
  production channels \cite{lepto_weaktev} involve the weak coupling
  constant and more than two final state particles; their cross
  sections are significantly smaller than the squark-antisquark
  production via an $s$ channel gluon.}. In the scenario under
consideration all other non-Standard Model particles are decoupled.
The Feynman diagrams for the Born cross section~\cite{my_leptohadro}
are the same as for the stop pair production,
Fig.~\ref{fig_stop_feyn}a:
\begin{equation}
q \bar{q}\, / \, gg \longrightarrow \sq \sqb 
\end{equation}

\begin{figure}[b] \begin{center}
\epsfig{file=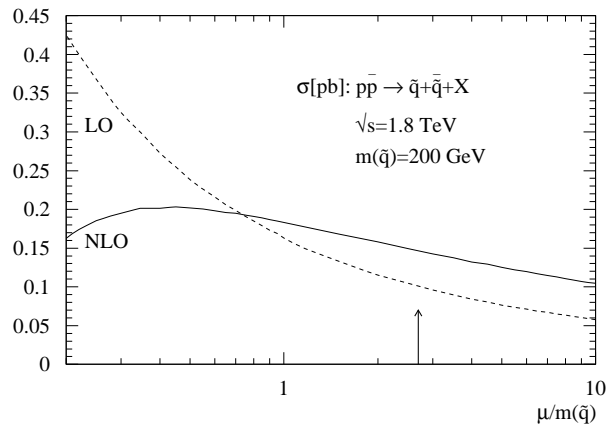,width=8cm}
\end{center}
\vspace*{-0.8cm}
\caption[]{\it Renormalization/factorization scale dependence 
  of the total cross section $p\bar{p}\to\sq\sqb$ at the 
  non-upgraded Tevatron.
  The arrow indicated the average invariant energy $<s>^{1/2}$
  in the hard subprocess, which was used in the original analysis.
  \label{fig_lepto_tevscale}}
\end{figure}

The Born cross section is given by eq.(\ref{eq_stop_lo}), the $t$
channel diagram for quark-antiquark collisions is in this case not
suppressed by the incoming state but absent, due to the decoupling of
the gluino. The strong coupling is independent of the flavor of the
light squark, \ie in case of hadroproduction all flavors look
identical, unlike Fig.~\ref{fig_lepto_hera}. The calculation for $R$
parity breaking squark production is the same as for the stop pairs,
the Feynman diagrams are given in Fig.~\ref{fig_stop_feyn} taking the
limit of decoupling gluino and removing the strong four-squark
coupling. This follows from appendix~\ref{chap_app_feynman}. The
partonic NLO cross section only depends on the partonic cm energy $s$
and on the mass of the squark, leading to the [in this case literal]
scaling functions for the Born, virtual and soft, hard, and scale
dependent contributions
\begin{equation}
\hat{\sigma}_{ij} = 
\frac{\alpha_s^2(\mu^2)}{\ms^2}
\left\{
f_{ij}^B(\eta) + 4\pi\alpha_s(\mu^2)
\left[ f_{ij}^{V+S}(\eta)
     + f_{ij}^H(\eta)
     + \bar{f}_{ij}(\eta) \log \left( \frac{\mu^2}{\ms^2} \right)
\right] \right\}
\end{equation}
where $i,j$ denote the initial state partons. The factorization and
the renormalization scale have been identified, and
$\eta=s/(4\ms^2)-1$. The numerical form and the structure of these
scaling functions are very similar to the stop case,
Fig.~\ref{fig_stop_scalingfct}, since the SUSY parameters except for
the outgoing mass influence the stop scaling functions only
marginally, and the four-squark vertex only contributes to the
numerically suppressed virtual corrections. This result for the $R$
parity violating squarks can also be obtained from the even more
general light-flavor squark production~\cite{roland} in the limit of a
large gluino mass. However, the occurrence of the gluino mass in the
Born term requires a numerically large gluino mass for the analysis.
And the four-squark coupling has to be removed as for the derivation
from the diagonal stop pair production.

Exactly as for the stops, the scale dependence
Fig.~\ref{fig_lepto_tevscale} in NLO leads to a maximum and an
increasing accuracy for the derivation of limits on the mass of the
particles from non-observation. Since in leading order any scale of
the process can be considered, choosing the invariant energy of the
final state $\mu=\sqrt{s}$ is possible.  Especially if NLO
calculations are not available the choice of the scale of the process
leads to considerable uncertainties, which is illustrated in this
example: In NLO the choice of the factorization scale is no longer free,
since it must be a parameter defined in terms of external variables,
which does not allow for $\mu=\sqrt{s}$. The choice of $\mu=\sqrt{s}$
in next-to-leading order leads to an inconsistence of the order
$\alpha_s$, independent of the order of perturbation theory, in which
the process has been calculated~\cite{collins}. Another scale for the
production of massive particles would be the final state mass $m$,
which turns out to be much smaller than $\sqrt{s}$ averaged with the
weight of the cross section. The different $K$ factors for both
choices of the scale are given in Fig.~\ref{fig_lepto_tevscale}.  For
the typical behavior of the leading and next-to-leading order cross
section the large scale, $\sqrt{s}$, is far from the point of maximal
convergence. This reflects the appearance of logarithms $\log
m^2/\mu^2$, which render the corrections unnaturally large.\smallskip

\begin{figure}[t] \begin{center}
\epsfig{file=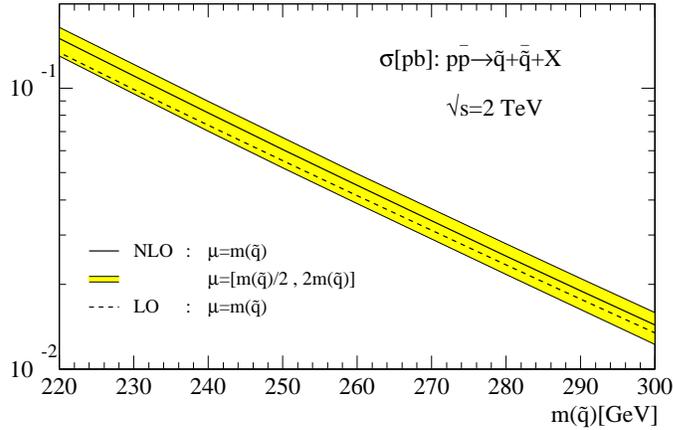,width=9cm}
\end{center}
\vspace*{-0.8cm}
\caption[]{\it Total hadronic cross section for the production of 
  $R$ parity violating squarks in hadron collisions
  $p\bar{p}\to\sq\sqb$. The leading order result is given for the
  renormalization/factorization scale $\mu=m$. In the first 
  case also the values of $\alpha_s$ and the parton densities have 
  been taken in leading order. The $K$ factor 
  for the smaller central scale $m$ is comparably small.
  \label{fig_lepto_tevmass}}
\end{figure}

The NLO cross section and the LO cross section for
the central mass scale for the upgraded Tevatron are given in
Fig.~\ref{fig_lepto_tevmass}. The variation of the cross section with
the scale decreases from $100\%$ to $30\%$ if the NLO result is used.
Compared to the $\mu=m$ leading order result this leads to a more
accurate and always higher mass limit, the improvement of the mass
bounds however stays below $10\gev$ for the central scale.

\subsubsection{Experimental Analyses}

One of the design features of the $ep$ collider HERA is the search for
leptoquark-like particles, \ie particles which carry electron and
quark quantum numbers, and can in case of scalars be identified with
$R$ parity violating squarks. They can occur either in the $s$ or in
the $t$ channel and lead to an excess in the $ep$ cross
section~\cite{hera_ano}. The decay channel of this squark strongly
depends on the masses of the supersymmetric scenario considered, but
may well include a high-$p_T$ electron. This signal is essentially
background free~\cite{dreiner}. The interpretation of the combined
data of ZEUS and H1 as $R$ parity violating squarks leads to combined
limits on the Yukawa coupling $\lambda'$ and the branching ratio $\br$
to the observed high-$p_T$ electron final state [The numbers are based
on the data analyzed by fall 1997.].
\begin{alignat}{9}
e^+ u \quad :& \qquad \lambda' \sqrt{\br} \sim 0.017 \cdots 0.025 
 \notag \\
e^+ d \quad :& \qquad \lambda' \sqrt{\br} \sim 0.025 \cdots 0.033 
 \notag \\
e^+ s \quad :& \qquad \lambda' \sqrt{\br} \sim 0.15 \cdots 0.25 
\label{eq_lepto_measure}
\end{alignat}
Couplings to $\bar{d}$ and $\bar{u}$ would lead to a large enhancement
in the $e^-p$ run and are forbidden because of their non-observation.
Moreover, couplings involving a positron and an up-type quarks would
lead to electric charges, which do not occur for MSSM-type squarks.
The coupling $\lambda'$ has to be interpreted as entries into the
non-diagonal Yukawa coupling matrix, connecting down-type quarks to
squarks and electrons. The diagonal matrix element $\lambda'_{111}$,
which would lead to the production $e^+d \to \tilde{u}$, is excluded
by neutrinoless double beta decay.  Possible candidates for a
resonance production are
\begin{equation}
e^+ d \to \tilde{c}_L, \tilde{t}_L \qquad \qquad \qquad 
e^+ s \to \tilde{t}_L
\end{equation}\medskip 

As depicted in section~\ref{sect_susy_rpar}, atomic parity violation
yields strong limits on $\lambda'{}^2/m^2$ for any leptoquark
interacting with valence quarks. They can for a mass of $\sim 200\gev$
be translated into bounds on the Yukawa coupling matrix $\lambda'
\lesssim 0.055 \, [e^+d]$~\cite{r_theo,r_exp}.  Combined with the
measured values at HERA, eq.(\ref{eq_lepto_measure}), this yields
lower limits on the branching ratio to the observed $eq$ final state
$\br \gtrsim 0.2 \cdots 0.4 \, [e^+d]$.  The limits obtained from
atomic parity violation are derived for the presence of only one
particle being responsible for the possible deviation from the
Standard Model. More than one $R$ parity violating squark influences
this analysis, the result depending on the sign of interference terms
and thereby on the quantum numbers. The assumed left handed stop quark
is in general a superposition of two states with different mass and
equal electroweak properties. This strengthens the bound on the
branching ratio:
\begin{equation}
\br \longrightarrow 
\br \left( 1 + \tan^2 \tmix \, \frac{\mse^2}{\msz^2} \right)
\end{equation}
At LEP, the obtained limits on $\lambda'$ are relevant for sea quarks
only~\cite{r_theo,r_exp}. By the same token as for atomic parity
violation they start from $\lambda' \lesssim 0.6 \, [e^+s]$ and give
$\br \gtrsim 0.05 \cdots 0.2 \, [e^+s]$.\medskip

The data from the search at the Tevatron can be written --- given the
mass of $\sim 200\gev$ from the HERA analysis --- as an upper bound in
the branching ratio $\br \lesssim 0.5 \cdots 0.7$. Theoretically the
competition between supersymmetric $R$ parity conserving, and $R$
parity violating decays makes it possible to vary the branching ratio
into the $eq$ mode with the mass of the particles forming the decay
chains.

In the case of $ed \to \tilde{c}_L$ the most important MSSM-like decay
modes are $c \nnj$ and $s \cpj$. Assuming the gauginos being heavy
[$\mce > 200\gev$] insures that the branching ratio lies in the region
of $\br \sim 1/2$. The higgsino decays suffer from the small strange
quark mass. For $ed \to \tilde{t}_L$ the strong decay channels are
forbidden, as depicted in section \ref{chap_decay}. The decays into $t
\nnj$ and $b \cpj$ can be suppressed by large gaugino and higgsino
mass parameters.  However, suppressing the whole set of possible stop
decays, eq.(\ref{eq_dec_channels}), yields some fine tuning of masses
and mixing also in the sbottom sector. The coupling $\lambda'$ for the
$es \to \tilde{t}_L$ production channel is comparably large, which
renders the different Yukawa coupling and gauge coupling mediated
decay channels of similar size, and thereby prevents from any fine
tuning.\smallskip 

The interpretation of the HERA excess as $R$ parity violating squarks
is therefore not ruled out by the bounds set by other experimental and
theoretical analyses, but give an impression how different collider
experiments and non-collider experiments like atomic parity violation
and search for neutrinoless beta decay can altogether constrain the
parameters in the same model. Incorporating all available data, HERA
itself is the only experiment able to remove this interpretation by
non-confirming the excess.

\chapter{Conclusions}

In this work supersymmetric QCD corrections to decays involving scalar
top quarks and to the hadroproduction of neutralinos/charginos and
stops are presented. The decay widths as well as the production cross
sections calculated in perturbation theory exhibit an unphysical
dependence on the renormalization and/or factorization scales. In
leading order this dependence is in general strong. Compared to a
central scale, which could be the mass $m$ for the decay width and the
hadro--production cross section of massive particles, this leads to
variations up to a factor of two for scales between $m/2$ and $2m$. In
next-to-leading order this dependence is considerably weaker, \ie in
addition to the $K$ factor, the next-to-leading order results always
improve the precision of the theoretical prediction used for the
experimental analysis.\medskip

The basic properties of the scalar top sector are investigated by
calculating the supersymmetric QCD corrections to the strongly and
weakly coupling decays, including a stop either in the initial or in
the final state. An elegant definition of a running mixing angle in
next-to-leading order is given, in order to restore the Born type
symmetries between the stops in next-to-leading order observables. The
mixing angle counter term is compared to other renormalization
schemes. Although the phenomenological motivations for the various
schemes are different, the numerical differences are shown to be
small.

The different stop decay widths obey a strong hierarchy, starting with
rare decay channels, and then proceeding towards weak and strong
two-body decays for an increasing mass of the decaying stop. The
strong decay will be dominant for a heavy stop state. The
next-to-leading order corrections to the heavy stop state decaying
into a gluino are large $\sim 30\%$ and always positive, while for the
gluino decay into a top squark they turn out to be small and negative
$\sim -5\%$. This feature also arises for the light-flavor squarks and
is due to interference between different color structures and the
different analytical continuation of logarithms. The dependence of all
decay widths on the mixing angle can be described by a $K$ factor,
which stays constant for varying angle. The scale dependence of the
decay widths is reduced from a factor of two to about $50\%$ in
next-to-leading order.

The weak decays of a light stop into a neutralino and a chargino are
analyzed, to illustrate the running mixing angle. The next-to-leading
order supersymmetric QCD corrections are small compared with typical
strong decays. Apart from special mass scenarios and threshold effects
they are $\lesssim 10\%$.  In contrast to the strong decays the sign
of the correction is not fixed, it is strongly dependent on the mass
scenario considered. The same holds for decays of heavy neutralinos.
They can be produced at $e^+e^-$ linear colliders and will in
supergravity inspired scenarios be higgsino-like. Therefore the decay
induced by the top-stop Yukawa coupling can give large contributions,
whereas the light-flavor final states are strongly suppressed. The
next-to-leading order corrections to these widths are moderate: $\sim
10\%$.\medskip

The next-to-leading order production cross section for neutralinos and
charginos can be used to derive mass limits at the upgraded Tevatron
and at the LHC. This yields an improvement of the mass bounds for
these particles obtained at LEP2.  Although mass and mixing parameters
could be derived from cascade decays of strongly interacting
particles, the only way to keep maximal independence of the choice of
the model is the direct search.  Similar to the case of gluino
production, on-shell and off-shell intermediate particle contributions
have to be distinguished. This is done in a manner, which naturally
coincides with the experimental analyses. The different higgsino-like
and gaugino-like contributions to the production cross section can be
analyzed and give a smooth picture of the next-to-leading order
corrections. Although the scale dependence in next-to-leading order is
not as much improved as for strongly interacting particles, where the
dependence on the running QCD coupling arises in leading order, it
stays below few percent in next-to-leading order. The $K$ factor for
all possible final state neutralinos/charginos is nearly constant for
varying masses and corrects the leading order result by +20$\%$ to
+50$\%$. However, strong cancelations between different diagrams may
lead to large $K$ factors in the mixed production channel, strongly
dependent on the mass and mixing parameters chosen.\medskip

The search for scalar top quarks is naturally the next step after the
search for light-flavor squarks and gluinos at the upgraded Tevatron
and at the LHC. Since the QCD type couplings are invariant under
chiral transformations, and cannot distinguish between the right and
the left stop, the production cross section in leading order depends
only on the mass of the produced particles. The mixing angle as well
as the mass of the light-flavor squarks and the gluino only enters
through the virtual corrections. This dependence is found to be much
smaller than the scale dependence and thereby negligible. The
corrections to the diagonal stop pair production are different for
quark and gluon incoming states.  Whereas they are small and negative
$\sim -5\%$ for incoming quarks, they are large and positive $\sim
50\%$ for gluons. Since even at the Tevatron only very heavy stops are
produced mainly in quark-antiquark collisions, this mass dependent $K$
factor leads to an increase of the mass bounds derived in leading
order. In addition to the total cross section the differential cross
sections are analyzed: The rapidity distribution at the upgraded
Tevatron can be described using a $K$ factor, the transverse momentum
distribution however is shifted to a softer regime in next-to-leading
order.\smallskip

The search for mixed stop final states could be a means to measure the
stop mixing angle directly at hadron colliders. However, the cross
section is based on a one-loop amplitude, and the calculation in the
limit of decoupled gluinos shows, that it is much smaller than the
cross section for the production of two heavy stop particles. The
direct measurement of the stop mixing angle will be possible only in
the process $e^+e^-\to \ste\stzb$.\medskip

The calculation of the stop production cross section can be adapted to
the search for supersymmetric $R$ parity violating squarks at the
Tevatron. Similar to the HERA production process, the next-to-leading
order calculation for the hadroproduction gives rise to improved
bounds on the mass and on the branching ratio into the observed $eq$
channel.  However, this is not sufficient to close the window in the
branching ratio between the atomic parity violation and LEP on one
side and the Tevatron on the other side, before the excess completely
vanishes by accumulating more HERA data.\bigskip

\begin{figure}[p] \begin{center}
\epsfig{file=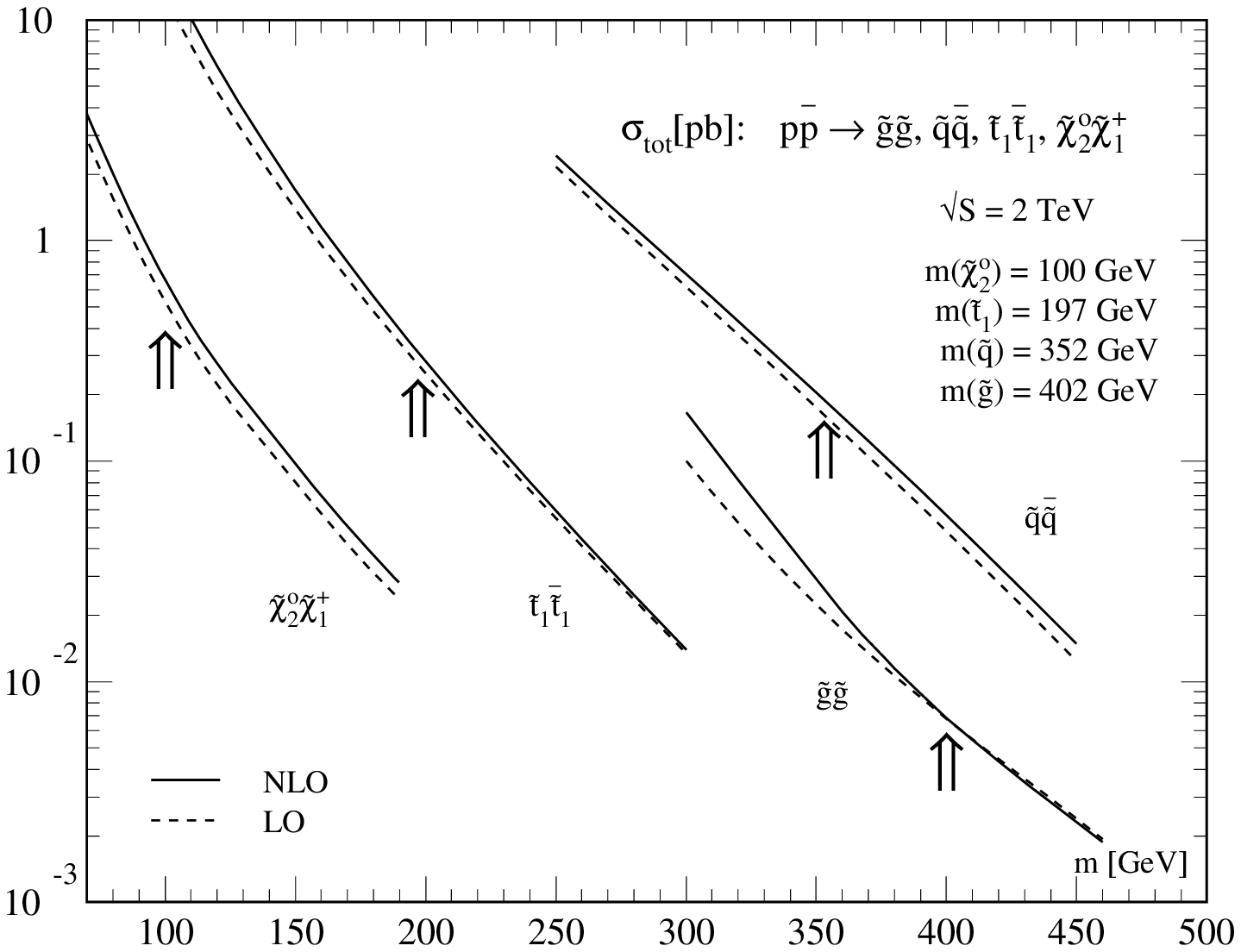,width=12cm}
\epsfig{file=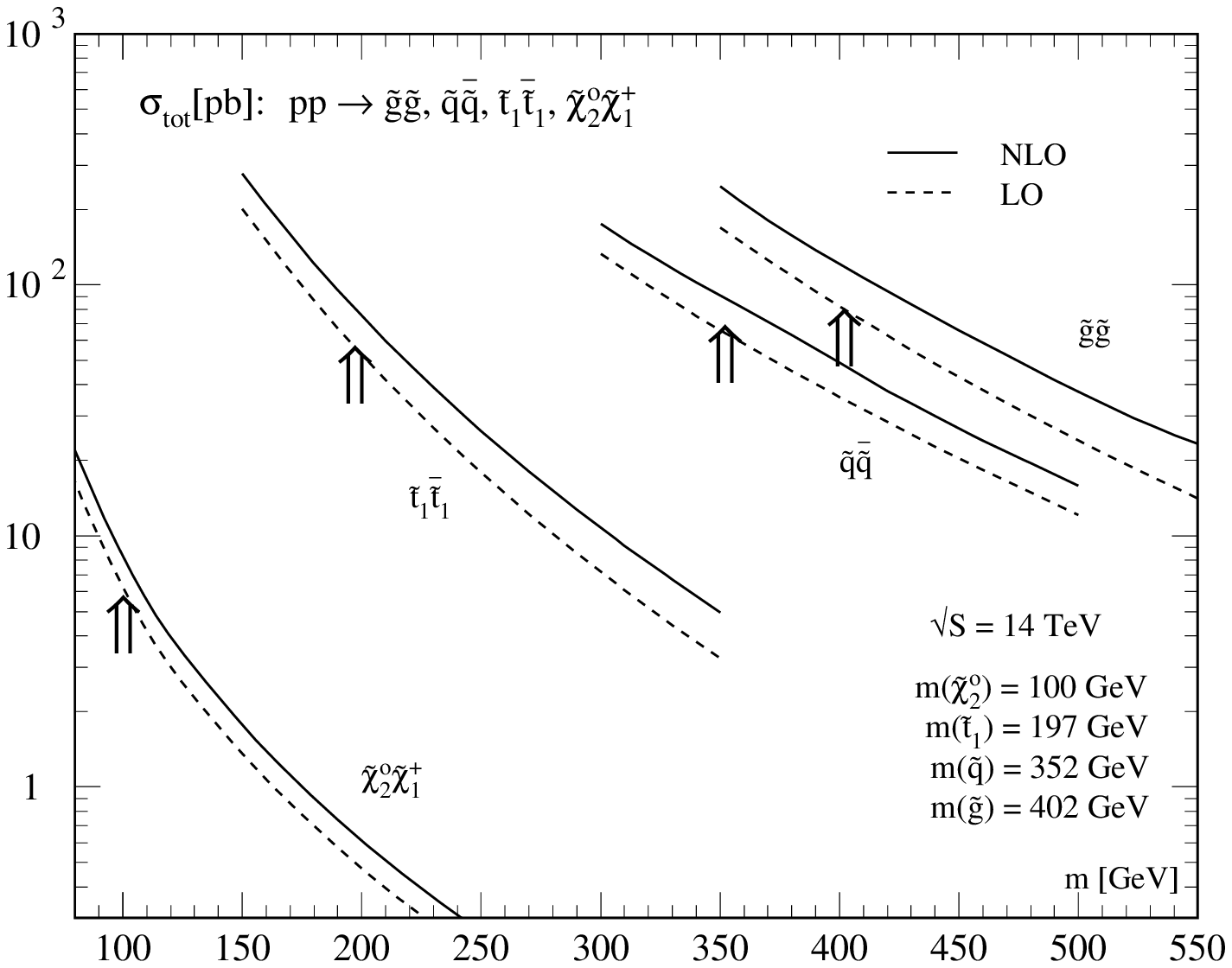,width=12cm}
\end{center}
\caption[]{\it The total cross section for pairs of squarks, gluinos, 
  stops, and neutralinos/charginos as a function of the mass of the 
  produced particles. The usual SUGRA mass spectrum
  $m_0=100\gev, m_{1/2}=150\gev, A_0=300\gev, \tan \beta=4, \mu>0$ 
  is denoted by the arrows. The masses of $\nnz$ and $\cme$ differ 
  only by a few $\gev$ in the small $m_{1/2}$ regime. For the 
  neutralino/chargino production cross section the masses are 
  consistently varied with $m_{1/2}$. 
  The cross sections are given for the 
  upgraded Tevatron and for the LHC.
  \label{fig_con}}
\end{figure}

All calculations have been performed in a supergravity inspired GUT
scenario. This connects not only the different gaugino masses, but
also the masses of the squarks and the stops. All these particles can
be searched for at the upgraded Tevatron and at the LHC. In
Fig.~\ref{fig_con} the relevant cross sections are given as a function
of the mass of the produced particles. The weakly interacting
particles are strongly suppressed at the LHC; however, the search for
leptonic events at hadron colliders is completely different from the
hadronic final states.  Assuming a similar efficiency for the search
for stops and light-flavor squarks the search for both of them in
parallel seems to be promising, in particular, since the search for
squarks is a two dimensional, the search for stops a one dimensional
problem. The smaller stop cross section, due to the missing $t$
channel gluino contribution and the fixed non-degenerate flavor, is
compensated by the small mass of the light stop state in typical
scenarios. For gluinos, like-sign leptons in the final state should
improve the efficiency considerably, yielding all production processes
given in Fig.~\ref{fig_con} very promising at the upgraded Tevatron
and the LHC.

\newpage 

{\Large {\sc Acknowledgments}}\bigskip 

I would like to thank my advisor, Prof.~Peter~M.~Zerwas, for the
continuous support of this work and for his enthusiasm and
encouragement.\smallskip 

Furthermore, I would like to thank Roland H\"opker, Wim Beenakker and
Michael Kr\"amer for a pleasing collaboration and encouraging
discussions. Also I would like to thank Michael Spira for his guidance
and countless clarifying discussions during our collaboration in the
last five years.\smallskip

Finally I would like to thank Michael Pl\"umacher, Gudrun Hiller,
Thomas Gehrmann, Aude Gehrmann-De~Ridder, Axel Krause, Hubert
Spiesberger, Christoph J\"unger, and all members of the DESY theory
group for numerous discussions and an enjoyable working
atmosphere.\smallskip

\setlength{\parindent}{0pt}
\begin{appendix}
\chapter{SUSY Lagrangean}
\label{chap_app_feynman}

\section{Feynman Rules for Supersymmetric QCD}

In this appendix we give a complete set of Feynman rules, as used in
 the calculation of the various processes. Since fermion number
violating processes have to be considered, the rules make use of a
continuous fermion flow~\cite{denner_fermion}, which has to be fixed
once for any process. The Dirac trace has to be evaluated in the
opposite direction of this fermion flow. If not stated otherwise,
we assume the fermion flow being identical to the Dirac fermion flow
\eg for the quarks.

Using these Feynman rules makes it needless to introduce charge
conjugation matrices. Moreover, the relative signs of different
diagrams contributing to the same process can be fixed: Any
permutation of the fermion flow of two external fermion lines gives
rise to a factor $(-1)$ for the matrix element, due to Fermi
statistics and Wick ordering.  To match the spinors for processes
including two external Majorana and two external Dirac fermions, all
diagrams have to be evaluated with two different directions of the
fermion flow. We have checked explicitly that the two possible
orientations of the fermion flow of two combined diagrams lead to the
same result.

By changing the orientation of the fermion flow, the signs of vertices
with different types of couplings [$S,P,V,A$] change:

\begin{table}[h] \begin{center} \begin{tabular}{|l|rl|c|}
\hline 
coupling \rule[-2mm]{0mm}{6mm}
&&& sign \\ \hline
scalar ($S$) \rule[-2mm]{0mm}{7mm}
& $ C\;1\;C^{-1}\;=$&$ 1^T $
& $+$ \\
pseudo-scalar ($P$) \rule[-2mm]{0mm}{6mm}
& $ C\;\gamma_5\;C^{-1}\;=$&$ \gamma_5^T $
& $+$ \\
vector ($V$) \rule[-2mm]{0mm}{6mm}
& $ C\;\gamma_\mu\;C^{-1}\;=$&$ -\gamma_\mu^T $
& $-$ \\
axial-vector ($A$) \rule[-3mm]{0mm}{7mm} 
& $ C\;(\gamma_5\gamma_\mu)\;C^{-1}\;=$&$ (\gamma_5\gamma_\mu)^T $
& $+$ \\ \hline
\end{tabular} \end{center} 
\caption[]{\it Transformation of couplings with the orientation of the 
  fermion flow. $C$ is the charge conjugation matrix.}
\end{table}

\subsubsection{Standard Model Feynman Rules}

\begin{figure}[ht] \begin{center}
\pspicture(0,0)(16,8)
\rput[cl]{0}(0,7){\epsfig{file=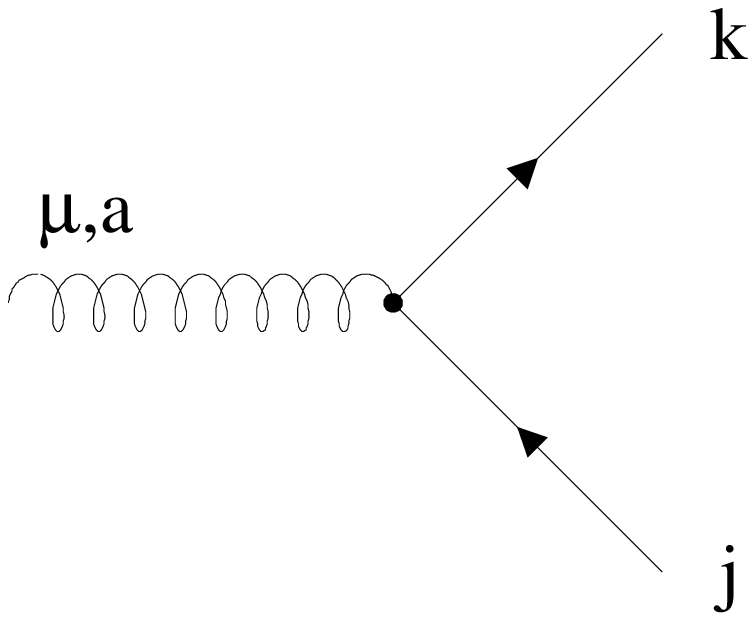,width=3cm}}
\rput[cl]{0}(3,7){$-i g_s\;T^a_{kj}\;\gamma^\mu$} 

\rput[cl]{0}(8,7){\epsfig{file=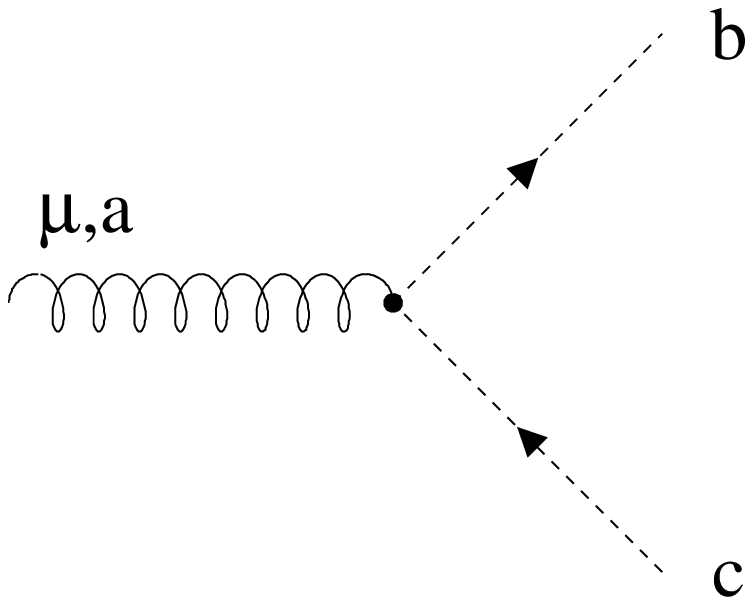,width=3cm}}
\rput[cl]{0}(11,7){$g_s\;f_{abc}\;p_b^\mu$} 

\rput[cl]{0}(0,4){\epsfig{file=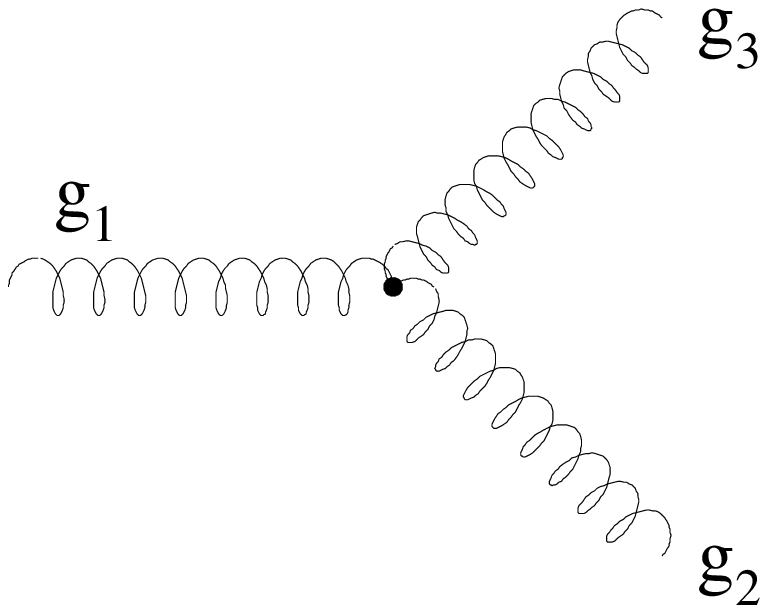,width=3cm}}
\rput[cl]{0}(  3,  4){$-g_s\;f_{a_1a_2a_3}
                     (k_1\!-\!k_2)^{\mu_3}\;g^{\mu_1\mu_2}$} 
\rput[cl]{0}(3.3,3.3){+ cycl.}

\rput[cl]{0}(8,4){\epsfig{file=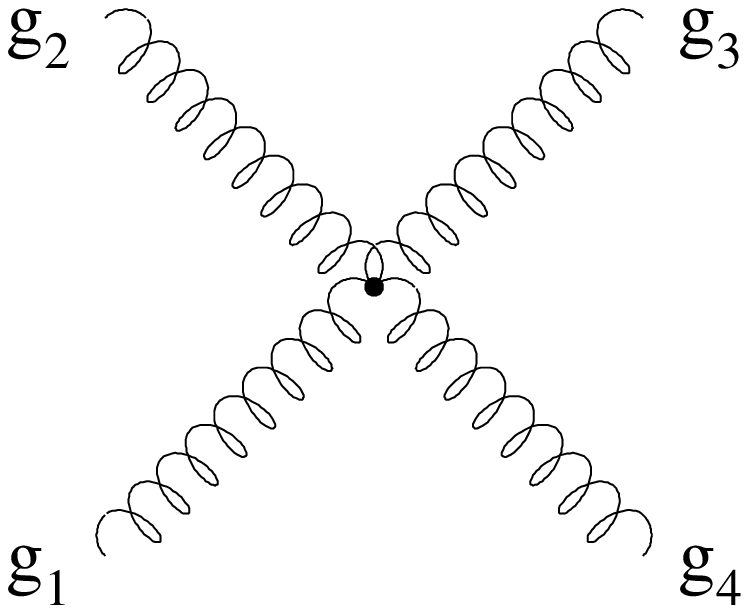,width=3cm}}
\rput[cl]{0}(  11,  4){$-i g_s^2\;f_{a_1a_2b}\;f_{a_3a_4b}$}
\rput[cl]{0}(11.3,3.3){$(g^{\mu_1\mu_3} g^{\mu_2\mu_4}
                        -g^{\mu_1\mu_4} g^{\mu_2\mu_3})$}
\rput[cl]{0}(11.3,2.6){+ cycl.}

\rput[cl]{0}(0,1){\epsfig{file=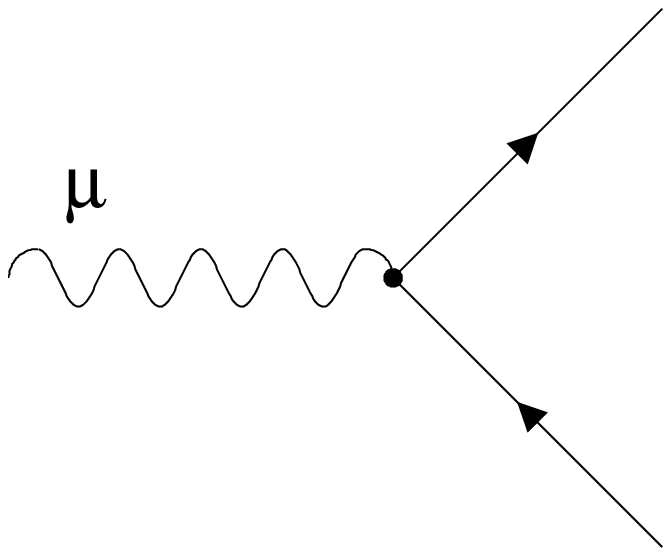,width=3cm}}
\rput[cl]{0}(3,1){$-i \gamma^\mu\;(\ell P_L + r P_R)$}
\endpspicture
\end{center}
\caption[]{\it Feynman rules for Standard Model quarks, gluons, 
  ghosts and weak gauge bosons, the dotted lines for scalars refer to
  Fadeev-Popov ghosts. The generic couplings $\ell,r$ are defined in 
  Tab.~\ref{tab_app_feynqcd}; $P_{LR}$ are the chiral projectors 
  $(1\mp \gamma_5)/2$
  \label{fig_app_feynqcd}}
\end{figure}

All momenta in the Feynman rules are defined incoming. The Standard
Model couplings of quarks, gluons, Fadeev-Popov ghosts and weak gauge
bosons are given in Fig.~\ref{fig_app_feynqcd} in the Feynman gauge.
The generators of SU(3)$_C$ obey the relations 
\begin{equation}
{\rm Tr} \left( T^a T^b \right) = \frac{1}{2} \delta^{ab} \qquad \qquad
\left[ T^a,T^b \right] = i f^{abc} T_c
\end{equation}

\begin{table}[b] \begin{center} \begin{tabular}{|l|c|c|}
\hline 
\rule[-1mm]{0mm}{6mm}
& $\ell$
& $r$ \\ \hline
$q \bar{q} \gamma$ \rule[-1mm]{0mm}{7mm}
& $Q e$
& $\ell$ \\
$q \bar{q} Z$ \rule[-1mm]{0mm}{7mm}
& $\displaystyle \frac{e}{s_w c_w} \left( T_3 - Q s_w^2 \right)$
& $\ell \; [T_3=0]$ \\
$u \bar{d} W^+_{\rm out}$ \rule[-6mm]{0mm}{12mm}
& $\displaystyle \frac{e}{\sqrt{2} s_w}$
& 0 \\ \hline 
\end{tabular} \end{center}
\caption[]{\it Couplings of quarks to weak gauge bosons as used in 
  Fig.~\ref{fig_app_feynqcd}. 
  \label{tab_app_feynqcd}}
\end{table}

The fermion propagators are defined as $i/(\slash{p}-m+i\varepsilon)$,
where $p$ is the momentum in the direction of the fermion flow. The
fermion {\sl number} flow does not occur. The gluon propagator is
$-i g^{\mu\nu}/(p^2+i\varepsilon)$.

\subsubsection{Supersymmetric QCD Feynman Rules}

The Feynman rules for supersymmetric QCD include, besides the Standard
Model particles, the gluino ($\gt$), the light flavor squarks
($\sql,\sqr$), and the mixing top squarks ($\ste,\stz$). Like in
Fig.~\ref{fig_app_feynqcd}, all momenta are defined incoming. The
coupling $q \sq \gt$ preserves the helicity of the quark and its
scalar partner as well as the flavor. In higher orders it has to be
modified to restore the supersymmetric Ward identity, as described in
chapter \ref{sect_susy_ward}. The same holds for the $\sq \sq g$
coupling. In leading order we use $g_s$ for all strong Standard Model
and their supersymmetry transformed couplings.

\begin{figure}[ht] \begin{center}
\pspicture(0,0)(16,8)
\rput[cl]{0}(0,7){\epsfig{file=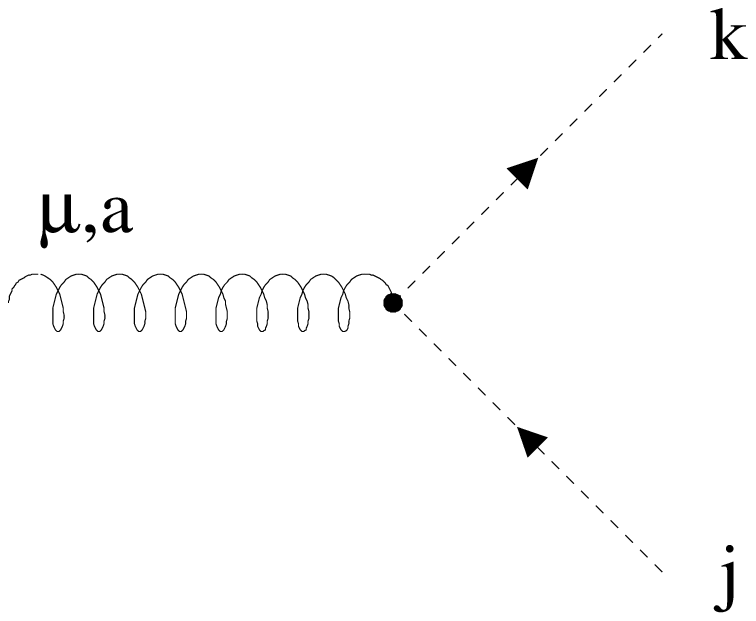,width=3cm}}
\rput[cl]{0}(3,7){$-i g_s\;T^a_{kj}\;(p_j-p_k)^\mu$}

\rput[cl]{0}(8,7){\epsfig{file=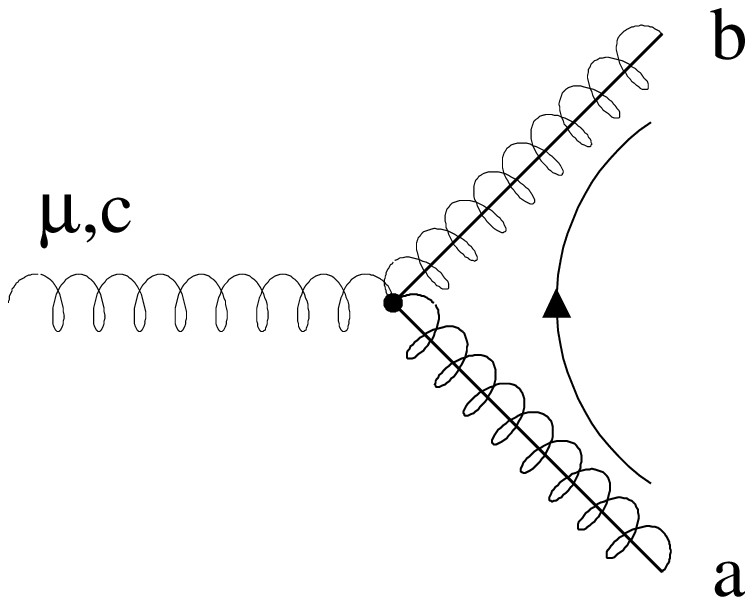,width=3cm}}
\rput[cl]{0}(11,7){$-g_s\;f_{abc}\;\gamma^\mu$}

\rput[cl]{0}(0,4){\epsfig{file=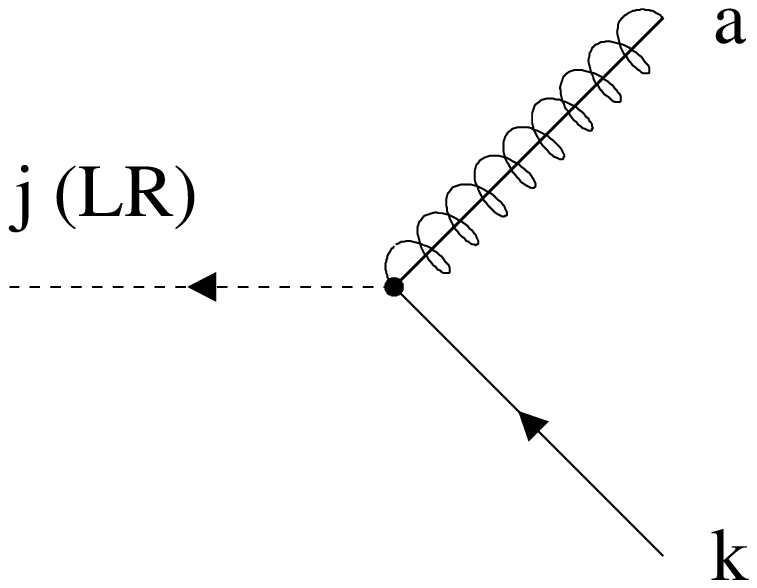,width=3cm}}
\rput[cl]{0}(3,4){$-i g_s \sqrt{2}\;T^a_{jk}\;(\pm P_{LR})$}
\rput[cl]{0}(8,4){\epsfig{file=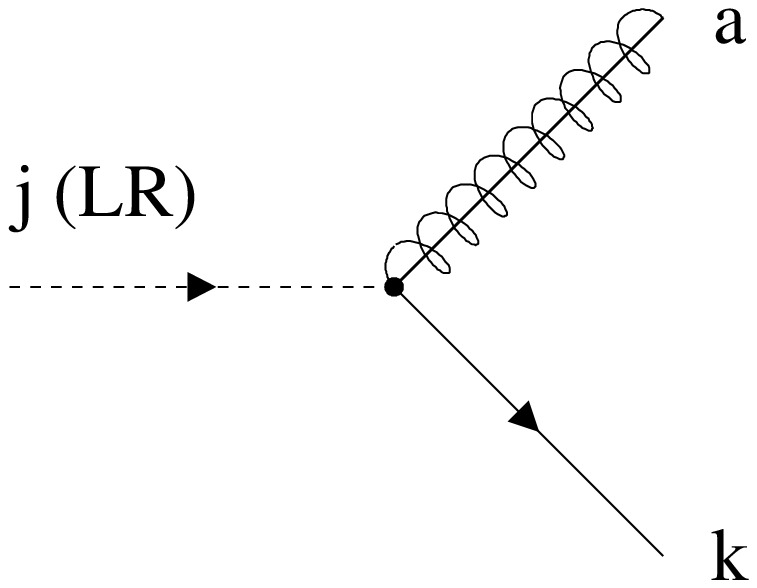,width=3cm}}
\rput[cl]{0}(11,4){$-i g_s \sqrt{2}\;T^a_{kj}\;(\pm P_{RL})$}

\rput[cl]{0}(0,1){\epsfig{file=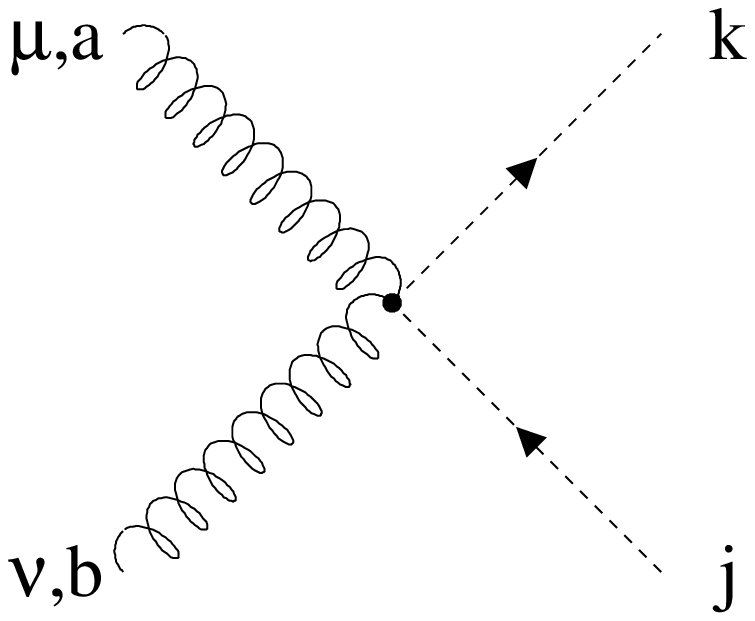,width=3cm}}
\rput[cl]{0}(3,1){$i g_s^2\;\{T^a,T^b\}\;g^{\mu \nu}$}
\rput[cl]{0}(8,1){\epsfig{file=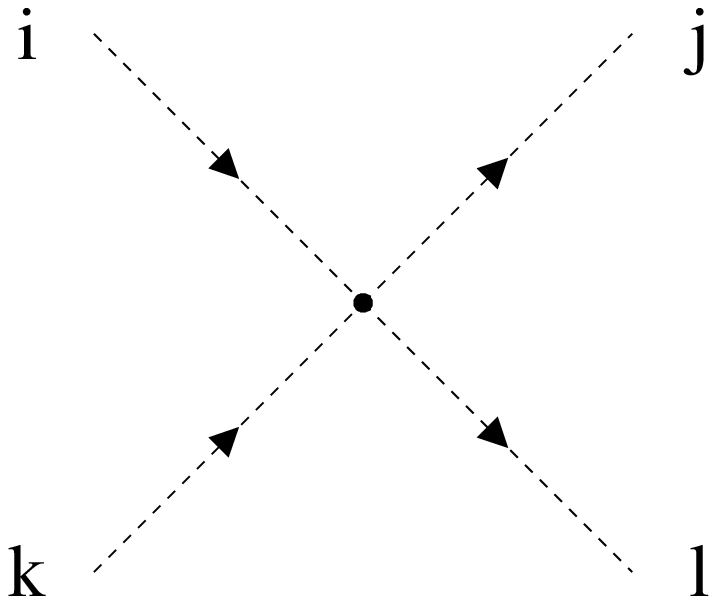,width=3cm}}
\rput[cl]{0}(11,1){$\displaystyle i g_s^2\;\frac{S_{ijkl}}{2}$}
\endpspicture
\end{center}
\caption[]{\it Feynman rules for supersymmetric QCD. The dotted 
  lines denote the squark $\sql,\sqr$. The tensors 
  $S_{ijkl}$ are defined 
  in Tab.~\ref{tab_app_feynsusyqcd}; $P_{LR}$ are the chiral projectors 
  $(1\mp \gamma_5)/2$
  \label{fig_app_feynsusyqcd}}
\end{figure}

The $\sq \sq g (g)$ vertices preserve the flavor and the 'helicity' of
the squark. Since only two squarks are present, this coupling cannot
mix the mass eigenstates.

The $q \sq \gt$ vertex Feynman rule is given for a light-flavor
$\sqlr$ with the corresponding sign and projector. To obtain the rules
for $\ste$ one has to add the contributions of the helicity
eigenstates and multiply the $L$ term with $\sin \tmix$ and the $R$
term with $\cos \tmix$. The mixing of the scalar top quark is
described in detail in chapter \ref{sect_susy_stop}.  The relevant
terms for stop mixing in the Lagrangean include the coupling to the
gluino and the four squark coupling which arises from the $D$ term in
the scalar potential, described in
eq.(\ref{eq_susy_scalarpot})\footnote{This coupling is not fixed by
  the requirement that the scalar top quark should carry a fundamental
  representation SU(3) charge.}. They can be expressed using the 
permutation operator 
\begin{equation}
\PX \quad : \quad  [\ste \leftrightarrow \stz;\ 
\cet\to-\set,\ \set\to\cet]
\label{eq_app_perm}
\end{equation}
which links the vertices including $\ste$ and $\stz$ 
\begin{alignat}{7}
  &{\cal L}_3 \quad &=\ & - \sqrt{2}\,g_s T^a_{ij} 
   \left( 1 + \PX \right) 
      \bar{\gt}_a \Big[ \cet\,P_L -\set\,P_R \Big] t_j\,\ste{}_i^* 
      + \text{h.c.} \notag \\[1mm]
  &{\cal L}_4 \quad &=\ & - \frac{g_s^2}{8}\,\left( 1 + \PX \right)
      \ste{}_i^*\,\ste{}_j\,
      \bigg\{ \cztq\,S_2^{ijkl}\,\ste{}_k^*\,\ste{}_l 
            + 2\,\Big[ \sztq\,S_2^{ijkl}-S_1^{ijkl} \Big]\,\stz{}_k^*\,\stz{}_l
       \notag \\
  &&& \hphantom{- \frac{g_s^2}{8}\,\left( 1+\PX \right)\ste{}_i^*\,\ste{}_j aA}
          {}+ 4\,\czt\,S_1^{ijkl} \sum_{\sq\neq\st} 
              \big( \sql{}_k^*\,\sql{}_l - \sqr{}_k^*\,\sqr{}_l \big)
      \bigg\}   
\label{eq_app_l3l4} 
\end{alignat}
The sine/cosine of an angle $\beta$ is as usually denoted by
$s_\beta,c_\beta$. However, these vertices describe only processes
which are essentially diagonal either in $\ste$ or in $\stz$. Four
squark vertices, mixing $\ste$ and $\stz$, can be derived from the
Lagrangean
\begin{alignat}{7}
  &{\cal L}'_4 \quad &=\ & \frac{g_s^2}{4}\,\szt\,
      \big( \ste{}_i^*\,\stz{}_j + \stz{}_i^*\,\ste{}_j \big)\,
      \bigg\{ \czt\,S_2^{ijkl}\,\big( \ste{}_k^*\,\ste{}_l 
                                   - \stz{}_k^*\,\stz{}_l \big) 
       \notag \\
  &&& \hphantom{\frac{g_s^2}{4}\,\szt\,\big( \ste{}_i^*\,\stz{}_j 
                + \stz{}_i^*\,\ste{}_j \big)\, A}
          {}+ 2\,S_1^{ijkl} \sum_{\sq\neq\st} 
              \big( \sql{}_k^*\,\sql{}_l - \sqr{}_k^*\,\sqr{}_l \big)
      \bigg\}   
\label{eq_app_l4p}
\end{alignat}

The structure of the four squark coupling is given in terms of the
flavor $f_j$ and the helicity of the Standard Model partner $h_j$.
The two tensors used in Tab.~\ref{tab_app_feynsusyqcd} are
\begin{alignat}{3}
&S_{ijkl}^{(1)} = \left( \delta_{il} \delta_{jk} 
                       - \frac{1}{N} \delta_{ij} \delta_{kl} \right) 
\notag \\
&S_{ijkl}^{(2)} = \frac{N-1}{N} \left( \delta_{il} \delta_{jk} 
                                     + \delta_{ij} \delta_{kl} \right)
\phantom{HHHHAAAASSSSSSS}
\label{eq_app_feyn_susyqcd}
\end{alignat}

\begin{table}[h] \begin{center} \begin{tabular}{|l|r|}
\hline 
$\sq_i \sq_j \sq_k \sq_l$ \rule[-2mm]{0mm}{6mm} 
& $S_{ijkl}$ \\ \hline
$\sq \sq \sq \sq \quad
[ h_i = h_j \; h_j = h_l \; f_i = f_j \; f_j = f_l ]$
& $-S_{ijkl}^{(2)}$ \rule[-2mm]{0mm}{7mm} \\
$\sq \sq \sq \sq \quad
[ h_i = h_j \; h_j = h_l \; f_i \neq f_j \; f_j \neq f_l ]$ 
& $-S_{ijkl}^{(1)}$ \rule[-2mm]{0mm}{7mm} \\
$\sq \sq \sq \sq \quad  
[ h_i \neq h_j \; h_j \neq h_l ]$ 
& $+S_{ijkl}^{(1)}$ \rule[-2mm]{0mm}{7mm} \\
$\ste \ste \sqlr \sqlr$
& $\mp \czt \; S_{ijkl}^{(1)}$ \rule[-2mm]{0mm}{7mm} \\
$\stz \stz \sqlr \sqlr$
& $\pm \czt \; S_{ijkl}^{(1)}$ \rule[-2mm]{0mm}{7mm} \\
$\ste \stz \sqlr \sqlr \quad \stz \ste \sqlr \sqlr$
& $\pm \szt \; S_{ijkl}^{(1)} $ \rule[-2mm]{0mm}{7mm} \\
$\ste \ste \ste \ste \qquad \; \; \stz \stz \stz \stz$
& $- \czt^2 \; S_{ijkl}^{(2)}$ \rule[-2mm]{0mm}{7mm} \\
$\ste \ste \ste \stz \qquad \; \; \ste \ste \stz \ste$
& $+ \szt \czt \; S_{ijkl}^{(2)}$ \rule[-2mm]{0mm}{7mm} \\
$\stz \stz \stz \ste \qquad \; \; \stz \stz \ste \stz$
& $- \szt \czt \; S_{ijkl}^{(2)}$ \rule[-2mm]{0mm}{7mm} \\
$\ste \ste \stz \stz$
& $S_{ijkl}^{(1)} - \szt^2 \; S_{ijkl}^{(2)}$ \rule[-2mm]{0mm}{7mm} \\
$\ste \stz \ste \stz \qquad \; \; \stz \ste \stz \ste$
& $- \szt^2 \; S_{ijkl}^{(2)}$ \rule[-4mm]{0mm}{9mm} \\
\hline 
\end{tabular} \end{center}  
\caption[]{\it Tensors arising in the generic four squark coupling in
  Fig.~\ref{fig_app_feynsusyqcd}. The tensors $S^{(1)}$ and $S^{(2)}$
  are defined in 
  eq.(\ref{eq_app_feyn_susyqcd}). \label{tab_app_feynsusyqcd}}
\end{table}

\section{Neutralinos and Charginos}

\subsubsection{Diagonalization for Neutralinos}

The diagonalization procedure for Neutralinos is described \eg
in~\cite{gunion_haber}. The four Majorana neutralinos ($\nnj$) are
defined as the mass basis of all neutral higgsinos, the photino, and
the zino interaction eigenstates. They are four component Majorana
spinors, therefore the mass matrix is symmetric. It is possible to
start from a non-diagonal matrix in the $(\tilde{B} \tilde{W}_3)$
basis or in the $(\tilde{\gamma} \tilde{Z})$ basis. We denote the two
possible mass matrices as $\M$ and $\M'$ for the two component states:
\begin{eqnarray}
\M = 
\left(  \begin{array}{cccc} 
 \mbin    &   0 &
 -\mz s_w c_\beta &   \mz s_w s_\beta  \\
 0                &   \mwin &
  \mz c_w c_\beta &  -\mz c_w s_\beta  \\
 -\mz s_w c_\beta &   \mz c_w c_\beta &
  0               &  -\mu              \\
  \mz s_w s_\beta &  -\mz c_w s_\beta &
 -\mu             &  0
        \end{array}  \right)  
\label{eq_app_neutmass}
\end{eqnarray}

\begin{figure}[b] \begin{center}
\pspicture(0,0)(16,5)
\rput[cl]{0}(0,4){\epsfig{file=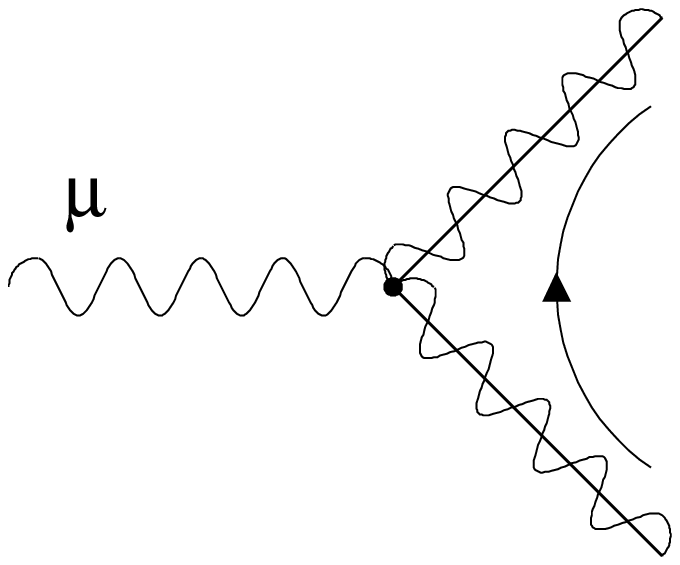,width=3cm}}
\rput[cl]{0}(3,4){$-i \gamma^\mu \left( L P_L + R P_R \right)$}
\rput[cl]{0}(0,1){\epsfig{file=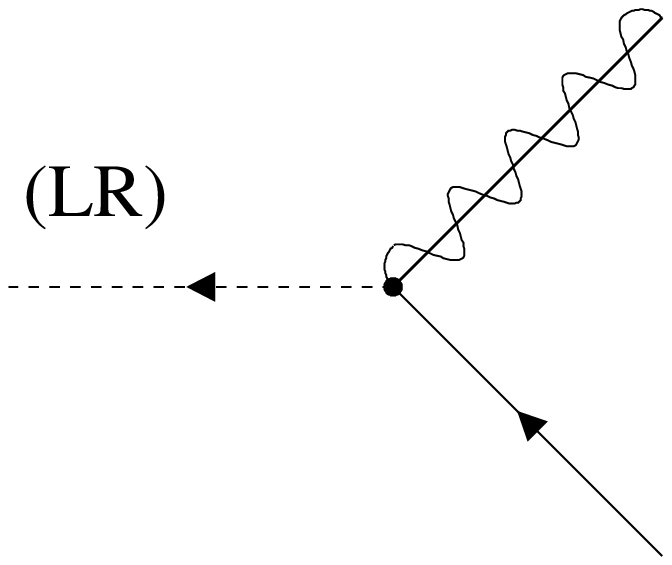,width=3cm}}
\rput[cl]{0}(3,1){$-i \sqrt{2} (\pm A^*_{LR}P_{LR}+B^*_{LR}P_{RL}) $}
\rput[cl]{0}(8,1){\epsfig{file=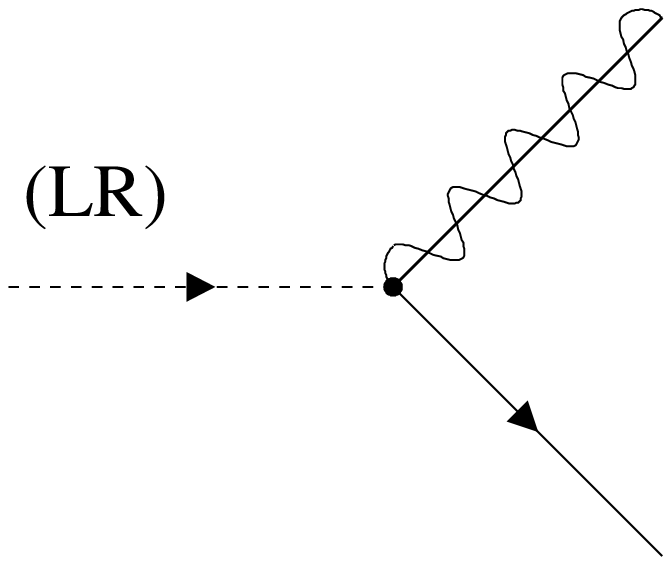,width=3cm}}
\rput[cl]{0}(11,1){$-i \sqrt{2} (\pm A_{LR}P_{RL}+B_{LR} P_{LR})$}
\endpspicture
\end{center}
\caption[]{\it Feynman rules for neutralinos and charginos. The dotted 
  lines denote a left or right scalar quarks $\sql,\sqr$. $L,R$ are 
  defined in Tab.~\ref{tab_app_feynneut1}, $A,B$ in 
  Tab.~\ref{tab_app_feynneut2}. Note that the $q\sq \tilde{\chi}$
  vertices for incoming and outgoing squarks are linked. 
  \label{fig_app_feynneut}}
\end{figure} 

The corresponding unitary mixing matrices $N,N'$ diagonalize these
mass matrices, $N$ in the $(\tilde{B} \tilde{W}_3)$ and $N'$ in the
$(\tilde{\gamma} \tilde{Z})$ basis\footnote{Any complex symmetric
  matrix can be diagonalized to a real diagonal matrix by a unitary
  transformation $U^TAU$ where $U^\dagger=U^{-1}$. The diagonal matrix
  can be chosen real since phase factors can be absorbed into the
  unitary mixing matrix.}. In case the mass matrix is real, the mixing
matrices $N,N'$ are also chosen to be real, to keep the couplings from
becoming complex \ie in this special case the in general complex
unitary transformation becomes real and orthogonal.  However, for CP
invariant observables the typical coupling factors must be purely real
or imaginary for any complex unitary mixing matrix.
\begin{alignat}{5}
  & N^* \; \M \; N^{-1} = \M_{\rm diag} && \notag \\
  & N'{}^* \; \M' \; N'{}^{-1} = \M_{\rm diag} \qquad \qquad \qquad &&
  {N'}_{j1} = N_{j1} c_w + N_{j2} s_w \notag \\
  &&&{N'}_{j2} = -N_{j1} s_w + N_{j2} c_w \notag \\
  &&&{N'}_{j3} = N_{j3} \notag \\
  &&&{N'}_{j4} = N_{j4}
\end{alignat}

The mixing matrix is defined in terms of the arbitrary-sign
eigenvalues of the mass matrix, $m_i$:
\begin{alignat}{9}
\frac{N_{i2}}{N_{i1}} = & \, 
 - \frac{c_w}{s_w}
   \frac{\mbin - m_i}{\mwin - m_i} \notag \\
\frac{N_{i3}}{N_{i1}} = & \, 
 \frac{ \mu ( \mbin - m_i ) ( \mwin - m_i )
       - \mz^2 s_\beta c_\beta [( \mbin - \mwin ) c_w^2
                                + \mwin - m_i ] }
      { \mz s_w ( \mwin - m_i )
        ( \mu c_\beta + m_i s_\beta )} \notag \\ 
\frac{N_{i4}}{N_{i1}} = & \, 
 \frac{-m_i ( \mbin - m_i ) ( \mwin - m_i )
      - \mz^2 c_\beta^2 [( \mbin - \mwin ) c_w^2
                                + \mwin - m_i ] }
      { \mz s_w ( \mwin - m_i )
        ( \mu c_\beta + m_i s_\beta )} \notag \\[2mm] 
N_{i1} = & \, \left[ 1 + \left( \frac{N_{i2}}{N_{i1}} \right)^2
                       + \left( \frac{N_{i3}}{N_{i1}} \right)^2
                       + \left( \frac{N_{i4}}{N_{i1}} \right)^2
              \right]^{-1/2}
\end{alignat}

\begin{table}[t] \begin{center} \begin{tabular}{|l|c|c|}
\hline 
\rule[-1mm]{0mm}{6mm}
& $L$
& $R$ \\ \hline
$\nni \nnj Z$ \rule[-6mm]{0mm}{12mm}
& $\displaystyle \frac{e}{2 s_w c_w}
  \left( \NY_{i3} \NY^*_{j3} - \NY_{i4} \NY^*_{j4} \right)$
& $L^* \; [\NY_{k4} \leftrightarrow -\NY_{k3}]$\\
$\cpi \cmj \gamma$ \rule[-6mm]{0mm}{12mm}
& $e \delta_{ij}$ 
& $L^*$ \\
$\cpi \cmj Z$ \rule[-6mm]{0mm}{12mm}
& $\displaystyle \frac{e}{s_w c_w}
  \left(   V_{i1}V^*_{j1}
         + \frac{1}{2}V_{i2}V^*_{j2} - \delta_{ij}s_w^2 \right)$
& $L^* \; [V \to U]$ \\
$\nni \cpj W^+_{\rm out}$ \rule[-6mm]{0mm}{12mm}
& $\displaystyle \frac{e}{\sqrt{2} s_w} 
  \left(\NY_{i4}V^*_{j2} 
    - \sqrt{2} \left(s_w\NY_{i1}+c_w\NY_{i2}\right) V^*_{j1}\right)$
& $L^* \; [V \to U, \NY_{i4} \to -\NY_{i3}]$ \\
$\nni \cmj W^+_{\rm in}$ \rule[-6mm]{0mm}{12mm}
& $\displaystyle \frac{e}{\sqrt{2} s_w} 
  \left(\NY^*_{i4}V_{j2} 
    - \sqrt{2} \left(s_w\NY^*_{i1}+c_w\NY^*_{i2}\right) V_{j1}\right)$
& $L^* \; [V \to U, \NY_{i4} \to -\NY_{i3}]$ \\
\hline 
\end{tabular} \end{center}
\caption[]{\it Couplings of neutralinos and charginos to weak gauge 
  bosons as used in Fig.~\ref{fig_app_feynneut}. The mixing matrix 
  $\NY$ is defined in the photino-zino basis. For charginos, the 
  fermion flow is assumed to follow the $\cpi$ fermion number
  flow. 
  \label{tab_app_feynneut1}}
\end{table}

The entries of $\M_{\rm diag}$ are not necessarily positive, if the
mixing matrix is kept real, \ie the eigenvalues are only equal to the
physical masses $\mnj$ up to a sign. It is possible to work with a
Lagrangean including negative mass eigenvalues $m_j$. In the final
expression these eigenvalues have to be substituted by their absolute
values $m_j \to \pm \mnj$, in order to express the analytical result
in terms of physical masses.  An equivalent way of introducing these
phases is to define a complex mixing matrix [$\NX,\NY$], in which a
row is multiplied by $i$, if the corresponding eigenvalue is negative:
\begin{eqnarray}
N_{kl} = \begin{cases}
  \NX_{kl} \quad [l=1,...,4] & 
     \text{if eigenvalue $m_k$ positive} \\[1mm]
i \NX_{kl} \quad [l=1,...,4] & 
     \text{if eigenvalue $m_k$ negative} \\
         \end{cases} 
\end{eqnarray}
The re-definition of $N' \to \NY$ is defined in analogy; in typical
scenarios one of the higgsino eigenvalues $m_k \; [k=3,4]$ turns out
to be negative, whereas the re-rotation $\NX \to \NY$ only affects the
gaugino part of the mass matrix.  Using this matrix $\NX,\NY$ one can
always stick to the positive mass values.
\begin{alignat}{5}
\NX^* \; \M \; \NX^{-1} &= \; 
   {\rm Diag}\left( \mnj \right)
\notag \\
\NY^* \; \M' \; \NY^{-1} &= \; 
   {\rm Diag}\left( \mnj \right) \qquad j=1,2,3,4
\end{alignat}
The masses of the four neutralinos are re-ordered by their size after
diagonalization, where $\nne$ is defined being the lightest of the
four. Combining the complex couplings including the matrix $\NX,\NY$
leads to exactly the same analytical results for CP invariant
observables as using the matrix $N,N'$; the phase factors from the
negative masses now enter by collecting factors of $i^2$ in the
typical combinations of the couplings and by anti-commuting the Dirac
matrices.  One advantage of the latter strategy is, that the
neutralino mass matrix is not fixed to real values by first
principles~\cite{kane_complex}.

\subsubsection{Diagonalization for Charginos} 

Charginos ($\cmj,\cpj$) are the mass eigenstates of charged winos and
higgsinos. The positive and negative charge particles mix
independently. Since the charginos are no Majorana particles the mass
matrix is not symmetric. Nevertheless the Dirac-chargino vertices can
be fermion number violating.
\begin{eqnarray}
\M = 
\left(  \begin{array}{cc} 
 \mwin        & \sqrt{2} \mw s_\beta \\
 \sqrt{2} \mw c_\beta & \mu 
        \end{array}  \right)  
\label{eq_app_charmass}
\end{eqnarray}

The unitary diagonalization matrices for the positive and negative
winos and higgsinos are $V$ and $U$, and the eigenvalues of the
diagonalized mass matrix can in general assumed to be
real\footnote{Any complex matrix can be diagonalized to a real and
  positive diagonal matrix using two unitary matrices $U^TAV$. If the
  matrix $A$ is real the matrices $U$ and $V$ can be chosen real.}.
The mixing matrices themselves are only real, if $\mu$ is chosen to be
real.
\begin{equation}
U^* \; \M \; V^{-1} = {\rm Diag}\left( \mnj \right) \qquad j=1,2 
\end{equation}

\subsubsection{Neutralino/Chargino Feynman Rules}

The Feynman rules for the neutralinos and charginos are given in
Fig.~\ref{fig_app_feynneut}. The $\sq q \tilde{\chi}$ vertex is given for
left and right squarks, the coupling to the mixing scalar top quark is
a superposition of both couplings, as it is for the gluino case.

\begin{table}[t] \begin{center} \begin{tabular}{|l|c|c|c|c|}
\hline 
\rule[-1mm]{0mm}{6mm}
& $A_L$
& $A_R$
& $B_L$
& $B_R$ \\ \hline
$\tilde{u} u \nnj$ \rule[-6mm]{0mm}{12mm}
& $\displaystyle g s_w Q \NY_{j1}+
     \frac{g}{c_w} \NY_{j2} \left( T_3-Q s_w^2 \right)$
& $A_L^* \; [T_3=0]$
& $\displaystyle \frac{g m_u}{2 \mw s_\beta} \NY^*_{j4}$ 
& $B_L^*$ \\
$\tilde{d} d \nnj$ \rule[-6mm]{0mm}{12mm}
& $\displaystyle g s_w Q \NY_{j1}+
     \frac{g}{c_w} \NY_{j2} \left( T_3-Q s_w^2 \right)$
& $A_L^* \; [T_3=0]$
& $\displaystyle \frac{g m_d}{2 \mw c_\beta} \NY^*_{j3}$ 
& $B_L^*$ \\
$\tilde{d} u \left( \cmj \right)_{\rm out}$ \rule[-6mm]{0mm}{12mm}
& $\displaystyle \frac{g}{\sqrt{2}} U_{j1}$ 
& $0$
& $\displaystyle -\frac{g m_u}{2 \mw s_\beta} V^*_{j2}$
& $\displaystyle -\frac{g m_d}{2 \mw c_\beta} U_{j2}$ \\
$\tilde{u} d \left( \cpj \right)_{\rm out}$ \rule[-6mm]{0mm}{12mm}
& $\displaystyle \frac{g}{\sqrt{2}} V_{j1}$ 
& $0$
& $\displaystyle -\frac{g m_d}{2 \mw c_\beta} U^*_{j2}$
& $\displaystyle -\frac{g m_u}{2 \mw s_\beta} V_{j2}$ \\
\hline 
\end{tabular} \end{center}
\caption[]{\it Couplings of neutralinos and charginos to squarks 
  and quarks as used in Fig.~\ref{fig_app_feynneut}. The mixing matrix
  $\NY$ is defined in the photino-zino basis.. 
  \label{tab_app_feynneut2}}
\end{table}

\section{$\mathbf{R}$ Parity breaking Squarks}
\label{chap_app_rpar}

The breaking of $R$ parity only adds new interaction terms to the
Lagrangean, eq.(\ref{eq_susy_rviol}), which are not related to any of
the Standard Model gauge symmetries. For any of these scenarios
including non-MSSM squark couplings, the QCD Feynman rules are still
completely fixed by the requirement, that any squark should be part of
the fundamental representation of SU(3)$_C$, \ie carries quark-type
color charge~\cite{leptoquark_intro}.

Under the simplifying assumption of one light squark flavor, one can
integrate out all heavy strongly interacting supersymmetric particles
and regard the $R$ parity violating scenario as the extension of the
Standard Model by one leptoquark-like squark. Another difference
between a scalar leptoquark model and the MSSM squark sector occurs:
The MSSM four-squark coupling originates from the $D$ terms in the
scalar potential eq.(\ref{eq_susy_scalarpot}). In the most general
effective model this term need not be present. The four-squark
coupling will then be proportional to the weak coupling constant, and
quadratic divergences of scalar masses occur.\smallskip

The coupling to any pair of Dirac and Majorana fermions can be
identified with the most general parameterization of the $\ste t \nnj$
coupling. However, the $q\sq e$ coupling constant is not fixed by any
gauge coupling, but results from the superpotential term
eq.(\ref{eq_susy_rviol})
\begin{equation}
{\cal L}_{\rm int} = \lambda' \bar{e} q \sq + {\rm h.c.}
\end{equation}
Since its Dirac structure is fixed by the helicity of the quark and
the electron, the coupling can be parameterized as $-i (A_L P_l + A_R
P_R)$. For unpolarized particles this yields $(A_L^2+A_R^2)$ in the
HERA production cross section and the decay width, likewise for
leading and next-to-leading order, this means the coupling completely
drops out of the next-to-leading order QCD $K$ factors.

Because the coupling $\lambda'$ connects two strongly interacting
particles, it has to be renormalized, \eg in $\msbar$.  Furthermore,
it runs in complete analogy to the running QCD coupling constant
\begin{equation}
{\lambda'}^2(\mu_R) 
= \frac{{\lambda'}^2(M)}
       {1+\frac{\alpha_s}{\pi}\log\frac{\mu_R^2}{M^2}} 
\end{equation}

\chapter{Radiative Corrections}
\label{chap_app_rad}

\section{Phase Space and Partonic Cross Sections}
\label{chap_app_phase}

\subsubsection[]{$\mathbf{(2\to1)}$ Production Cross Section}

Assume the production of one final state particle in NLO
\begin{equation}
q(k_1) + \bar{q}(k_2) \to X_1(p_1) [ + g(k_3) ]
\end{equation}
with massless partons $k_j [j=1,2,3]$ and a massive particle in the
final state $p_1^2=m^2$. The invariants are the usual Mandelstam
variables
\begin{equation}
s=2(k_1 k_2) \qquad t_1=2(k_1 k_3) \qquad u_1=2(k_2 k_3)
\end{equation}
These Mandelstam variables can be expressed in terms of the rescaled
gluon emission angle $\theta$
\begin{alignat}{9}
 y    =& \; \frac{1}{2} \left( 1 + c_\theta \right) \in [0,1] \notag \\
 t    =& \; - s ( 1 - \tau ) y \notag \\
 u    =& \; - s ( 1 - \tau ) ( 1 - y ) 
\end{alignat}
where $\tau = m^2/s$ and $c_\theta = \cos \theta$. The formula for the
cross section becomes~\cite{aem}:
\begin{equation}
s \frac{d \hat{\sigma}^R }{d y} = 
K_{ij} \frac{\pi (4 \pi)^{-2+\epsilon}}{\Gamma(1-\epsilon)}
       \left( \frac{m^2}{\mu^2} \right)^{-\epsilon}
       \frac{ \tau^\epsilon (1-\tau)^{1-2\epsilon}}
            { y^\epsilon (1-y)^\epsilon }
\sum \left| \M^R \right|^2 
\end{equation}
$K_{ij}$ are the spin-color averaging factors, for the HERA process
$K_{eq}=1/(4N)$. For the general squark production at HERA the
integration over the gluon emission angle can be performed
analytically. Poles in $\epsilon$ appearing in the soft/collinear
phase space region cancel with the divergences arising from the
virtual gluon contributions and with the mass factorization, \ie the
renormalization of the parton densities described in section
\ref{sect_neut_massfac}.

\subsubsection[]{$\mathbf{(1\to2)}$ Decay Widths}

The calculation of the different decays has been performed in four
dimensions. The general decay mode is
\begin{equation}
X_1(p_1) \to X_2(p_2) + X_3(p_3) [ + g(k) ]
\end{equation}
where all particles are assumed massive $p_j^2=m_j^2 [j=1,2,3]$ and
$k^2=\lambda^2$. The Born and  real gluon emission decay width is
\footnote{As usual $\Lambda(x,y,z)=x^2+y^2+z^2-2(xy+xz+yz)$}
\begin{alignat}{5}
 \Gamma^B =& \; K_i \frac{1}{16 \pi m_1} 
 \frac{\Lambda^{1/2}(m_1^2,m_2^2,m_3^2)}{m_1^2} 
 \sum \left| \M^B \right|^2 \notag \\
 \Gamma^R =& \; K_i \frac{1}{64 \pi m_1}
 \int \frac{d^3p_2}{2p_2^0} \frac{d^3p_3}{2p_3^0} \frac{d^3k}{2k^0}
      \frac{1}{\pi^2} \delta^4(p_1+p_2+p_3+k)
      \sum \left| \M^R \right|^2 
\end{alignat}
$K_{\sq} = 1/N$ is the spin-color averaging factor  for the decay
of the squark. The integrals
\begin{equation}
I^{i_1 \cdots i_n}_{j_1 \cdots j_m} =  
\frac{1}{\pi^2} 
 \int \frac{d^3p_2}{2p_2^0} \frac{d^3p_3}{2p_3^0} \frac{d^3k}{2k^0} \;
 \delta^4(p_1+p_2+p_3+k) \;
 \frac{2(kp_{i_1}) \cdots 2(kp_{i_n})}
      {2(kp_{j_1}) \cdots 2(kp_{j_m})}
\label{eq_app_dennerint}
\end{equation}
can be found in the literature~\cite{denner}\footnote{Note that in the
  formulae (D.11) and (D.12) of~\cite{denner} the indices of the
  $m_{0,1}^4$ term have to be interchanged}.

\subsubsection[]{$\mathbf{(2\to2)}$ Production Cross Section}

The partonic production process of two massive particles can be
written as
\begin{equation}
q(k_1) + \bar{q}(k_2) \to X_1(p_1) + X_2(p_2) [ + g(k_3) ]
\end{equation}
The $k_j [j=1,2,3]$ are assumed massless, the $p_j [j=1,2]$ will in
general have different masses $m_1,m_2$. Replacing quarks by gluons
and vice versa does not have any effect on the kinematics. We
introduce the invariants:
\begin{alignat}{9}
  s  =& \; 2 (k_1 k_2) \qquad \notag \\ 
  t_1=& \; 2 (k_1 p_1) \qquad 
& u_1=& \; 2 (k_2 p_1) \qquad
& s_4=& \; 2 (k_3 p_1) \notag \\
  t_2=& \; 2 (k_2 p_2) \qquad 
& u_2=& \; 2 (k_1 p_2) \qquad
& s_3=& \; 2 (k_3 p_2) \notag \\
  t' =& \; 2 (k_2 k_3) \qquad
& u' =& \; 2 (k_1 k_3)  
& s_5=& \; 2 (p_1 p_2) + m_1^2 + m_2^2 
\end{alignat}
All momenta are chosen incoming: $k_1+k_2+k_3+p_1+p_2=0$. Only five
invariants are independent:
\begin{alignat}{5}
s_3 =& \; -s_5 - t_2 - u_2 - m_2^2 + m_1^2 \notag \\
s_4 =& \; -s_5 - t_1 - u_1 - m_1^2 + m_2^2 \notag \\
s_5 =& \; +s + t' + u' \notag \\
u'  =& \; - s - t_1 - u_2 \notag \\
t'  =& \; - s - t_2 - u_1  
\end{alignat}
From the relation $s_4=s+t_2+u_2-m_1^2+m_2^2$ one can see that the
limit $s_3,s_4 \to 0$ is equivalent to the Born kinematics. The
differential cross section in the Born approximation therefore
includes a factor $\delta(s_4)$:
\begin{equation}
s^2 \frac{d^2 \hat{\sigma}^B}{dt_2 ds_4} =
K_{ij} \frac{\pi (4 \pi)^{-2+\epsilon}}{\Gamma(1-\epsilon)}
\left( \frac{t_2 u_2 - s m_2^2}{s \mu^2} \right)^{-\epsilon}
\delta(s_4)
\sum \left| \M^B \right|^2  
\end{equation}
with the $n$ dimensional spin-color averaging factors for different
strongly interacting  incoming states 
\begin{equation}
K_{q\bar{q}}=\frac{1}{4 N^2} \qquad
K_{gg}=\frac{1}{4(1-\epsilon)^2(N^2-1)^2} \qquad
K_{qg}=K_{g\bar{q}}=\frac{1}{4(1-\epsilon)N(N^2-1)}
\end{equation}

The matrix element for the real gluon emission corresponds to a
differential cross section in four variables. 
\begin{equation}
s^2 \frac{d^2 \hat{\sigma}^R}{dt_2 ds_4 } =
K_{ij} \frac{\mu^{2\epsilon} (4 \pi)^{-4+2\epsilon}}
            {2 \Gamma(1-2\epsilon)}
\left( \frac{t_2 u_2 - s m_2^2}{s \mu^2} \right)^{-\epsilon}
\frac{s_4^{1-2\epsilon}}{(s_4+m_1^2)^{1-\epsilon}}
\int d\Omega \sum \left| \M^R \right|^2  
\end{equation}
The total partonic  cross section is defined as 
\begin{alignat}{9}
\hat{\sigma} =& \; \int_{t_2^{\rm min}}^{t_2^{\rm max}}
                   \int_0^{s_4^{\rm max}}
\frac{d^2 \hat{\sigma}}{dt_2 ds_4}& \qquad \qquad
&  s_4^{\rm max} =
 \; s + t_2 + m_2^2 - m_1^2 + \frac{s m_2^2}{t_2} \notag \\
&&&t_2^{\rm min/max} =
 -\frac{s+m_2^2-m_1^2 \pm \Lambda^{1/2}(s,m_1^2,m_2^2)}{2}
\end{alignat}

To integrate over the angle of the final state gluon analytically, one
chooses different parameterizations of the phase space. The angular
integration is performed in the center-of-mass frame of $p_1$ and
$k_3$. One of  the three dimensional components of $k_2,k_1$ or
$p_2$ is taken parallel to the $z$ axis. In the first case the momenta
are~\cite{bible}:
\begin{alignat}{7}
k_1 =& \; \left( -w_1,...,0,-p s_\psi,-p c_\psi+w_2 \right) \notag \\
k_2 =& \; \left( -w_2,...,0,0,-w_2 \right) \notag \\
k_3 =& \; \left( w_3,...,w_3s_1s_2,w_3s_1c_2,w_3c_1 \right) \notag \\
p_1 =& \; \left( E_1,...,-w_3s_1s_2,-w_3s_1c_2,-w_3c_1 \right) \notag \\
p_2 =& \; \left( E_2,...,0,p s_\psi, p c_\psi \right) 
\end{alignat}
$s_i,c_i$ are defined as the sine and cosine of the angles
$\theta_1,\theta_2,\psi$. The angles $\theta_j$ are connected to the
angular integration for the additional gluon:
\begin{equation}
\int d \Omega = \int_0^\pi d \theta_1 s^{n-3}_1
                \int_0^\pi d \theta_2 s^{n-4}_2
\end{equation} 
The non-invariant variables can be expressed in the invariants:
\begin{alignat}{9}
 w_1 =& \; \frac{s+u_2}{2\sqrt{s_4+m_1^2}} \qquad 
&w_2 =& \; \frac{s+t_2}{2\sqrt{s_4+m_1^2}} \qquad
&w_3 =& \; \frac{s_4}{2\sqrt{s_4+m_1^2}} \notag \\
 E_1 =& \; \frac{s_4+2m_1^2}{2\sqrt{s_4+m_1^2}} \qquad
&E_2 =& \; \frac{t_2+u_2+2m_2^2}{2\sqrt{s_4+m_1^2}} \notag \\
 p   =& \; \frac{\sqrt{(t_2+u_2)^2-4sm_2^2}}{2\sqrt{s_4+m_1^2}} \qquad
&c_\psi =& \; \frac{t_2(s_4-m_2^2+m_1^2)-s(u_2+2m_2^2)}
                   {(s+t_2)\sqrt{(t_2+u_2)^2-4sm_2^2}}
\end{alignat}
Inserting these equalities into the matrix element after partial
fractioning leads to the typical integral~\cite{bible}
\begin{equation}
\int d \Omega \; \frac{1}{(a+bc_1)^k} \frac{1}{(A+Bc_1+Cs_1c_2)^l}
\qquad k,l \, \epsilon \, {\mathbb N}
\label{eq_app_angular}
\end{equation}
Note that it is not necessary to perform this integration
analytically, if the divergent regions of phase space are regularized
\eg by the subtraction method~\cite{subtraction}.

\subsubsection{Cut-off Method} 
\label{chap_app_cutoff}

The radiation of on-shell gluons leads to divergences in the phase
space integration in two limiting cases: (i) the gluon is soft, all
components of its four vector vanish $k_\mu \to 0 \, [\mu=0,1,2,3]$;
(ii) the gluon is collinear to another massless particle with the
momentum $p$, \ie $k \sim p+k_\perp$ with $k_\perp \to 0$. In both
cases the invariant vanishes, $(pk)\to 0$, leading to infrared
divergences in the matrix element. In the limit of a soft gluon the
invariants for the three body process approach
\begin{alignat}{9}
  s_3 \to& \; 0 \qquad 
& t'  \to& \; 0 \qquad
& t_2 \to& \; t_1 + m_1^2 - m_2^2 \notag \\
  s_4 \to& \; 0 \qquad 
& u'  \to& \; 0 \qquad
& u_2 \to& \; u_1 + m_1^2 - m_2^2 \qquad
& s_5 \to& \; s 
\end{alignat}
The integration of real gluon matrix element is split into two
regions, corresponding to soft and hard gluons~\cite{bible}:
\begin{alignat}{9}
\int_0 ds_4 \frac{d^2\hat{\sigma}}{dt_2 ds_4} 
&= 
 \left( \int_0^\Delta + \int_\Delta \right)
   ds_4 \frac{d^2\hat{\sigma}}{dt_2 ds_4} \notag \\
&=
 \int_0^\Delta ds_4 \frac{d^2\hat{\sigma}}{dt_2 ds_4}\Bigg|_{\rm approx}
 + \int_\Delta ds_4 \frac{d^2\hat{\sigma}}{dt_2 ds_4}
\end{alignat}
In the second integrand the limit $\Delta \to 0$ has to be checked
numerically.  The soft gluon matrix element in the first integrand is
evaluated in the eikonal approximation, where the gluon momentum is
neglected compared to any other variable [$\delta(s_4)$].  The angular
integral is analytically evaluated using the angular integrals
eq.(\ref{eq_app_angular}). In addition to soft poles in $\epsilon$
logarithms $\log^j \Delta [j=1,2]$ appear. These have to be added to
the final expression. To cancel the $\Delta$ dependence of the hard
gluon part, these terms are rewritten as
\begin{alignat}{9}
\int_0^{s_4^{\rm max}} ds_4 \; 
  \log \left( \frac{\Delta}{\mu^2} \right) \; \delta(s_4) =& \; 
\int_\Delta^{s_4^{\rm max}} ds_4 \; 
  \left[ \; \frac{\log (s_4^{\rm max}/\mu^2)}{s_4^{\rm max}-\Delta}
           -\frac{1}{s_4} \; \right] \notag \\
\int_0^{s_4^{\rm max}} ds_4 \; 
  \log^2 \left( \frac{\Delta}{\mu^2} \right) \; \delta(s_4) =& \; 
\int_\Delta^{s_4^{\rm max}} ds_4 \; 
  \left[ \; \frac{\log^2 (s_4^{\rm max}/\mu^2)}{s_4^{\rm max}-\Delta}
           -\frac{2 \log(s_4/\mu^2)}{s_4} \; \right] 
\end{alignat}

In addition to the soft singularities in the gluon emission matrix
element collinear poles arise in next-to-leading order. As long as the
final state particles are massive, these have to be absorbed
completely into the renormalization of the parton densities, as
described in chapter \ref{sect_neut_massfac}. The additional terms are
of the form
\begin{equation}
s^2 \frac{d^2\hat{\sigma}^{MF}_{ij}}{dt_2 ds_4} \simeq
\frac{\alpha_s}{2\pi} 
\left[ \frac{1}{\epsilon} - \gamma_E
       + \log \left( \frac{4\pi\mu^2}{Q_F^2} \right) \right]
\int_0^1 \frac{dx}{x} \,
P_{li}(x) \; \left( s^2 \frac{d^2 \hat{\sigma}^B_{lj}}{dt_2 ds_4}
             \right)_{x k_l} 
\end{equation}
The splitting function $P_{ij}$ describes an incoming parton $i$
changing to $j$\footnote{The color factors for the SU(3) are
  $N=3,C_A=3,C_F=4/3,T_R=1/2$}
\begin{alignat}{9}
P_{gg}(x) =& \; 
 2 C_A \left[ \; 
       \left( \frac{x}{1-x} \right)_+ + \frac{1-x}{x} -1 + x(1-x) 
    \; \right] 
 + \delta(1-x) \left[ \; \frac{11}{6} C_A - \frac{2}{3} T_R n_f 
            \; \right] \notag \\
P_{gq}(x) =& \; C_F \frac{1+(1-x)^2}{x} \notag \\
P_{qg}(x) =& \; T_R \left( x^2+(1-x)^2 \right) \notag \\
P_{qq}(x) =& \; C_F \left( \frac{1+x^2}{1-x} \right)_+
\label{eq_app_ap}
\end{alignat}
with the + distribution defined as
\begin{equation}
\left( F(x) \right)_+ = F(x) \; \theta(1-x-\beta)
 - \delta(1-x-\beta) \int_0^{1-\beta} d\xi F(\xi)
\label{eq_app_plus}
\end{equation}
in the limit $\beta \to 0$; the parameter $\beta$ separates soft from
hard gluons. Numerically the + distributions can be evaluated
following
\begin{equation}
\int_\tau^{1} dx 
  \left[ f(x) \left( L(x) \right)_+ + g(x) \right] = 
 \int_\tau^{1-\beta} dx 
  \left[ f(x) L(x) + g(x) \right] 
-f(1) \int_\tau^{1-\beta} dx 
  \left[ L(x) + \frac{\bar{L}}{1-\tau} \right] \; 
\end{equation}
where $\bar{L}=\int_0^\tau dx L(x)$ has to be calculated analytically
and $\beta \to 0$. The second part of the + distribution
eq.(\ref{eq_app_plus}) contributes to the soft gluon part of the
splitting function. After performing the integration over the momentum
shift $x$ using the $\delta(1-x)$ term, the result follows the Born
kinematics and cancels the collinear divergences of the virtual
corrections. The separating parameter can be linked to the cut-off
$\Delta$ via $\beta=\Delta/(s+t_2)$ and $\beta=\Delta/(s+u_2)$ for a
shifted momentum $k_2$ or $k_1$.

For hard gluons the four momentum conservation $\delta(s_4)$ included
in the Born differential cross section can be used to remove the
integration over the shift in the four momentum, \ie the mass
factorization term cancels the collinear divergences arising from the
virtual and from the real corrections.

\subsubsection{Subtraction Method}
\label{chap_app_subtract}

The subtraction method~\cite{subtraction} is used to calculate the
neutralino production cross section, \ie there are no massive strongly
interacting  particles present in the final state and $q\bar{q}$
is the only incoming state.\smallskip 

Given a divergent real gluon matrix element, a subtraction matrix
element $\sigma^{A,3}$ is constructed for the three particle final
state, in order to remove the soft and collinear singularities
point-wise from the gluon emission phase space.
\begin{equation}
s^2 \frac{d^2\hat{\sigma}^R}{dt_2 ds_4} 
- s^2 \frac{d^2\hat{\sigma}^{A,3}}{dt_2 ds_4} \; \sim \;
\sum \left| \M^R \right|^2 - \D^{13,2} - \D^{23,1} 
\end{equation}
In case of only two initial state partons $k_1,k_2$ and one
emitted gluon $k_3$ the dipole terms
\begin{alignat}{9}
\D^{13,2} =& \; \frac{1}{2 x (k_1k_3)} \; V^{13,2}(x) \; 
 \sum \left| \M^B \right|^2_{\rm subst 1} \notag \\
\D^{23,1} =& \; \frac{1}{2 x (k_2k_3)} \; V^{23,1}(x) \; 
 \sum \left| \M^B \right|^2_{\rm subst 2} 
\end{alignat}
have to be added to $\sum \left| \M^R \right|^2$.  The rescaling
factor for the split incoming momentum is given as
$x=(k_1k_2+k_3k_1+k_3k_2)/k_1k_2$ and re-defines the kinematics of the
Born matrix element $k'_i,p'_j$ \eg for $\D^{13,2}$
\begin{alignat}{9}
k'_1 =& \; x \; k_1 \notag \\
k'_2 =& \; k_2 \notag \\ 
p'_j =& \; p_j - \frac{2p_jK+2p_jK'}{(K+K')^2} \; (K+K')
               + \frac{2p_jK}{K^2} \; K
\end{alignat}
with $K=k_1+k_2+k_3$ and $K'=k'_1+k'_2$.  The result for $\D^{23,1}$
is obtained by interchanging the incoming quarks. The kernels
$V^{13,2}$ and $V^{23,1}$ are of the form
\begin{equation}
V^{qg,q}(x) 
= 8 \pi \mu^{2\epsilon} \alpha_s P_{qq}(x)
= 8 \pi \mu^{2\epsilon} \alpha_s C_F
  \left[ \; \frac{2}{1-x} - (1+x) + \epsilon (1-x) \; \right]
\end{equation}
Since this subtracted real gluon emission matrix element is finite, $d
\hat{\sigma}^R$ and $d\hat{\sigma}^{A,3}$ can be evaluated in $n=4$
dimensions. However, this does not hold for the virtual corrections
matrix elements, where one has to use the dipole terms
$d\hat{\sigma}^{A,2}$, which arise from the exact integration of the
subtraction term $d\hat{\sigma}^{A,3}$ over the additional gluon
emission phase space. This results in soft and collinear poles in
$\epsilon$, therefore the dipole moments for the calculation of the
virtual subtraction term have to be evaluated in $n$ dimensions. The
integrated soft and collinear gluon subtraction yields
\begin{equation} 
s^2 \frac{d\hat{\sigma}^{A,2}}{dt_2} = \int_0^1 dx \; I(x,\epsilon) \; 
\left[ \; \left( s^2 \frac{d\hat{\sigma}^B}{dt_2} \right)_{x k_1}
        + \left( s^2 \frac{d\hat{\sigma}^B}{dt_2} \right)_{x k_2}
\right]
\end{equation}
The integration of $V^{qg,q}(x)$ and the mass factorization leads to
the integration kernels for quark-antiquark and quark-gluon incoming
states:
\begin{alignat}{9}
I_{qq}(x,\epsilon) = \; \frac{\alpha_s}{2 \pi} \Bigg[&
\delta(1-x) \; \frac{C_F}{\Gamma(1-\epsilon)} 
 \left( \frac{4\pi\mu^2}{s} \right)^\epsilon 
 \left( \; \frac{1}{\epsilon^2}+\frac{3}{2\epsilon}+5-3\zeta_2 \;\right)
\notag \\
& - P_{qq}(x) \log \left( \frac{Q_F^2}{xs} \right)
  + C_F \left( 4 \left( \frac{\log(1-x)}{1-x} \right)_+
              - 2 (1-x) \log(1-x) + 1-x \right)  
\Bigg] \notag \\
I_{qg}(x) = \; \frac{\alpha_s}{2 \pi} \Bigg[&
 - P_{qg} \log \left( \frac{Q_F^2}{(1-x)^2s} \right)
 + 2 \, T_R \, x (1-x) \Bigg]
\end{alignat}

Since using the subtraction method the gluon emission angles are
integrated numerically the mass factorization terms are not added
analytically to the real gluon matrix element. Instead of using the
$\delta(s_4)$ distribution to remove the integration over the momentum
shift $x$, it is kept for the phase space integration. The convolution
is then performed numerically. The separation into virtual-soft and
hard gluon contributions is by no means unique.

\section{Hadronic and Differential Cross Sections}

For a general hadronic cross section the partons are treated as parts
of a massive hadron $h_j [j=1,2]$. The hadronic variables are referred
to as capital letters.
\begin{equation}
h_1(K_1) + h_2(K_2) \to X_1(p_1) + X_2(p_2) [ + g(k_3) ]
\end{equation}
 The additional gluon $k_3$ is again assumed massless, the $p_j
[j=1,2]$ will in general have different masses $m_1,m_2$. We introduce
the invariants:
\begin{alignat}{9}
  S  =& \; 2 (K_1 K_2) \qquad 
& T_1=& \; 2 (K_1 p_1) \qquad 
& U_1=& \; 2 (k_2 p_1) \notag \\
&&T_2=& \; 2 (K_2 p_2) \qquad 
& U_2=& \; 2 (K_1 p_2) 
\end{alignat}
All momenta are chosen incoming. After integration of the gluon
emission angles the real emission matrix element together with the
parton densities is a differential cross section with respect to four
variables including some general function $F$:  
\begin{alignat}{9}
\sigma_{\rm tot} = \;
 \int_{x_1^{\rm min}}^1 dx_1 \int_{x_2^{\rm min}}^1 dx_2
 \int_{t_2^{\rm min}}^{t_2^{\rm max}} dt_2 &\int_0^{s_4^{\rm max}} ds_4
 \; F \notag \\
&x_1^{\rm min} =
 \; \frac{(m_1+m_2)^2}{4S} \notag \\
&x_2^{\rm min} = 
 \; \frac{(m_1+m_2)^2}{4Sx_1} \notag \\
&s_4^{\rm max} =
 \; s + t_2 + m_2^2 - m_1^2 + \frac{s m_2^2}{t_2} \notag \\
&t_2^{\rm min/max} =
 -\frac{s+m_2^2-m_1^2 \pm \Lambda^{1/2}(s,m_1^2,m_2^2)}{2}
\end{alignat}
This can be rewritten for the on-shell subtraction in the $s_4$
channel, when the final state particle $X_1$ is produced via decay of
one on-shell squark~\cite{roland}:
\begin{alignat}{9}
\sigma_{\rm tot} = \;
 \int_{x_1^{\rm min}}^1 dx_1 &\int_{x_2^{\rm min}}^1 dx_2 
 \int_0^{s_4^{\rm max}} ds_4 \int_{t_2^{\rm min}}^{t_2^{\rm max}} dt_2 
 \; F \notag \\
&s_4^{\rm max} =
 \; s + m_2^2 - m_1^2 - 2 \sqrt{s m_2^2} \notag \\
&t_2^{\rm min/max} =
 -\frac{s-s_4+m_2^2-m_1^2 
  \pm \sqrt{(s-s_4+m_2^2-m_1^2)^2-4sm_2^2}}{2} 
\end{alignat}
The integration variables $x_1,x_2$ are only used to compute the total
hadronic cross section. Therefore the subtraction in $s_3$ involving
the particle $X_2$ can be obtained by exchanging $p_1 \leftrightarrow
p_2$ in the subtraction matrix element and in the phase space. This
holds only if the gluon emission angle $d \Omega$ is integrated out
completely.\smallskip 

To be able to compute differential cross sections with respect to the
transverse momentum of one of the final state particles or the
rapidity
\begin{equation}
p_T^2 = \frac{T_2U_2-Sm_2^2}{S} = \frac{t_2u_2-sm_2^2}{s} \qquad \qquad
y = \frac{1}{2} \log \left( \frac{T_2}{U_2} \right)
\end{equation}
the integration has to be ordered differently
\begin{alignat}{9}
\sigma_{\rm tot} = \;
 \int_0^{p_T^{\rm max}} dp_T \int_{-y^{\rm max}}^{y^{\rm max}} dy
 \int_0^{s_4^{\rm max}} ds_4 &\int_{x_1^{\rm min}}^1 dx_1
 \; 2p_TS \; \frac{x_1 x_2}{x_1S+T_2} \; F \notag \\
&p_T^{\rm max} =
 \; \frac{1}{2\sqrt{S}} \; \Lambda^{1/2}(S,m_1^2,m_2^2) \notag \\
&y^{\rm max} = 
 \; {\rm arcosh}\left( \frac{S+m_2^2-m_1^2}{2\sqrt{S(p_T^2+m_2^2)}}
                \right) \notag \\
&s_4^{\rm max} =
 \; S + T_2 + U_2 + m_2^2 - m_1^2 \notag \\
&x_1^{\rm min} =
 \frac{s_4-T_2-m_2^2+m_1^2}{S+U_2}
\label{eq_app_os_sub1}
\end{alignat}
with $x_2=(s_4-x_1U_2-m_2^2+m_1^2)/(x_1S+T_2)$.

The subtraction of differential cross sections in the $s_3$ channel
requires a substitution of one of the two angular integrations in $d
\Omega$~\cite{roland}, which is implicitly included in
eq.(\ref{eq_app_os_sub1}).  We refer to the integration boundaries 
therein as $s_4^*,x_1^*$.
\begin{alignat}{9}
\int_0^{s_4^*} ds_4 \int_{x_1^*}^1 dx_1 
\int d \Omega 
&= \int_0^{s_4^*} ds_4 \int_{x_1^*}^1 dx_1 
   \int_0^\pi d\theta_1 \int_0^\pi d\theta_2 s_1 \notag \\
&= \int_0^{s_3^{\rm max}} ds_3 
   \int_{x_1^{\rm min}}^1 dx_1 
   \int_{s_4^{\rm min}}^{s_4^{\rm max}} ds_4
   \int_0^\pi d\theta_2 
   \frac{2(s_4+m_1^2)}{s_4\sqrt{s-s_4-m_1^2+m_2^2}} 
\end{alignat}
The integration borders are 
\begin{alignat}{9}
s_3^{\rm max} &=&& \frac{S+T_2+U_2+m_2^2-m_1^2}{2(S+T_2+U_2+m_2^2)}
                   \left( -T_2-U_2-2m_2^2+\sqrt{(T_2+U_2)^2-4m_2^2S}
                   \right) \notag \\
x_1^{\rm min} &=&& \frac{1}{2(S+U_2)(m_2^2S+m_2^2U_2+s_3U_2)}
    \Bigg[ -2m_2^4S + 2m_2^2m_1^2S - 2m_2^2Ss_3 - Ss_3^2 -2m_2^2ST_2 
      \notag \\ &&&
           - Ss_3T_2 - 2m_2^4U_2 + m_2^2m_1^2U_2 - 3m_2^2s_3U_2 
           + m_1^2s_3U_2 - s_3^2U_2 - 2m_2^2T_2U_2 - 2s_3T_2U_2
      \notag \\ &&& 
   - s_3 \Big( 4m_2^4S^2 - 4m_2^2m_1^2S^2 + 4m_2^2S^2s_3 + S^2s_3^2
             + 4m_2^2S^2T_2 + 2S^2s_3T_2 + S^2T_2^2 + 4m_2^4SU_2
      \notag \\ &&& \qquad 
             - 4m_2^2m_1^2SU_2 + 6m_2^2Ss_3U_2 - 2m_1^2Ss_3U_2 
             + sSs_3^2U_2 + 2m_2^2ST_2U_2 + 2m_1^2ST_2U_2 
      \notag \\ &&& \qquad 
             + 2Ss_3T_2U_2
             + m_2^4U_2^2 - 2m_2^2m_1^2U_2^2 + m_1^4U_2^2 
             + 2m_2^2s_3U_2^2 - 2m_1^2s_3U_2^2 + s_3^2U_2^2 \Big)^{1/2}
    \Bigg] \notag \\
s_4^{\rm min/max} &=&& \frac{s_3}{2(m_2^2x_1S + m_2^2T_2 + s_3T_2)}
    \Bigg[ - m_2^2T_2 - m_1^2T_2 - s_3T_2 - 2m_2^2x_1S - x_1Ss_3 
           - x_1^2SU_2 \notag \\ &&&
    \mp \Big( ( - m_2^2T_2 - m_1^2T_2 - s_3T_2 - 2m_2^2x_1S 
                - x_1Ss_3 - x_1^2SU_2 )^2
     \notag \\ &&& \qquad 
                - 4m_1^2(x_1S+T_2)
                        (m_2^2x_1S + m_2^2T_2 + s_3T_2) \Big)^{1/2} 
    \Bigg] \quad \text{for} \; s_3 \le s_3^* \notag \\
s_4^{\rm max} &=&& \; x_1(S+U_2) + T_2 + m_2^2 - m_1^2 
    \hspace{4cm} \text{for} \; s_3 > s_3^* \notag \\ 
s_3^* &=&& \frac{x_1S + T_2 + x_1U_2 + m_2^2 - m_1^2}
                {2(x_1S + T_2 + x_1U_2 + m_2^2)}
    \Big( - T_2 - x_1U_2 - 2m_2^2 - 
          \sqrt{(T_2+x_1U_2)^2-4m_2^2x_1S} \Big) \notag \\
\end{alignat}

\section{Scalar Integrals}
\label{chap_app_virt}

\subsubsection{Dimensional Regularization}

The analytical expressions for the virtual corrections contain scalar
integrals which are multiplied by polynomials including the Mandelstam
variables. The scalar integrals $A,B,C,D$ in case of dimensional
regularization~\cite{dim_reg,pass_velt} are defined in $n=4-2
\epsilon$ dimensions and are used to calculate the production
processes of neutralinos/charginos and stops.
\begin{alignat}{5}
 C(\{p_i\};\{m_j\})
&= \int \frac{d^nq}{i (2\pi)^n}
    \frac{\mu^{4-n}}{[q^2-m_1^2][(q+p_1)^2-m_2^2]
             [(q+p_{12})^2-m_3^2]} \notag \\
 D(\{p_i\};\{m_j\})
&= \int \frac{d^nq}{i (2\pi)^n}
    \frac{\mu^{4-n}}{[q^2-m_1^2][(q+p_1)^2-m_2^2]
             [(q+p_{12})^2-m_3^2][(q+p_{123})^2-m_4^2]} \notag \\
 p_{ijk} &= p_i + p_j + p_k 
\label{eq_app_scalarint}
\end{alignat}
The definition of the one and two point functions $A,B$ follows the
same conventions.  Using this integration measure, the non-absorptive
virtual contributions are real and the integrals have got integer
dimension.  The infrared and collinear divergences occur as poles
$1/\epsilon^k\;[k=1,2]$~\cite{bible,roland}. $\mu$ is the
renormalization scale of the process.

As long as only CP conserving observables are calculated, the typical
combinations of couplings in front of the scalar integrals are real,
therefore only the real part of the scalar integrals is needed. If, as
described in appendix  \ref{chap_app_feynman}, we chose the
parameters in the Lagrangean as being complex we also need the
imaginary parts of the scalar integrals.

The expressions for finite integrals have been taken from the
literature~\cite{nierste,denner}. The divergent integrals have been
calculated using Cutkosky cut rules and dispersion
relations~\cite{wim}, and most of them are also present in the
literature~\cite{roland}.

A typical singular scalar three point function occuring in the
neutralino/chargino production cross section is:
\begin{alignat}{5} 
&C(p_1,k_1;m_x,0,0) = \frac{C_\epsilon}{t_1} \Bigg[  && 
- \frac{1}{\epsilon} \log \left( \frac{-t_x}{M_1^2} \right)
+ {\rm Li}_2 \left( \frac{t}{m_x^2} \right)
- {\rm Li}_2 \left( \frac{m_1^2}{m_x^2} \right)  \notag \\  &&&
+ \log^2 \left( \frac{-t_x}{m_x^2} \right)
- \log^2 \left( \frac{M_1^2}{m_x^2} \right) 
+ \log \left( \frac{m_x^2}{M^2}  \right) 
  \log \left( \frac{-t_x}{M_1^2} \right)
\Bigg]  \notag \\ 
&t_x = t-m_x^2 && \notag \\  
&M_i^2 = m_x^2 - m_i^2 && 
\end{alignat}
The definition of the momenta and Mandelstam variables follows
appendix~\ref{chap_app_phase}. The factor in front contains the
typical $\msbar$ terms
\begin{equation} 
C_\epsilon = \frac{1}{16 \pi^2} e^{-\epsilon \gamma_E}
             \left( \frac{4 \pi \mu^2}{M^2} \right)^\epsilon  
\end{equation} 
The scale $M$ is either chosen as the mass of the outgoing particle
or, for different masses in the final state, as the averaged mass. The
factor $C_\epsilon$ is a common factor of all virtual gluon
contributions and can be pulled out of all renormalization
contributions, \ie it occurs as an over-all factor in the regularized
and renormalized matrix element and can then be evaluated in the limit
$n=4$.

One scalar four point function is 
\begin{alignat}{5} 
&D(k_2,k_1,p_1;0,0,0,m_x) = \frac{C_\epsilon}{st_x} \Bigg[  && 
  \frac{1}{\epsilon^2}
- \frac{1}{\epsilon} 
        \left(  \log \left( \frac{-s}{M^2} \right)
              + \log \left( \frac{-t_x}{M_1^2} \right)
              + \log \left( \frac{-t_x}{M_2^2} \right) \right)
 \notag \\ &&&
- 2 {\rm Li}_2 \left( 1+\frac{M_1^2}{t_x} \right)
- 2 {\rm Li}_2 \left( 1+\frac{M_2^2}{t_x} \right)
- {\rm Li}_2 \left( 1+\frac{M_1^2 M_2^2}{s m_x^2} \right)
 \notag \\ &&&
- \log \left( 1+\frac{M_1^2 M_2^2}{s m_x^2} \right)
 \notag \\ &&& \quad
  \left[ \log \left( \frac{-M_1^2 M_2^2}{s m_x^2} \right)
       - \log \left( \frac{M_1^2}{M^2} \right)
       - \log \left( \frac{M_2^2}{M^2} \right)
       + \log \left( \frac{-s m_x^2}{M^4} \right) \right]
 \notag \\ &&&
+ \frac{1}{2} \log^2 \left( \frac{-s}{M^2} \right)
- \frac{1}{2} \log^2 \left( \frac{-s}{m_x^2} \right)
+ 2 \log \left( \frac{-s}{M^2} \right) 
    \log \left( \frac{-t_x}{m_x^2} \right) 
 \notag \\ &&&
- \log \left( \frac{M_1^2}{M^2} \right)
  \log \left( \frac{M_1^2}{m_x^2} \right) 
- \log \left( \frac{M_2^2}{M^2} \right)
  \log \left( \frac{M_2^2}{m_x^2} \right) 
\Bigg]
- \frac{3}{2} \zeta_2 
\end{alignat}
The $\log(1+M_1^2M_2^2/(sm_x^2))$ multiplied with the terms in
brackets originates from the analytical continuation of the
Dilogarithm~\cite{wim}. This method of calculating scalar integrals
omits the typical roots, which appear by using the Feynman
parameterization after partial fractioning. Thereby, large
cancelations are absent in our results, which improves the numerical
accuracy.

\subsubsection{Massive gluon regularization}

For the massive gluon regularization scheme, as it is used for the
calculation of the decay processes, the conventions concerning the
scalar integrals are exactly the same as for dimensional
regularization in the limit $n=4$. In this case we obtain divergences
in form of logarithms of the gluon mass $\log \lambda^2$~\cite{wim}.
For linear divergences the $\log \lambda^2$ description can naively be
translated into $1/\epsilon$, whereas for higher divergences this is
more involved.

For the gluino decay the divergent scalar integral 
\begin{alignat}{5}
&C(p_1,p_2;m_1,\lambda,m_2) = 
\frac{C_\epsilon x_s}{m_1 m_2 (1-x_s^2)} \Bigg[  && 
\log (x_s) \left[ - \log \left( \frac{\lambda^2}{m_1m_2} \right)
                  - \frac{1}{2} \log (x_s) 
                  + 2 \log \left( 1-x_s^2 \right) \right]
 \notag \\ &&&
+ {\rm Li}_2 \left( 1 - x_s \frac{m_1}{m_2} \right)
+ {\rm Li}_2 \left( 1 - x_s \frac{m_2}{m_1} \right)
+ {\rm Li}_2 \left( x_s^2 \right) 
 \notag \\ &&&
+ \frac{1}{2} \log^2 \left( \frac{m_1}{m_2} \right)
- \zeta_2 
\Bigg]
\label{eq_app_massgluon}
\end{alignat}
is used in the massive gluon regularization scheme. The factor
$C_\epsilon$ is taken in the limit $n=4$, and 
\begin{alignat}{5}
& p_1^2 \; =&& \; m_1^2 \qquad \qquad 
  p_2^2 \; = \; m_2^2 \notag \\
& x_s   \; =&& \; \frac{ \sqrt{1-\frac{4m_1m_2}{s-(m_1-m_2)^2}} -1}
                       { \sqrt{1-\frac{4m_1m_2}{s-(m_1-m_2)^2}} +1}
\end{alignat}
For the stop production the same integral is needed in the case of
equal masses and regularized dimensionally. There the divergent
logarithm has to be replaced by a linear pole $\log \lambda^2 \to
1/\epsilon + \log \mu_R^2$
\begin{alignat}{5}
&C(p_1,p_2;m_1,0,m_2) = 
\frac{C_\epsilon x_s}{m_1 m_2 (1-x_s^2)} \Bigg[  && 
\log (x_s) \left[ - \frac{1}{\epsilon}
                  - \log \left( \frac{M^2}{m_1m_2} \right)
                  - \frac{1}{2} \log (x_s) 
                  + 2 \log \left( 1-x_s^2 \right) \right]
 \notag \\ &&&
\cdots \Bigg]
\end{alignat}
The same relation holds for the other scalar integrals used
calculating stop decay widths. 

\section{Counter Terms}
\label{chap_app_counter}

Renormalization of the external masses in the on-shell scheme
preserves the Slavnov-Taylor identity~\cite{bible}
\begin{equation}
k_1^\mu \; \M_{\mu\nu} = \M_\nu^{\rm ghost} \propto k_{2 \nu}
\label{eq_app_ward}
\end{equation}
where $\M_{\mu\nu}$ is the matrix element for the production of two
massive SU(3) charged particles in gluon fusion and $\M_\nu^{\rm
  ghost}$ the ghost contribution, both of which are present in
LO and NLO. This identity has been used to calculate the ghost
contribution of the stop pair production, as described in chapter
\ref{sect_susy_stop}. The mass counter terms for a heavy quark $t$ and
a squark $\sq$ are
\begin{alignat}{9}
\mt^{(0)} = \; \mt \Bigg[ 1 + \frac{\alpha_s C_F}{4 \pi}
& \Big( -\frac{3}{\epsilon} + 3 \gamma_E - 3\log (4\pi)
        - 4 - 3 \log \frac{\mu_R^2}{\mt^2} 
 \Big) \Bigg] \notag \\
\ms^{(0)} = \; \ms  \Bigg[ 1 + \frac{\alpha_s C_F}{4 \pi}
& \Bigg( \Big( -\frac{1}{\epsilon} + \gamma_E - \log (4\pi)
               - \log \frac{\mu_R^2}{\ms^2} \Big)
        \frac{2 \mg^2}{\ms^2} \notag \\
& \quad - 1 - \frac{3 \mg^2}{\ms^2} 
        + \frac{\ms^2-2\mg^2}{\ms^2} \log \frac{\ms^2}{\mg^2}
        - \frac{(\ms^2-\mg^2)^2}{\ms^4}
          \log \left| \frac{\ms^2-\mg^2}{\mg^2} \right|
\Bigg) \Bigg] \notag \\
\end{alignat}
The strong coupling is renormalized in the $\msbar$ scheme 
\begin{alignat}{9}
g_s^{(0)} = \; g_s(\mu_R) \Bigg[ 1 + \frac{\alpha_s}{4\pi} 
&\Bigg( \Big( - \frac{1}{\epsilon} + \gamma_E - \log (4\pi) 
              + \log \frac{\mu_R^2}{M^2} \Big) 
        \frac{\beta_0}{2} \notag \\
& \quad
 - \frac{N}{3} \log \frac{\mg^2}{\mu_R^2}
 - \frac{n_f-1}{6} \log \frac{\ms^2}{\mu_R^2}
 - \frac{1}{12} \log \frac{\mse^2}{\mu_R^2}
 - \frac{1}{12} \log \frac{\msz^2}{\mu_R^2}
 - \frac{1}{3} \log \frac{\mt^2}{\mu_R^2}
 \Bigg) \Bigg] \notag \\
\text{where} \quad
\beta_0 = & \; \frac{11}{3} N - \frac{2}{3} N 
             - \frac{2}{3} n_f - \frac{1}{3} n_f 
\label{eq_app_coupren}
\end{alignat}
$\mu_R$ is the renormalization scale of the process, $n_f=6$ the
number of quark and squark flavors. The $\beta$ function gets
contributions from the gluons, the gluinos, the quarks, and the
squarks. For the MSSM the sign of this coefficient is the same as for
the Standard Model. To decouple the heavy particles from the running
of $\alpha_s$ the massive $\log m^2$ terms have to be subtracted \ie
the strong coupling constant is effectively evaluated as the usual low
energy Standard Model QCD coupling.

In addition to these Standard Model like renormalization constants,
which remove the UV divergences from the virtual correction matrix
element, finite counter terms have to be added to restore the
supersymmetric Ward identity, as described in chapter
\ref{sect_susy_ward}. These are no counter terms observables arising
from divergent vacuum fluctuations in field theory, \ie even the weak
coupling will contain a new counter term $\propto \alpha_s$ although
physically $G_F$ should stay unrenormalized in supersymmetric QCD.

\chapter{Analytical Results for the Stop Decay Width} 
\label{chap_app_dec}

As an example, the analytical results for the decay of a light stop
$\ste$ into top and gluino are given.  The decay width in
next-to-leading order may be split into the following components:
\begin{equation}
\Gamma_{\text{NLO}} = \Gamma_{\text{LO}} 
                    + \real \left( \Delta \Gamma_t 
                    + \Delta \Gamma_{\tilde{g}}
                    + \Delta \Gamma_{11}
                    + \Delta \Gamma_v 
                    + \Delta \Gamma_r
                    + \Delta \Gamma_c
                    + \Delta \Gamma_f                    
                    + \Delta \Gamma_{\text{dec}} \right)
\label{eq_app_dec}
\end{equation}
To allow for more compact expressions we first define a few short-hand
notations:
\begin{equation}
\mu_{abc} = m_a^2 + m_b^2 - m_c^2  \qquad
\xzt = \mt \mg \sin(2 \tmix) \qquad
\fac = \frac{\Lambda^{1/2}(\mse^2,\mg^2,\mt^2)}
             {16 \pi \mse^3 N} 
\end{equation}
where $a,b,c = \gt,t,j$ with $j$ representing $\stj$. The different
contributions defined in eq.(\ref{eq_app_dec}) are listed
below:\bigskip

lowest-order decay width\footnote{The Casimir invariant for the gauge
  group SU(3) is $C_F=4/3$}:
\begin{equation}
\Gamma_{\text{LO}} = 8 N C_F \pi\as \left( - \mugte + 2 \xzt \right) 
                   \,\fac
                   \equiv \fac \MB
\end{equation}

top self-energy contribution: 
\begin{alignat}{4}
   \Delta \Gamma_t &=&
        \frac{2 \fac C_F \pi \as \MB}{\mt^2} &\Big\{
            2 (1-\ee) A(\mt) 
          + 2 A(\mg)             
          - A(\mse)
          - A(\msz)                 \notag \\[-2mm] &&&
          + \muteg B(\pt;\mg,\mse) 
          + \mutzg B(\pt;\mg,\msz)          \notag \\ &&&
          - 4 \mt^2 \xzt \big[ \Bp(\pt;\mg,\mse)  
                           - \Bp(\pt;\mg,\msz) \big]  \notag \\ &&&
          + 2 \mt^2 \big[ \mugte \Bp(\pt;\mg,\mse) 
                    + \mugtz \Bp(\pt;\mg,\msz) 
                    - 4 \mt^2 \Bp(\pt;\ml,\mt) \big]  
          \Big\} \notag \\ 
       &+& \frac{16 \fac N C_F^2 \pi^2 \as^2}{\mt^2}
            \czt^2 & \; \mugte \Big\{
            A(\msz) 
          - A(\mse)   
          - \mugte B(\pt;\mg,\mse) 
          + \mugtz B(\pt;\mg,\msz) 
          \Big\} \notag \\
\end{alignat}
where the definitions of the scalar integrals are given in
eq.(\ref{eq_app_scalarint}). When renormalizing external masses in the
on-shell scheme the scalar function $\dot{B}(p;m_a,m_b) = \partial
B(p;m_a,m_b)/\partial p^2$ appears.

gluino self-energy contribution (for $n_f=6$ quark flavors):
\begin{alignat}{4}
   \Delta \Gamma_{\gt} &=
         \frac{4 \fac \pi \as \MB}{\mg^2}\, && \ntp \ntp \ntp
          (n_f\ntp -\ntp 1) \Big\{ \ntp
          - \ntp \ntp A(\ms)
          + ( \ms^2
            + \mg^2 ) B(\pg;\ms,0) 
          + 2 \mg^2 ( \mg^2 - \ms^2 ) \Bp(\pg;\ms,0) 
          \Big\} \notag \\ 
      &+ \frac{2 \fac \pi \as \MB }{\mg^2}\, && \ntp \ntp \ntp \Big\{
            2 A(\mt)
          - A(\mse)
          - A(\msz) 
          + \muegt B(\pg;\mse,\mt) 
          + \muzgt B(\pg;\msz,\mt) \notag \\[-2mm] &&&
          - 4 \mg^2 \xzt \big[ \Bp(\pg;,\mse,\mt)
          - \Bp(\pg;\msz,\mt) \big] \notag \\ &&&
          + 2 \mg^2 \big[ \mugte \Bp(\pg;\mse,\mt)  
                        + \mugtz \Bp(\pg;\msz,\mt) \big]
           \Big\} \notag \\ 
      &+ \frac{4 \fac N \pi \as \MB}{\mg^2} && \Big\{
            (1-\ee) A(\mg)
          - 4 \mg^4 \Bp(\pg;\ml,\mg) \Big\}
\end{alignat}

diagonal stop self-energy:
\begin{alignat}{4}
   \Delta \Gamma_{11} &=&
         8 C_F \fac \pi \as \MB \Big\{&
            B(\ps;\mg,\mt)
          - B(\ps;\ml,\mse) 
          + 2 \xzt \Bp(\ps;\mg,\mt) \phantom{PLATTZZ} \notag \\ &&&
          - \mugte \Bp(\ps;\mg,\mt) 
          - 2 \mse^2 \Bp(\ps;\ml,\mse) 
          \Big\}
\end{alignat}

[The off-diagonal mixing contribution 
\begin{equation}
   \Delta \Gamma_{12} =
        \frac{128 \fac N C_F^2 \pi^2 \as^2}{\mse^2-\msz^2}
        \, \xzt \, \czt^2 \, \left\{
          A(\msz) 
        - A(\mse) 
        + \frac{4 \mt^2 \mg^2}{\xzt}\, B(\ps;\mg,\mt) 
        \right\} 
\end{equation}
is absorbed into the renormalization of the mixing angle for 
$\tmix = \tmix(\mse^2)$, as described in chapter \ref{sect_susy_stop} ]

vertex corrections:
\begin{alignat}{4}
   \Delta \Gamma_v &=&
      64 \fac \pi^2 \as^2 N C_F^2 & 
       \left[ F^F_1 
            + \xzt F^F_2 
            + \xzt^2 F^F_3 \right] \notag \\ 
  &+& 32 \fac \pi^2 \as^2 N^2 C_F & 
       \left[ F^A_1 
            + \xzt F^A_2 
            + \xzt^2 F^A_3 \right]
   + 8 \fac \pi \as \MB & F^B
\end{alignat}
with:
\begin{alignat}{2}
   F^F_1 =&
            2 ( \mt^2 
              + \mg^2 ) B(\ps;\mg,\mt) 
          + ( \mse^2 
            + \mt^2
            + \mg^2 ) B(\ps;\ml,\mse) \notag \\ 
         +& 2 ( \mg^2 
              - \mse^2 ) B(\pt;\ml,\mt) 
          - 2 \mt^2 B(\pt;\mg,\msz) 
          - 4 \mg^2 B(\pg;\mt,\mse)  \notag \\ 
         +& 4 \mg^2 ( \mse^2 
                    - \mg^2 ) C(\ps,\pt;\mt,\mg,\mse) 
          + 2\mt^2 ( \mse^2 + \msz^2 - 2 \mt^2 ) C(\ps,\pt;\mt,\mg,\msz)
          \notag \\ 
   F^F_2 =&
          - 2 B(\ps;\ml,\mse)
          - 2 B(\pt;\ml,\mt) 
          - 4 B(\ps;\mg,\mt)
          + 2 B(\pt;\mg,\mse) \notag \\
         +& 4 B(\pg;\mt,\mse) 
          + 4 \mugte C(\ps,\pt;\mt,\mg,\mse) 
          + 2 (\mse^2-\msz^2) C(\ps,\pt;\mt,\mg,\msz)
          \notag \\
   F^F_3 =&
       \frac{1}{\mg^2 \mt^2} \Big\{
            2 \mt^2 [ B(\pt;\mg,\msz) 
                    - B(\pt;\mg,\mse) ] 
          + \mugte ( \mugte 
                   - 4 \mt^2 ) C(\ps,\pt;\mt,\mg,\mse) \notag \\[-2mm]
         & \qquad \quad
          - ( \mugte \mugtz 
            - 2 \mt^2 \mugte
            - 2 \mt^2 \mugtz ) C(\ps,\pt;\mt,\mg,\msz)  
          \Big\} \notag
\end{alignat}
\begin{alignat}{2}
   F^A_1 =&
          - 2 \ee \,\mugte B(\ps;\mg,\mt) 
          + 4 ( \mt^2 
              - \mse^2 ) B(\pg;\ml,\mg) \notag \\
         +& 2\mt^2 [ B(\pt;\mg,\msz) 
                  - B(\pt;\mg,\mse) ] 
          + 4 \mg^2 B(\pg;\mt,\mse) \notag \\
         +& 4 \mg^2 ( \mg^2 
                    - \mse^2 ) C(\ps,\pt;\mt,\mg,\mse) 
          - 2\mt^2 ( \mse^2 + \msz^2 - 2 \mt^2 ) C(\ps,\pt;\mt,\mg,\msz)
           \notag \\ 
   F^A_2 =&
            4 \ee \,B(\ps;\mg,\mt) 
          - 4 B(\pg;\ml,\mg)
          - 4 B(\pg;\mt,\mse) \notag \\
         -& 4 \mugte C(\ps,\pt;\mt,\mg,\mse)
          - 2 (\mse^2-\msz^2) C(\ps,\pt;\mt,\mg,\msz) 
          \notag \\ 
   F^A_3 =&
       \frac{1}{\mg^2 \mt^2} \Big\{
            2 \mt^2 [ B(\pt;\mg,\mse) 
                    - B(\pt;\mg,\msz) ] 
          + \mugte ( 4 \mt^2 
                   - \mugte ) C(\ps,\pt;\mt,\mg,\mse) \notag \\[-2mm]
         & \qquad \quad
          + ( \mugte \mugtz 
            - 2 \mt^2 \mugte
            - 2 \mt^2 \mugtz ) C(\ps,\pt;\mt,\mg,\msz) 
          \Big\} \notag \\ 
   F^B =&
          N \big[
            \muteg C(\ps,\pt;\mse,\ml,\mt) 
          - \mugte C(\ps,\pt;\mg,\mt,\ml) 
          - \muegt C(\ps,\pt;\ml,\mse,\mg) \big] \notag \\
         -& 2 C_F \muteg C(\ps,\pt;\mse,\ml,\mt)
\end{alignat}

corrections from real-gluon radiation:
\begin{alignat}{9} 
\Delta \Gamma_r &= 
         \frac{\as \MB}{4 \pi^2 \mse} \left[ 
                                       ( \mse^2   
                                       - \mt^2 ) I_{\ste \gt}  
                                       - \mt^2 I_{\ste t}  
                                       - \mg^2 I_{\gt \gt}  
                                       - I_{\gt}   \right] \notag \\
       &+ \frac{\as C_F \MB}{4 \pi^2 \mse N} \left[  
                                       {}- \mse^2 I_{\ste \ste}  
                                       - \mt^2 I_{t t}  
                                       + \muteg I_{\ste t}  
                                       + I_{\ste}  
                                       - I_t \right] 
       + \frac{\as^2 C_F^2}{\pi \mse} I^{\gt}_t  
       + \frac{\as^2 N C_F}{\pi \mse} I^{\ste}_{\gt} \notag \\
\end{alignat}

renormalization of the coupling constant as defined in
eq.(\ref{eq_app_coupren}):
\begin{equation}
\Delta \Gamma_c =-\frac{\fac \as \MB}{4 \pi}
                  \left[ \frac{1}{\ee} 
                       - \gamma_E 
                       + \log(4 \pi) 
                       - \log\left(\frac{\mu_R^2}{\mu^2}\right) \right] 
                                          \left( \frac{11}{3}N 
                                               - \frac{2}{3}N 
                                               - \frac{2}{3}n_f 
                                               - \frac{1}{3}n_f \right) 
\end{equation}

finite shift of the Yukawa coupling relative to the gauge coupling in
$\msbar$, as described in chapter \ref{sect_susy_ward}:
\begin{equation}
\Delta \Gamma_f = \frac{\fac \as \MB}{4 \pi}
                 \left( \frac{4}{3} N - C_F \right)
\end{equation}

decoupling of the heavy flavors from the running strong coupling
constant:
\begin{alignat}{4}
\Delta \Gamma_{\text{dec}} &=& \frac{\fac \as}{\pi} \MB \left\{ 
      \vphantom{\log\left(\frac{\mu_R^2}{\ms^2}\right)} \right.&
      \frac{n_f-1}{12} \log\left(\frac{\mu_R^2}{\ms^2}\right) 
    + \frac{1}{24} \log\left(\frac{\mu_R^2}{\mse^2}\right)   
    + \frac{1}{24} \log\left(\frac{\mu_R^2}{\msz^2}\right) 
    \phantom{ Roland ist albern !! } \notag \\ &&& \left. \! 
    + \frac{1}{6} \log\left(\frac{\mu_R^2}{\mt^2}\right)    
    + \frac{N}{6} \log\left(\frac{\mu_R^2}{\mg^2}\right) 
    \right\} 
\end{alignat}

\end{appendix}

\bibliographystyle{plain}

\begin{thebibliography}{99}
\bibitem{spythia}
  S.~Mrenna 
   ANL-HEP-PR-96-63, \hep{9609360}.\footnote{This implementation 
   of the formulae given 
   in Ref.~\cite{drees_martin}, as it is used in this work, 
   is not correct: The masses of all strongly interacting
   particles are underestimated by $\lesssim 10 \%$, and the 
   stop mixing angle is given with a wrong sign. The
   numerical changes do not affect the phenomenological analysis.}
\end{thebibliography}

\end{document}